\documentclass[12pt,a4paper]{article}
\usepackage{amsmath}
\usepackage{graphicx}
\usepackage{cite}
\usepackage{amsfonts}
\usepackage{amssymb}

\usepackage[cp1251]{inputenc}
\usepackage[english,russian]{babel}

\usepackage{cite}
\usepackage{myart}

\begin{document}

\selectlanguage{english}

\title{\textbf{Polyakov's String: Twenty Five Years After}}
\author{Edited by A.Belavin and Al.Zamolodchikov}
\maketitle

\begin{center}
{\huge \textbf{Proceedings}}

\vspace{1.0cm}

{\Large \textbf{International Workshop} }

\vspace{0.4cm}

{\Large Chernogolovka, June 23--25, 2005}
\end{center}

\vspace{1.0cm}

{\large \textbf{Table of contents}}

\vspace{1.0cm}

\begin{itemize}
\item {\large Preface}\hfill
{page \phantom{0}1 }

\item \textit{{\large Valery Pokrovsky}}\newline {\large Hydden Sasha
Polyakov's life}\newline in Statistical and Condensed Matter Physics\hfill
{page \phantom{0}2 }

\item {\large \textit{Alexander Belavin} and \textit{Alexei Zamolodchikov}%
}\newline {\large Moduli Integrals, Ground Ring and Four-Point Function in
\newline Minimal Liouville Gravity}\hfill{page 16}

\item {\large \textit{Yukitaka Ishimoto} and \textit{Alexei Zamolodchikov}%
}\newline {\large Massive Majorana Fermion Coupled to 2D Gravity and Random
\newline Lattice Ising Model}\hfill{page 47}

\end{itemize}

\setcounter{page}{0}
\setcounter{section}{0}

\newpage
\selectlanguage{russian}
\hbox{}
\vspace{-2.5cm}
\begin{center}
{\huge Предисловие}
\end{center}

В Июне 2005 года в Черноголовке состоялась Международная
Конференция, посвя\-щенная ``Струне Полякова''\footnote{В наши дни,
естественно, нет необходимости пояснять содержание этого термина.}.
По идее оргкомитета, кроме  25-летней годовщины появле\-ния Струны
Полякова \cite{string}, эта конференция была приурочена также к
35-летию открытия Конформной Инвариантности \cite{conf}, 30-летию
Монополя \cite{mono} и Инстантонов \cite{inst} и, наконец, 20-летию
Конформной Теории Поля \cite{cft}. И, по редкому совпадению, к
круглой годовщине со дня рождения самого Саши, перечисленный выше
список достижений которого далеко не исчерпывает его вклада в
Теоретическую Физику 20-го века.

И хотя нам не удалось собрать достаточно представительную
конференцию (соот\-ветственно, доклады, приведенные в настоящем
Сборнике, отражают развитие идей Полякова в высшей степени
фрагментарно), нам представляется важным, что она про\-изо\-шла именно в
Черноголовке, месте, где много лет, в Институте Теоретической Физики
Л.Д.Лан\-дау, работал Александр Поляков.

\vspace{0.6cm}
\selectlanguage{english}

The International Workshop dedicated to an anniversary of the Polyakov's String
(of course today there is no need to remind the meaning and the role of this
theory) was held in Chernogolovka in June 2005. Apart from the 25-th anniversary
of the first appearance of the Polyakov's String theory \cite{string}, this conference,
to our mind, might be also thought of as a 35 years from the discovery of the
Conformal Invariance \cite{conf}, 30 years of the Monopole \cite{mono} and the
Instantons \cite{inst} and, finally, as the 20-th anniversary of the CFT \cite{cft}.
A case of mysterious coincidence, this year is also a jubilee of Sasha
himself, whose contribution to the Theoretical Physics of 20-th century is far
from being exhausted by the achievements listed above.

Although we haven't managed to bring together a really representative conference
(consequently, the talks delivered give only a fragmentary and incomplete picture
of the developments of Sasha's ideas), we find it significant, that it took place
Chernogolovka, where during many years Alexander Polyakov worked in the Landau
Institute of Theoretical Physics.

\bigskip
\selectlanguage{russian}
{\large Оргкомитет}

\selectlanguage{english}

\setcounter{section}{0}
\setcounter{footnote}{0}
\newpage

\begin{center}
\bigskip{\huge Hidden Sasha Polyakov's life }

{\Large in Statistical and Condensed Matter Physics}

\bigskip

\textit{\ Valery Pokrovsky,}

\textit{Department of Physics, Texas A\&M\ University}

\textit{and }

\textit{Landau Institute for Theoretical Physics}\footnote{Department of
Physics, Texas A\&M University, College Station, TX 77843-4242,
\par
USA and Landau Institute for Theoretical Physics, Chernogolovka, Moscow
\par
region 142432, Russia}
\end{center}

\bigskip

\section{\bigskip Introduction}

History of Science (is it a Science itself?) shows that the greatest field and
elementary particle theorists felt an insistent necessity to make something in
a more earthly subjects. I do not speak about such universal giants as Bethe,
Fermi and Landau, for whom there was no separation between different parts of
physics. But their younger and more specialized contemporaries also time to
time did seminal works in statistical or condensed matter physics. An example
is the brilliant work by Feynman on vortices in a superfluid and his
unpublished work on vortices in 2D XY-magnet repeated independently in a
famous articles by Berezinskii and by Kosterlitz and Thouless. A fundamental
contribution into Statistical Physics was done by Gell-Mann and Bruckner in
their work on the virial expansion of weakly interacting Fermi-gas. Lee and
Yang theorem on the distribution of nodes of the partition function became one
of the cornerstones of the phase transition theory. C.N. Yang probably has the
longest list of fundamental works in statistical physics, which includes,
besides of the mentioned work, the derivation of the Onsager formula for the
magnetization of the Ising magnet (1952, theory of 1D interacting Fermi-gas
(together with C.P. Yang, 1965), virial expansion for weakly interacting Bose
gas (together with T.D. Lee) and recent works on Bose-Einstein condensation.

The response to these works was always very vivid and eager. Apart of the
undoubted merits of the works it can be explained by a simple psychological
effect: it is flattering to occur in touch with the deepest minds of humanity.

Even in this company, Sasha Polyakov's works are probably most popular in the
earthly physics. Despite Sasha's carelessness about experimental consequences
of his theories and real figures his works occurred very close to real
experiments, most of them in Condensed Matter Physics. They also played very
important role for development of Statistical Physics and Condensed Matter
Theory. Below I give a brief review of the life of Sasha's ideas, sometimes
unexpected, sometimes rediscovered in other terms. Most of them were conceived
as field-theoretical works with no intended application to Condensed Matter Physics.

\section{Phase Transitions and Critical Phenomena}

No surprise that the Quantum Field Theory can be applied in Statistical
Physics. As it was demonstrated by many people the two subjects are almost
identical: statistics is a Euclidian field theory. In his book
\cite{polyakov-book} Sasha suggested that this analogy is more than a formal
trick, that it may stem from a deep, not yet understood physics associated
with the nature of time. Leaving this philosophical question for future, we
turn to the Polyakov's works.

\subsection{Bootstrap theory of Phase Transitions}

The first Sasha's intervention into the Phase Transition theory was his
article of 1968 \cite{bootstrap}, just at the end of his "Aspirantura". In
this work he moved opposite to the standard approach replacing the Euclidian
field theory by pseudoeuclidean one. The purpose of this trick was to employ
the unitarity condition and avoid a rather doubtful subtraction procedure,
which Patashinskii and I applied in our earlier work on the same topic
published in 1964 \cite{PP}. At that moment we already understood something is
wrong in our work, but we did not know what. In Sasha's work the wrong point
was explicitly found and reformulated correctly. It was so exciting that, when
Sasha by my request visited me in a hospital where I stayed already for a
while, I immediately have felt myself recovered, an unrecorded case of
miraculous healing. To my surprise and enjoyment it occurred that a large part
of our work (scaling) was correct.

Although the bootstrap equations in principle determined critical exponents,
they were too complicated to allow numerical calculations, at least at that
time. I believe that now it would be possible by employing a modification of
the bootstrap equations proposed by Sasha Migdal \cite{aamigdal}
(independently and with very small difference in time) and three-point
correlators derived by AP in 1970 (see subsection Conformal invariance).
Unfortunately, the train has departed in 1972 bringing the Nobel Prize with it.

\subsection{Algebra of fluctuating fields}

In 1969 Sasha introduced a new concept: algebra of fluctuating fields
\cite{algebra} (independently the same discovery was made by Leo Kadanoff
\cite{kadanoff-algebra}). He suggested that any fluctuating field in a
vicinity of the phase transition point can be expanded in a basis (generally
infinite) of basic fields $A_{n}(\mathbf{x})$ characterized by their scaling
dimensionality $\Delta_{n}$. The product of two such fields taken at close
points can be also expanded in the same basis. The three-point coefficients of
this expansion describe the triple interaction between the fields and
completely determine the algebra. This idea not only simplified enormously the
structure of the phase transition theory, but also gave it a new dimension.
Till 1969 only the order parameter and the entropy (energy) density were
considered as basic fluctuating fields. The algebra revealed a multitude of
other fields.

\subsection{Conformal invariance}

In 1970 Sasha conjectured that symmetry of the effective field theory at phase
transition point is much more extensive than the global scaling group. In the
spirit of the gauge field theory by Yang and Mills, the scaling symmetry must
be local with the scale factor varying in space retaining local isotropy. The
transformations performing this job form the conformal group. Some its
implications were studied earlier in the quantum field theory, but Sasha was
the first to derive its consequences for correlation functions. For two-point
correlators in any dimension he found the orthogonality condition $\langle
A(\mathbf{x})B(\mathbf{x}^{\prime})\rangle=0$ if $\Delta_{A}\neq\Delta_{B}$.
Moreover, he showed that the conformal invariance completely determines the
3-point correlator and obtained a beautiful formula for it:
\begin{equation}
\langle A_{1}(\mathbf{x}_{1})A_{2}(\mathbf{x}_{2})A_{3}(\mathbf{x}_{3}%
)\rangle=\frac{\Gamma_{123}}{|\mathbf{x}_{1}-\mathbf{x}_{2}|^{\Delta
_{1}+\Delta_{2}-\Delta_{3}}|\mathbf{x}_{2}-\mathbf{x}_{3}|^{\Delta_{2}%
+\Delta_{3}-\Delta_{1}}|\mathbf{x}_{3}-\mathbf{x}_{1}|^{\Delta_{3}+\Delta
_{1}-\Delta_{2}}}%
\end{equation}

To my knowledge no attempts was made to measure 3-point correlators. All
scattering methods measure 2-point correlators. A reasonable way to find
3-point correlators would be to measure the dependence of a 2-point correlator
on weak oscillatory perturbations, for example, sound. The experiment looks
difficult since the temperature and pressure in critical measurements must be
fixed with very high precision. The perturbation should be very weak to retain
such a high precision. Nevertheless, such an experiment seems feasible.

In the same work Sasha noted that the conformal group in 2 dimensions is very
reach: it is well-known group of conformal transformation of a complex
variable $z=x+iy$ provided with any analytic function $w(z)$. Sasha
anticipated that irreducible representations of this group completely
determine all possible types of critical behavior (universality classes) in 2
dimensions. This program was realized only 13 years later in a famous work by
Belavin, Polyakov and Zamolodchikov \cite{bpz}. Enormous number of journal
articles and many books developed and reviewed this theory. I will not make
any attempt to describe it in this brief review. I only note that indeed all
known types of critical behavior including exactly solved Ising model,
8-vertex and Ashkin-Teller models, Potts model, their multicritical points and
an infinite class of model by Andrews and Baxter, Heisenberg model etc. took
their place as special representations of the Conformal Group.

\subsection{Multi-component vector model and real magnets}

In 1971-72 Vadim Berezinskii had published two seminal articles on
2-dimensional XY-model with global symmetry group SO(2) or U(1)
\cite{berezinskii}. He discovered the algebraic order in this system and its
destruction by vortices. In the first of these works he stated that
3-component and more generally n-component vector model with the non-abelian
symmetry group SO(n) also has algebraic order. In 1975 Sasha revised this
problem \cite{polyakov-n} and has found that strong interaction generated by
the non-abelian group and its topology completely changes the physical
properties of the vector fields. In particular, he found that there is no
phase transition in the model for $n\geq3$ and the ordering field acquires
mass at large distances at any temperature.

The n-vector model is a model of a classical vector field with the
Hamiltonian
\begin{equation}
H=\frac{J}{2}\int(\nabla\mathbf{n(x)})^{2}d^{2}x, \label{n-vector}%
\end{equation}
where $\mathbf{n(x)}$ is an n-component vector of unit length. Employing a
clever renormalization procedure conserving the length of the vector, Sasha
found a remarkable result for the dependence of the order parameter
$M=|\langle\mathbf{n}\rangle|$ on scale $L$:
\begin{equation}
M=\left(  1-\frac{(n-2)T}{4\pi J}\ln\frac{L}{a}\right)  ^{\frac{n-1}{2(n-2}}
\label{order-n}%
\end{equation}
This result known as Polyakov's renormalization has some limitations. Namely,
the ratio $T/4\pi J$ must be small and the value $M$ must be much larger than
$T/4\pi J$, though it can be much smaller than 1. At the distances $L\gg
R_{C}=a\exp(4\pi J/T)$ the expected behavior of $M(L)$ is $M=\exp(-L/R_{c})$.
The renormalization group does not work at such long distances. The conjecture
about exponential decay of correlations was confirmed later by exact
calculations (Polyakov and Wiegmann \cite{polwieg}). Physically the strong
renormalization stems from strong fluctuations generated by Goldstone modes
(spin waves). At $n=2$ the equation (\ref{order-n}) reproduces the
Berezinskii's algebraic order. At $n=3$ corresponding to isotropic
(Heisenberg) magnet the equation (\ref{order-n}) is strongly simplified:
\begin{equation}
M=1-\frac{T}{4\pi J}\ln\frac{L}{a} \label{order-3}%
\end{equation}

In real magnets the isotropy is slightly violated by spin-orbit and dipolar
interactions or by an external magnetic field. This violation fixes the length
scale $L_{A}=\sqrt{J/A}$, where $A$ is the amplitude of the symmetry-violating
field, for example a coefficient in the additional term $An_{z}^{2}$ in the
Hamiltonian. If the length $L_{A}$ is \noindent much less than the correlation
length $R_{c}$, the magnetization is given by equation (\ref{order-3}) with
$L=L_{A}$. The linear dependence of magnetization on temperature in weakly
anisotropic magnets is a firmly established experimental fact. It was found in
many materials by many authors. In Fig. 1 we demonstrate the experimental
dependence $M(T)$ for an ultrathin iron film on the surface Ag(111) found by
Z. Qiu \textit{et al.} \cite{qiu} by the measurement of the surface
magneto-optic Kerr effect (SMOKE). The linear dependence takes place at
$T/T_{c}$ from 0 to 0.95. Close to the Curie point they observed the critical
behavior $M\propto(T_{c}-T)^{1/8}$.

\begin{center}%
\begin{figure}
[ptb]
\begin{center}
\includegraphics[
height=2.0885in,
width=3.1929in
]%
{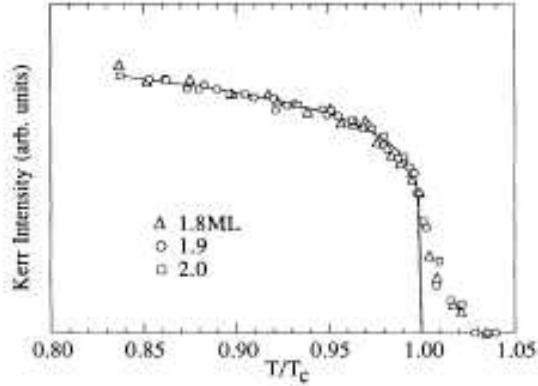}%
\caption{Magnetization vs. temperature }%
\end{center}
\end{figure}
\end{center}

\subsection{Quantum Phase Transitions}

In his book \cite{polyakov-book} published in 1987 Sasha considered probably
the first model displaying the Quantum Phase Transition in two space
dimensions. It was the lattice of quantum rotators. In this model plane
quantum rotators are placed at sites $\mathbf{a}$ of a regular lattice. Each
rotator is characterized by its rotation angle $\varphi_{\mathbf{a}}$ or
alternatively by its angular momentum $\hat{n}_{\mathbf{a}}$ with the standard
commutation relation $\left[  \hat{n}_{\mathbf{a}},\varphi_{\mathbf{a}%
^{\prime}}\right]  =-i\delta_{\mathbf{a,a}^{\prime}}$. The eigenvalues of
$\hat{n}_{\mathbf{a}}$ are integers. The Hamiltonian of the model reads:%
\begin{equation}
H=\frac{K}{2}%
{\displaystyle\sum\limits_{\mathbf{a}}}
\hat{n}_{\mathbf{a}}^{2}-J%
{\displaystyle\sum\limits_{(\mathbf{a,a}^{\prime})}}
\cos\left(  \varphi_{\mathbf{a}}-\varphi_{\mathbf{a}^{\prime}}\right)
\label{rotators}%
\end{equation}
The summation in the lust sum of equation (\ref{rotators}) proceeds over the
pairs of nearest neighbors. The competition between "kinetic" and "potential"
energy results in a phase transition at zero temperature from the "ordered"
state with $\left\langle \cos\varphi_{\mathbf{a}}\right\rangle \neq0$ and
indefinite $\hat{n}_{\mathbf{a}}$ at $J>K$ to the "disordered state with all
$\hat{n}_{\mathbf{a}}=0$ and $\left\langle \cos\varphi_{\mathbf{a}%
}\right\rangle =0$ at $J<K$. The tuning of the ratio $J/K $ could be produced
by pressure. The spin version of the model can be formulated for spins
$S\geq1$. This is the Heisenberg model with the strong anisotropy term
$KS_{z}^{2}$.

The problem of quantum phase transitions became rather popular last 15 years
in connection with High-T$_{c}$ superconductors, magnetic chains,
metal-insulator transitions etc. The detailed description of the state of art
is given in the book by Sachdev \cite{sachdev}.

\section{Topological Excitations}

In the beginning of 1970-th after the Berezinskii's work on vortices in
XY-magnets \cite{berezinskii}, Sasha has asked me what kind of objects are
vortices. Are they quasiparticles? I answered that in 3 dimensions the vortex
rings are indeed quasiparticles with the dispersion $E\sim\sqrt{p}$, but I do
not know what is the status of the vortex \ in 2 dimensions since its energy
is infinite. A week later Sasha told me that he finally realized what vortex
are: topological excitations or vacuum states with different topological
numbers. These conversations remains in my memory as a benchmark of a novel
and exciting Saha's studies of topological excitations, probably most popular
of his works. They included the discovery of the monopole solution in the
SU(2) gauge theory, rediscovery and deep study of skyrmions in 2D
$3-$component vector field theory and introduction of new notion and objects,
instantons. The monopole solution was independently found by
G. t'Hooft. Since my purpose is to review applications of Sasha's theories in
Condensed Matter Physics, I will speak below about skyrmions and instantons.

\subsection{Skyrmions}

A particle-like solution of the vector field theory was first found
by the nuclear theorist R.T.H. Skyrm \cite{skyrm}. Belavin and
Polyakov \cite{BP} in 1975 proved that this solution has a
nontrivial topological structure: it realizes the mapping of the
plane onto the surface of the unit sphere $S_{2}$. They introduced
the name Skyrmion and extended theory to the many-skyrmion
solutions. They discovered a deep analytical structure of Skyrmion
solutions and found that classical skyrmions do not interact: the
energy of $n-$skyrmion configuration is equal to $4\pi Jn$
independently on skyrmion radii and positions of their centers.
According to Belavin and Polyakov, the elementary skyrmion solution
can be conveniently described in terms of complex variables
$z=x+iy$, where $x$ and $y$ are coordinates in the plane, and
$w=\tan \frac{\theta}{2}e^{i\phi}$, $\theta$ and $\phi$ being
spherical coordinates of
the vector (magnetization):%
\[
w=\frac{R\cdot e^{i\alpha}}{z-z_{0}},
\]
where $R$ is the radius of the Skyrmion, $z_{0}$ determines position of its
center and $\alpha$ is a constant angular shift. The energy does not depend on
any of these four parameters (zero modes) and is equal to $4\pi J$. The
distribution of spins in a skyrmion is plotted in Fig. 2.%

\begin{figure}
[ptb]
\begin{center}
\includegraphics[
height=3.8268in,
width=4.7945in
]%
{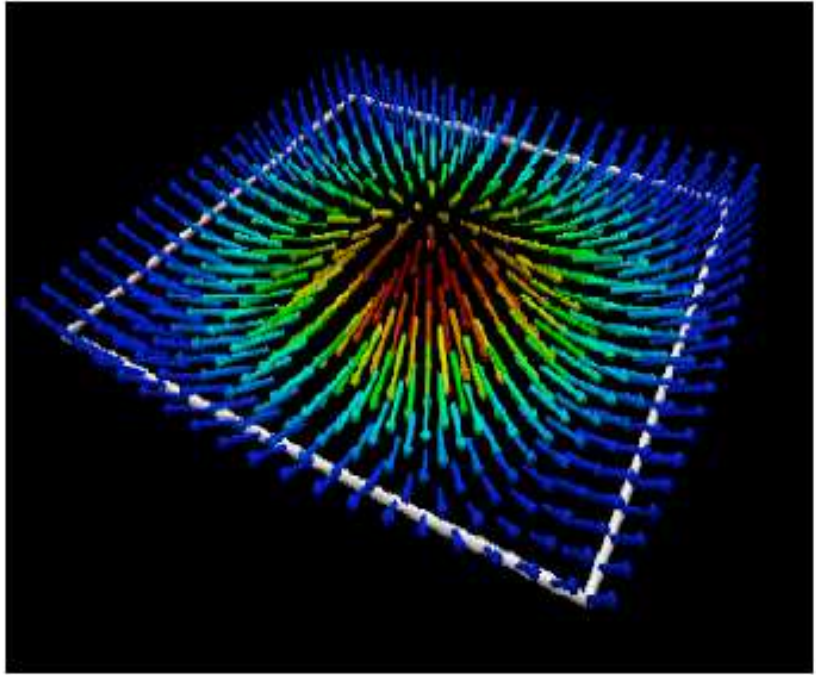}%
\caption{Distribution of spins \ in a skyrmion}%
\end{center}
\end{figure}

The presence of skyrmions was predicted theoretically and convincingly proved
experimentally for two-dimensional electron gas in the condition of the
Quantum Hall Effect. Theoretical prediction was made by Sondhi \textit{et al.}
\cite{sondhi}. They argued that the exchange interaction between electrons is
sufficiently large to make this system almost ideal ferromagnet. The Zeeman
energy is relatively small due to smallness of the gyromagnetic factor, which
can be reduced almost to zero by a comparatively small hydrostatic pressure.
The direct Coulomb interaction is also small in comparison to the exchange,
but it fixes the radius of the Skyrmion. In contrast to the case of the
Heisenberg ferromagnet the Skyrmion carries electric charge. These localized
objects has rather big spin proportional to its area. Participating in the
process of the spin relaxation in the NMR it increases dramatically the
relaxation time. Another effect is a sharp peak in the dynamic spin
polarization due to Skyrmions (see Fig. 3). The NMR measurements were
performed with the heterojunction Ga/GaAs/GaAlAs.

\begin{center}%
\begin{figure}
[ptb]
\begin{center}
\includegraphics[
height=3.0848in,
width=6.3027in
]%
{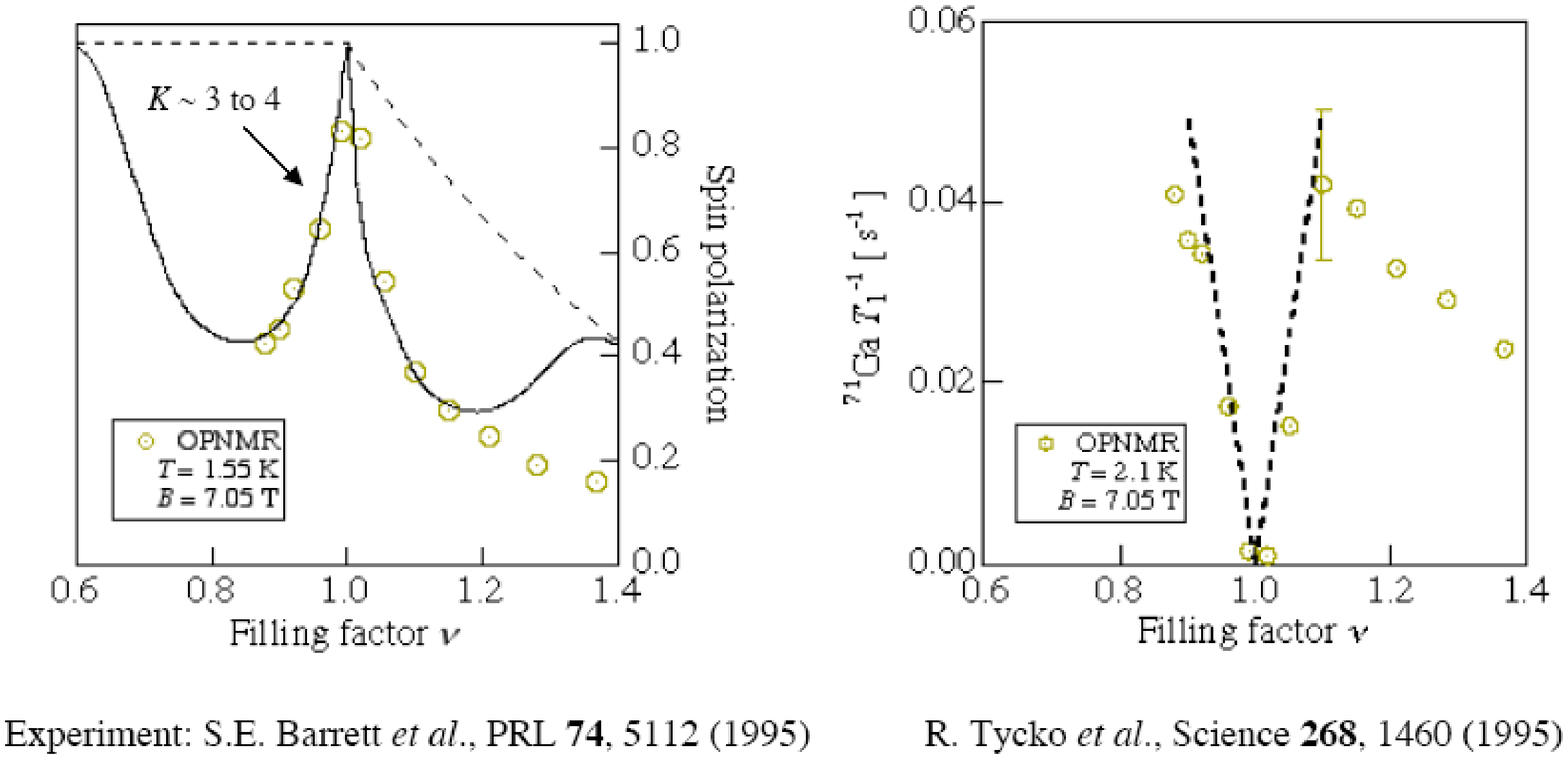}%
\caption{The left panel shows the spin polarization in the NMR experiment near
the filling factor $\nu=1$; the right panel shows the NMR relaxation rate vs.
the filling factor.}%
\end{center}
\end{figure}
\end{center}

\noindent These experiments allow to estimate the skyrmion radius or the total
spin of the skyrmion $S\approx16$. Many theoretical works were dedicated to
study of the phase diagram: do the skyrmion form a liquid gas or a crystal
structure and what is the magnetic state of the system as a whole (see, for
example \cite{iord-kash}), but experimentally it is not yet well established.

Indirect evidences of the skyrmion presence in antiferromagnets were indicated
by F. Waldner \cite{waldner}. He analyzed three types of experiments: elastic
neutron scattering for quasi-two dimensional compounds NTT \cite{shirane} and
Rb$_{2}$Mn$_{x}$Cr$_{1-x}$Cl$_{4}$ \cite{cowley} which gives the value of the
correlation length vs. temperature; the line broadening in the electron spin
resonance (ESR) in several quasi-two-dimensional compounds like (CH$_{2}%
$)$_{2}$(NH$_{3}$)$_{2}$MnCl$_{4}$ and similar organic magnets, K$_{2}%
$MnF$_{4}$ and Rb$_{2}$MnF$_{4}$ \cite{ESR-exp}; the NMR relaxation rate in
some of these compounds \cite{brinkmann}. In all cases he found the activation
exponent with the barrier energy equal to $AJS^{2}$ with the numerical
coefficient $A=4\pi$ within the precision of the experiment.

Why the skyrmion were not observed in weakly anisotropic ferromagnetic films
as permalloy, Fe and Ni on a smooth crystal substrate? The reason was
elucidated in the work \cite{abanov}. The authors studied what happens with
the skyrmion at small symmetry-breaking perturbations like anisotropy,
magnetic field or dipolar forces. Their conclusion is that the uniaxial
anisotropy itself leads to shrinking of the skyrmion to zero radius. However,
if the 4-order in derivatives term in the exchange interaction is positive, it
makes the skyrmion stable and fixes it radius $R\sim\sqrt{l_{dw}a}$, where
$l_{dw}$ is the domain wall width and $a$ is the lattice constant. For real
magnets this radius varies between 1 and 10 nm. This is a very small object
and its experimental observation requires very high resolution. Theory
predicts that in ferromagnetic insulators with localized spins the sign of the
mentioned correction to the exchange interaction is negative unless the
couplings between not-nearest neighbors are anomalously large. Thus, skyrmions
are absolutely unstable in these ferromagnets. In the itinerant ferromagnets
with the oscillating RKKY interaction between spins, the sign is positive and
we can expect stable, but very small skyrmions. This is a disadvantage for
their experimental discovery, but may be very useful in their application as
digits in a magnetic record.

Last several years experimentalists proposed a modification of the skyrmion
\cite{permalloy}. They deposit permalloy magnetic disks with the radius from
0.1 to 1 $\mu m$. The dipolar forces in such a disk put the spins into the
plane. At small radius the monodomain configuration with all spins parallel
(in-plane) is energetically stable, but at larger radius the vortex
configuration wins everywhere except of a small circle near the center where
spins go out of plane exactly as in skyrmion. The radius of this
pseudoskyrmion is $\sim\sqrt{J/M^{2}d}$, where $d$ is the thickness of the
disk. Despite of the seeming likeness between this problem and problem of
skyrmion the shape of this excitation is rather different from that of the skyrmion.

\subsection{Instantons}

The first classical instanton solution in the gauge field theory was obtained
by Belavin, Polyakov, Schwarz and Tyupkin in 1975 \cite{BPST}. It represents a
limited in time tunneling trajectory between two topologically different vacuum
states. In condensed matter physics such transitions are frequent phenomena.
In some situations they represent a dominant mechanism of dissipation. We will
describe couple of such cases.

\subsubsection{Phase slip centers}

\ The phase slip centers were proposed by Skocpol, Beasley and Tinkham
\cite{skocpol} as a mechanism of dissipation in thin superconducting wires.
Small transverse sizes of such wires make impossible the formation of vortices
and suppress the motion of normal carriers (quasiparticles). The paradox is
that the electric field penetrating in the wires should accelerate the Cooper
pairs and destroy the superconducting state. Instead they proposed that a new
node of the condensate (Ginzburg-Landau) wave function enters into a wire. The
phase changes by 2$\pi$ at passing such a center. This is the so-called
phase-slip center. Since it moves inside the wire, the phase changes with
time. According to Josephson the time dependent phase generates a voltage and
together with it dissipation.

Ivlev and Kopnin \cite{ivlev} were the first to recognize that the phase-slip
centers are typical instantons. The change of the phase by $2\pi$ can be
treated as a transition to a vacuum with another topological number. The field
and phase at a \ fixed point is time-dependent and also changes from point to
point. To ensure a state stationary in average the phase and field variations
must be periodic both in time and in space. Thus, the phase-slip centers form
a regular \ rectangular lattice in the space-time plane (see Fig. 4) and
physical values are double periodic.

\begin{center}%
\begin{figure}
[ptb]
\begin{center}
\includegraphics[
height=3.0476in,
width=3.4515in
]%
{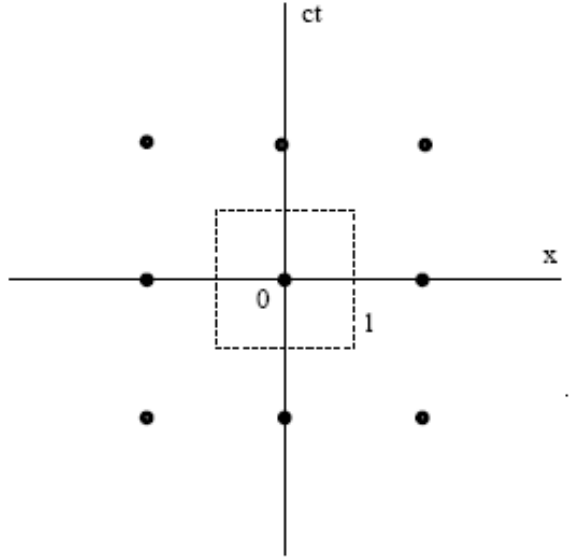}%
\caption{The lattice of phase-slip centers in space-time plane. The contour
$l$ encircles a unit cell of the lattice.}%
\end{center}
\end{figure}
\end{center}

Ivlev and Kopnin proposed a "quantization rule" for the electric field in the
two-dimensional space time $x_{0}=ct$, $x_{1}=x$. They introduced the
gauge-invariant potentials $\mathbf{Q=A-}\frac{\hbar c}{2e}\nabla\chi$ and
$\Phi=\varphi+\frac{\hbar}{2e}\frac{\partial\chi}{\partial t}$, where
$\mathbf{A}$ is the electromagnetic vector-potential, $\varphi$ is the scalar
potential and $\chi$ is the phase of the condensate wave function. In the wire
only a component along the wire $Q_{x}$ survives. Let introduce
relativistically-covariant 2-component vector $\mathbf{q=}\left(  \Phi
_{x},Q\right)  $ and consider its circulation around a contour $l$ surrounding
a phase-slip center and containing an elementary cell of the lattice. Since
the vector $\mathbf{q}$ has the periodicity of the lattice, this circulation
is equal to zero. The vector $\mathbf{q}$ can be represented in terms of the
standard "relativistic" 2-vector-potential $\mathbf{a=(\varphi,A})$ and the
2-gradient of the phase: $\mathbf{q=a-}\frac{\hbar c}{2e}\nabla\chi$. Zero
circulation of the vector $\mathbf{q}$ implies that $%
{\displaystyle\oint}
\mathbf{a\cdot} d\mathbf{r=\Phi}_{0}n;$ where $~\mathbf{\Phi}_{0}=\frac
{\pi\hbar c}{e}$ is the (superconducting) flux quantum and $n$ is an integer.
Using the Stokes theorem and the expression of electric field via potentials,
they found:%
\begin{equation}%
{\displaystyle\int\nolimits_{c}}
Edxcdt=\mathbf{\Phi}_{0}n \label{field quantization}%
\end{equation}
This equation determines the average electric field $\bar{E}$ in terms of
repeating phase-slip frequency $\omega$ and the distance between them $L$:
$\bar{E}=\frac{\hbar\omega}{2eL}$. In the dc regime the current drops each
time when the number of phase-slip centers in the wire changes by one.
Discontinuous current-voltage curves were first obtained in the experiment by
Skocpol \textit{et al}. \cite{skocpol}. In Fig. 5. we show the current voltage
characteristics extracted from the work by Jelia \textit{et al}. \cite{jelia}.%

\begin{figure}
[ptb]
\begin{center}
\includegraphics[
height=2.8452in,
width=3.4169in
]%
{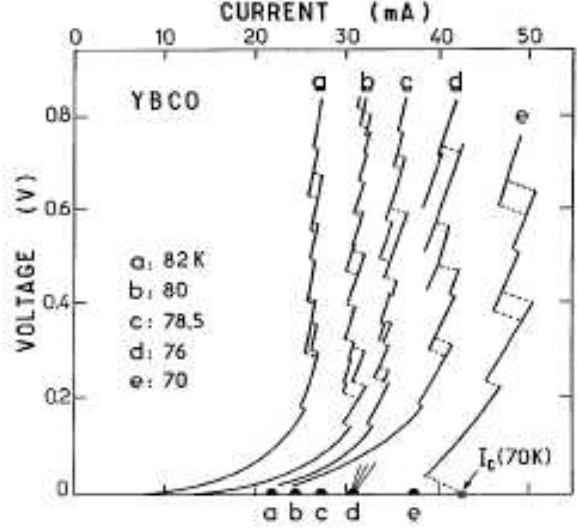}%
\caption{dc current-voltage characteristics of an YBCO bridge (length 200
$\mu$m, width 20 $\mu$m, thickness 90 nm) showing jumps of the current and
resistance.}%
\end{center}
\end{figure}

The resonance with an external ac radiation of proper frequency results in
Shapiro steps in current \ voltage characteristics of superconducting wires as
it is shown on Fig. 6.

\begin{center}%
\begin{figure}
[ptb]
\begin{center}
\includegraphics[
height=2.3981in,
width=3.8389in
]%
{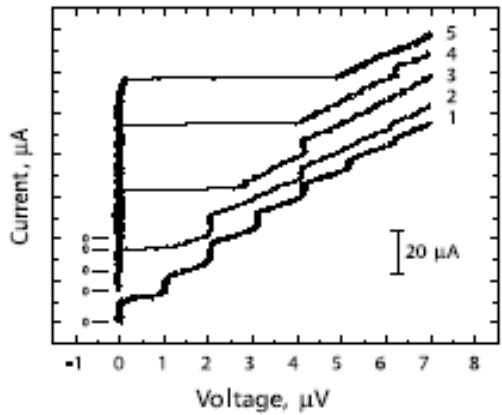}%
\caption{Current-voltage characteristics of a 15 $\mu$m wide Sn strip under
microwave irradiation with frequency $f$ 0.5 GHz (1), 1 GHz (2), 2 GHz (3),
3GHz (4), no microwaves (5). The graph is extracted from the article by
Sivakov \textit{et al}. \cite{ustinov} }%
\end{center}
\end{figure}
\end{center}

Phase-slip centers appear also in one-dimensional or quasi-one-dimensional
charge density waves (CDW) \cite{CDW}. Charge density waves are described by a
scalar complex order parameter. The change of its phase by $2\pi$ is the
instanton and has the same structure as in superconductors, but its physical
manifestation is different: it is caused by a voltage bias and causes the
charge transfer, i.e. the current. Therefore, in corresponding I-V
characteristics the current and voltage interchange each other in comparison
to the corresponding characteristics for superconductors.

\section{Thermodynamics of membranes}

In 1981 Polyakov made a fundamental work in the quantum string theory
\cite{polyakov-string}. He described their propagation as a random surface in
which not only the shape of the surface, but also the metrics fluctuates. In
his work \cite{polyakov-string} Sasha indicated the simple and reliable way
for practical computation of statistics: the triangulation of surfaces.
Besides of its direct impact on the string theory and quantum gravitation, it
had a deep influence onto statistical physics of membranes, in particular
biological membranes. Two reasons enforce me to be brief in this section: the
formalism is more complicated than in others sections and I am far from being
an expert in this problem. Therefore I will not try to reproduce complicated
fluctuating differential geometry. Instead I refer the reader to original
articles. However, we show some pictures to give the feeling what it is about.
One of the most important problems of this theory is the so-called crumpling
transition, i.e. the transition from a compact comparatively smooth state of
the membrane to a fractional state with sharp edges and peaks like in a
roughly folded sheet of paper \cite{kantor-nelson}. In a later work
\cite{polyakov-string1} Sasha indicated an useful analogy between the
Heisenberg model and his model of quantum surface: the normals to the surface
play the role of spins. The smooth state is analogous to a ferromagnet,
whereas the crumpled state is an analogue of a paramagnet. On Fig. 6,7 we
illustrate this transition by computational pictures extracted from the
P.Coddington's website http://www.cs.adelaide.edu.au/users/paulc/physics/randomsurfaces.html.%

\begin{figure}
[ptb]
\begin{center}
\includegraphics[
height=1.3474in,
width=1.3474in
]%
{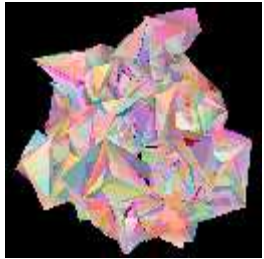}%
\caption{The smooth state of a membrane}%
\end{center}
\end{figure}

\begin{center}%
\begin{figure}
[ptb]
\begin{center}
\includegraphics[
height=1.2842in,
width=1.2842in
]%
{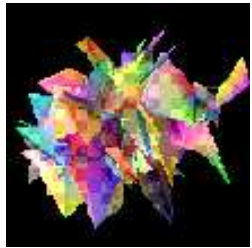}%
\caption{The crumpled state of a membrane. }%
\end{center}
\end{figure}
\end{center}

In a real experiment with the DMPC membranes (DMPC is an abbreviation for
dimyristol phosphatidylchlorine) the difference in shape looks not less
dramatic (see Fig. 8). The transition is driven with concentration of a
special reagent farnesol \cite{rowat}.%

\begin{figure}
[ptb]
\begin{center}
\includegraphics[
height=2.2502in,
width=5.1171in
]%
{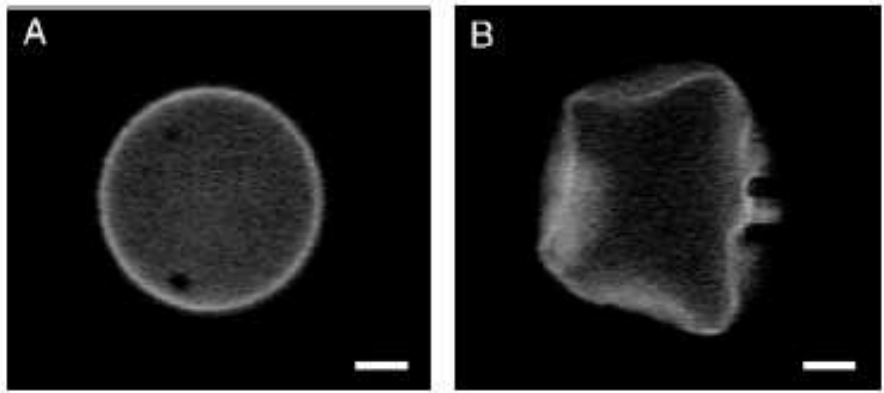}%
\caption{A DCMP vesicle before and after the crumpling transition. The scale
represents 5 $\mu$m. The transition proceeds at 20\% of the farnesol at
13$^{0}$C. }%
\end{center}
\end{figure}

\setcounter{figure}{0}
\setcounter{section}{0}
\setcounter{footnote}{0}

\newpage

\begin{center}
{\Large \textbf{Moduli Integrals, Ground Ring and Four-Point Function in
Minimal Liouville Gravity}}

\vspace{1.0cm}

{\large A.Belavin}

\vspace{0.2cm}

L.D.Landau Institute for Theoretical Physics RAS,

142432 Chernogolovka, Russia

\vspace{0.2cm}

and

\vspace{0.2cm}

{\large Al.Zamolodchikov}\footnote{On leave of absence from Institute of
Theoretical and Experimental Physics, B.Cheremushkinskaya 25, 117259 Moscow, Russia.}

\vspace{0.2cm}

Laboratoire de Physique Th\'eorique et Astroparticules\footnote{UMR-5207 CNRS-UM2}

Universit\'e Montpellier II

Pl.E.Bataillon, 34095 Montpellier, France
\end{center}

\vspace{1.0cm}

\textbf{Abstract}

Straightforward evaluation of the correlation functions in 2D minimal gravity
requires integration over the moduli space. For degenerate fields the higher
equations of motion of the Liouville field theory allow to turn the integrand
to a derivative and to reduce it to the boundary terms and the so-called
curvature contribution. The last is directly related to the expectation value
of the corresponding ground ring element. The action of this element on a
cohomology related to a generic matter primary field is evaluated directly
through the operator product expansions of the degenerate fields. This permits
us to construct the ground ring algebra and to evaluate the curvature term in
the four-point function. The operator product expansions of the Liouville
``logarithmic primaries'' are also analyzed and relevant logarithmic terms are
carried out. All this results in an explicit expression for the four-point
correlation number of one degenerate and three generic matter fields. We
compare this integral with the numbers coming from the matrix models of 2D
gravity and discuss some related problems and ambiguities.

\section{Introduction}

\textbf{1. Liouville gravity }(LG)\textbf{\ }is the term for the
two-dimensional quantum gravity whose action is induced by a ``critical''
matter, i.e., the matter described by a conformal field theory (CFT)
$\mathcal{M}_{c}$ with central charge $c$. This induced action is universal
and is called the Liouville action, because its variation with respect to the
metric is proportional to the Liouville (or constant curvature) equation
\cite{Polyakov1}. Let us denote $\left\{  \Phi_{i},\Delta_{i}\right\}  $ be
the set of primary fields and their dimensions in $\mathcal{M}_{c}$.

\textbf{2. Liouville field theory} (LFT) is constructed as the quantized
version of the classical theory based on the Liouville action. LFT is again a
conformal field theory with central charge $c_{\text{L}}$. It is convenient to
parameterize it in terms of variable $b$ or
\begin{equation}
Q=b^{-1}+b \label{Q}%
\end{equation}
as
\begin{equation}
c_{\text{L}}=1+6Q^{2} \label{cL}%
\end{equation}
Parameter $b$ enters the local Lagrangian
\begin{equation}
\mathcal{L}_{\text{L}}=\frac1{4\pi}\left(  \partial_{a}\phi\right)  ^{2}+\mu
e^{2b\phi} \label{LFT}%
\end{equation}
where $\mu$ is the scale parameter called the cosmological constant and $\phi$
is the dynamical variable for the quantized metric
\begin{equation}
ds^{2}=\exp\left(  2b\phi\right)  \widehat{g}_{ab}dx^{a}dx^{b} \label{ds}%
\end{equation}
Here $\widehat{g}_{ab}$ is the ``reference metric'', a technical tool needed
to give LFT a covariant form
\begin{equation}
\mathcal{A}_{\text{L}}=\int\left(  \frac1{4\pi}\widehat{g}^{ab}\partial
_{a}\phi\partial_{b}\phi+\frac Q{4\pi}\phi\widehat{R}+\mu e^{2b\phi}\right)
\sqrt{\widehat{g}}d^{2}x \label{Lcov}%
\end{equation}
with $\widehat{R}$ the scalar curvature of the background metric. Basic
primary fields are the exponential operators $V_{a}=\exp\left(  2a\phi\right)
$, parameterized by a continuous (in general complex) parameter $a$ in the way
that the corresponding conformal dimension is
\begin{equation}
\Delta_{a}^{\text{(L)}}=a(Q-a) \label{DL}%
\end{equation}
Liouville field theory is exactly solvable \cite{DO}. In particular the
three-point function $C_{\text{L}}(a_{1},a_{2},a_{3})=\left\langle V_{a_{1}%
}(x_{1})V_{a_{2}}(x_{2})V_{a_{3}}(x_{3})\right\rangle _{\text{L}}$ is known
explicitly for arbitrary exponential fields
\begin{equation}
C_{\text{L}}(a_{1},a_{2},a_{3})=\left(  \pi\mu\gamma(b^{2})b^{2-2b^{2}%
}\right)  ^{(Q-a)/b}\frac{\Upsilon_{b}(b)}{\Upsilon_{b}(a-Q)}\prod_{i=1}%
^{3}\frac{\Upsilon_{b}(2a_{i})}{\Upsilon_{b}(a-a_{i})} \label{CL}%
\end{equation}
where $a=a_{1}+a_{2}+a_{3}$ and $\Upsilon_{b}(x)$ is a special function
related to the Barnes double gamma function (see e.g. \cite{AAl}). Correlation
function (\ref{CL}) is consistent with the following identification of the
exponential fields with different values of $a$%
\begin{equation}
V_{a}(x)=R_{\text{L}}(a)V_{Q-a}(x) \label{Lrefl}%
\end{equation}
known as the reflection relations \cite{AAl}. Here
\begin{equation}
R_{\text{L}}(a)=\left(  \pi\mu\gamma(b^{2})\right)  ^{(Q-2a)/b}\frac
{\gamma(2ab-b^{2})}{\gamma(2-2ab^{-1}+b^{-2})} \label{Ra}%
\end{equation}
is called the Liouville reflection amplitude. The local structure of LFT is
determined completely by the general ``continuous'' operator product expansion
(OPE) of generic Liouville exponential fields
\begin{equation}
V_{a_{1}}(x)V_{a_{2}}(0)=\int^{\prime}\frac{dP}{4\pi}C_{a_{1},a_{2}%
}^{\text{(L)}Q/2+iP}(x\bar x)^{\Delta_{Q/2+iP}^{\text{(L)}}-\Delta_{a_{1}%
}^{\text{(L)}}-\Delta_{a_{2}}^{\text{(L)}}}\left[  V_{Q/2+iP}(0)\right]
\label{LOPE}%
\end{equation}
where the structure constant is expressed through (\ref{CL}) $C_{a_{1},a_{2}%
}^{\text{(L)}p}=$ $C_{\text{L}}(g,a,Q-p).$ The integration contour here is the
real axis if $a_{1}$ and $a_{2}$ are in the ``basic domain''
\begin{equation}
\left|  Q/2-\operatorname*{Re}a_{1}\right|  +\left|  Q/2-\operatorname*{Re}%
a_{2}\right|  <Q/2 \label{basic}%
\end{equation}
In other domains of these parameters an analytic continuation is implied. It
is equivalent to certain deformation of the contour due to the singularities
of the structure constant. This prime near the integral sign helps to keep in
mind this prescription.

In LG the parameter $b$ is chosen in the way that together with $\mathcal{M}%
_{c}$ LFT forms a joint conformal field theory with central charge
$c+c_{\text{L}}=26$. Technically it is also convenient to include the

\textbf{3. Reparametrization ghost field theory. }This is the standard
fermionic $BC$ system of spin $(2,-1)$%
\begin{equation}
A_{\text{gh}}=\frac1\pi\int(C\bar\partial B+\bar C\partial\bar B)d^{2}x
\label{Agh}%
\end{equation}
with central charge $-26$, which corresponds to the gauge fixing Faddeev-Popov
determinant. The matter+Liouville stress tensor $T$ is a generator of $26$
dimensional Virasoro algebra. Together with the ghost field theory this allows
to form a BRST complex with respect to the nilpotent BRST charge
\begin{equation}
\mathcal{Q}=\oint\left(  CT+C\partial CB\right)  \frac{dz}{2\pi i}
\label{brst}%
\end{equation}

\textbf{4. Correlation functions} is one of the most important problems in the
LG. In gravitational correlation functions the matter operators $\Phi_{i} $
are ``dressed'' by appropriate exponential Liouville fields $V_{a_{i}}$ in the
way to form either the $(1,1)$ form $U_{i}=\Phi_{i}V_{a_{i}}$ of ghost number
$0$ or the dimension $(0,0)$ operator $W_{i}=C\bar{C}U_{i}$ of ghost number
$1$. In both cases this requires
\begin{equation}
\Delta_{i}+a_{i}(Q-a_{i})=1 \label{Dbalance}%
\end{equation}
Invariant (or integrated) correlation functions are independent on any
coordinates and better called the correlation numbers. In the field theoretic
framework, a (genus $0$) correlation number $\left\langle U_{1}\ldots
U_{n}\right\rangle _{\text{G}}$ at $n\geq3$ is constructed as the integral
\begin{align}
\left\langle U_{1}\ldots U_{n}\right\rangle _{\text{G}}  &  =\int_{M_{n}%
}\left\langle W_{1}(x_{1})\ldots W(x_{n})\right\rangle \label{UG}\\
\  &  =\int\left\langle W_{1}(x_{1})W_{2}(x_{2})W_{3}(x_{3})U_{4}(x_{4}%
)d^{2}x_{4}\dots U(x_{n})d^{2}x_{n}\right\rangle \nonumber
\end{align}
The integration here is over the moduli space $M_{n}$ of the sphere with $n$
punctures. Technically it is equivalent to choose any $3$ of $W_{i}$ at
arbitrary fixed positions $x_{1}$, $x_{2}$ and $x_{3}$ and integrate the
$(1,1)$ forms $U_{i}(x_{i})d^{2}x_{i}$ inserted instead of $W_{i}$ at
$i=4,\ldots,n$. At $n<3$ the definition is slightly different. This is because
of non-trivial conformal symmetries of the sphere with $2$ and $0$ punctures.

The simplest case of \ref{UG} is the tree-point function, where the moduli
space is trivial and the result is factorized in a product of the matter,
Liouville and ghost three-point functions
\begin{equation}
\left\langle U_{1}U_{2}U_{3}\right\rangle _{\text{G}}=x_{12}\bar{x}_{12}%
x_{23}\bar{x}_{23}x_{31}\bar{x}_{31}\left\langle \Phi_{1}(x_{1})\Phi_{2}%
(x_{2})\Phi_{3}(x_{3})\right\rangle _{\text{CFT}}\left\langle V_{1}%
(x_{1})V_{2}(x_{2})V(x_{3})\right\rangle _{\text{L}} \label{U3}%
\end{equation}
The three-point functions $\left\langle \Phi_{1}(x_{1})\Phi_{2}(x_{2})\Phi
_{3}(x_{3})\right\rangle _{\text{CFT}}$, familiar also as the structure
constants of the operator product expansion (OPE) algebra, are known
explicitly in solvable matter CFT's. Thus eq.(\ref{U3}) gives the LG 3-point
correlation number in the explicit form. The two-point number and the
zero-point one (the partition sum) are simply read off from this expression of
the 3-point function.

\textbf{5. The four-point function} is the next step in the order of
complexity
\begin{align}
\ \left\langle U_{1}U_{2}U_{3}U_{4}\right\rangle _{\text{G}}  &  =\label{U4}\\
&  \ x_{12}\bar x_{12}x_{23}\bar x_{23}x_{31}\bar x_{31}\int\left\langle
\Phi_{1}(x_{1})\ldots\Phi_{4}(x_{4})\right\rangle _{\text{CFT}}\left\langle
V_{1}(x_{1})\ldots V_{4}(x_{4})\right\rangle _{\text{L}}d^{2}x_{4}\nonumber
\end{align}
This expression is much less explicit. First, it involves the integration over
$x_{4}$. Then, even if the matter 4-point function is known in any convenient
form, general representations for the Liouville four-point function are more
complicated. E.g., the ``conformal block'' decomposition \cite{AAl}
\begin{align}
\left\langle V_{1}(x_{1})\ldots V_{4}(x_{4})\right\rangle _{\text{L}}  &
=\label{dP}\\
&  \int\frac{dP}{4\pi}C_{\text{L}}(a_{1},a_{2},Q/2+iP)C_{\text{L}%
}(Q/2-iP,a_{3},a_{4})\mathcal{F}_{P}(a_{i},x_{i})\mathcal{F}_{P}(a_{i},\bar
x_{i})\nonumber
\end{align}
involves the so called general conformal block \cite{BPZ}
\begin{equation}
\mathcal{F}_{P}(a_{i},x_{i})=%
\raisebox{-0.3693in}{\includegraphics[
height=0.819in,
width=1.4399in
]%
{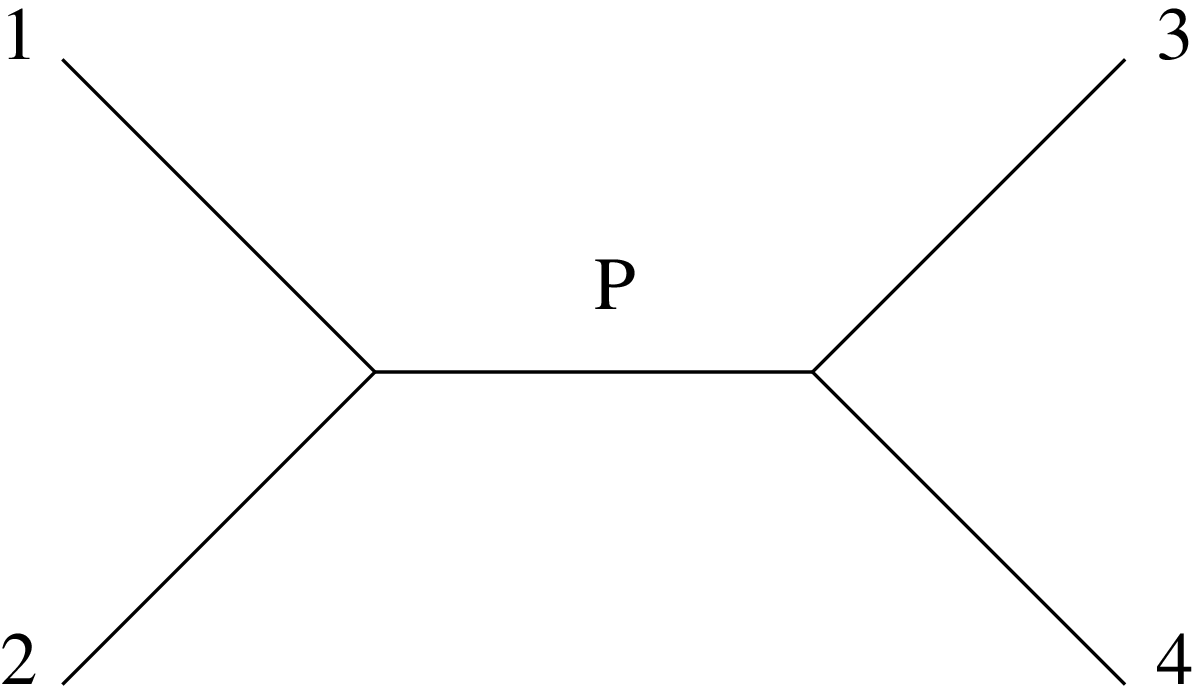}%
}%
=\mathcal{F}_{P}\left(  \left.
\begin{array}
[c]{cc}%
a_{1} & a_{3}\\
a_{2} & a_{4}%
\end{array}
\right|
\begin{array}
[c]{cc}%
x_{1} & x_{3}\\
x_{2} & x_{4}%
\end{array}
\right)  \label{block}%
\end{equation}
which is by itself a complicated function of its arguments, not to talk about
the integration over the ``intermediate momentum'' $P$ in (\ref{dP}). In the
present paper we take a preliminary step towards the evaluation of the
four-point integral in the special case of

\textbf{6. Minimal gravity }(MG). If the conformal matter $\mathcal{M}_{c}$ is
represented by a minimal CFT model (more precisely, a ``generalized minimal
model'' (GMM), see below) $\mathcal{M}_{b^{2}}$, we talk about the ``minimal
gravity'' (MG) (respectively generalized minimal gravity (GMG)). In GMG the
evaluation of the four-point integral is dramatically simplified in the case
when one of the matter operators $\Phi_{i}$ in the r.h.s. of (\ref{U4}) is a
degenerate field $\Phi_{m,n}$. This is due to the so called ``higher equations
of motion'' (HEM) which hold for the operator fields in LFT \cite{higher}. Let
$U_{4}=U_{m,n}$ where
\begin{equation}
U_{m,n}=\Phi_{m,n}\tilde V_{m,n} \label{Umn}%
\end{equation}
and $\tilde V_{m,n}$ is an appropriate Liouville dressing for $\Phi_{m,n}$.
Then HEM's allow to rewrite the integrand in (\ref{U4}) as a derivative
\begin{equation}
\ \left\langle U_{1}U_{2}U_{3}U_{m,n}\right\rangle _{\text{GMG}}=B_{m,n}%
^{-1}\int\partial\bar\partial\left\langle O_{m,n}^{\prime}(x)W_{1}(x_{1}%
)W_{2}(x_{2})W_{3}(x_{3})\right\rangle d^{2}x \label{dd}%
\end{equation}
($B_{m,n}$ is a numerical constant, see sect.3) and hence reduce the problem
to the boundary terms and the so-called curvature term. The last is directly
expressed in terms of the expectation value $\left\langle O_{m,n}W_{1}%
W_{2}W_{3}\right\rangle $ of the

\textbf{7. Ground ring} (GR) element $O_{m,n}$ associated with the field
$\Phi_{m,n}$. Therefore, we want to learn handle the ground ring algebra and
the correlation functions of its elements. This knowledge will also prove
instructive in the subsequent calculations of the

\textbf{8. Boundary terms.} To evaluate the boundary contributions we need to
know appropriate terms in the OPE of the field $O_{m,n}^{\prime}$ in
(\ref{dd}). This field is constructed as a ``logarithmic counterpart'' of the
element $O_{m,n}$ and satisfies the identity
\begin{equation}
\partial\bar{\partial}O_{m,n}^{\prime}=B_{m,n}U_{m,n}+\text{BRST exact}
\label{CHEM}%
\end{equation}
Analysis of this expansion allows to evaluate the boundary contributions and
finally the four-point integral (\ref{U4}) with $U_{4}=U_{m,n}$. This is the
main purpose of all the work presented below.

\section{Generalized minimal models}

Strictly speaking, minimal models of CFT $\mathcal{M}_{p/p^{\prime}}$
\cite{BPZ} are consistently defined as a field theoretic constructions only if
the ``parameter'' $p/p^{\prime}$ is an irreducible rational number so that $p
$ and $p^{\prime}$ are coprime integers. In this case the finite set of
$(p-1)(p^{\prime}-1)/2$ degenerate primary fields $\Phi_{m,n}$ with $1\leq m<p
$ and $1\leq n<p^{\prime}$ (modulo the identification $\Phi_{m,n}%
=\Phi_{p-m,p^{\prime}-n}$) form, together with their irreducible
representations, the whole space of $\mathcal{M}_{p/p^{\prime}}$.
``Canonical'' minimal models $\mathcal{M}_{p/p^{\prime}}$ are believed to be a
completely consistent CFT, i.e., to satisfy all standard requirements of
quantum field theory, except for (in most cases) the unitarity. They are also
considered exactly solvable as the structure of their OPE algebra is known
explicitly \cite{DF}.

There are many ways to relax some of the requirements leading to the set of
$\mathcal{M}_{p/p^{\prime}}$ as unique CFT structures. For example, in the
literature the ``parameter'' $p/p^{\prime}$ is often taken as an arbitrary
number (e.g., \cite{DF}). The algebra of the degenerate primary fields doesn't
close any more within any finite subset and rather the whole set $\{\Phi
_{m,n}\}$ with $(m,n)$ any natural numbers forms a closed algebra. Moreover,
some authors include local fields with dimensions different from the Kac
values, even continuous spectrum of dimensions. Although the consistency of
such constructions from the field theoretic point of view remains to be
clarified, these extensions prove to be a convenient technical tool. Moreover,
statistical mechanics offers a number of examples where either a
generalization of $\mathcal{M}_{p/p^{\prime}}$ for non-integer values
$p/p^{\prime}$ is essentially necessary or non-degenerate primary operators
appear as observables (both generalization are sometimes required).

In this paper we denote $b^{2}$ the parameter $p/p^{\prime}$ and admit the
notion of GMM in the most wide sense as a conformal field theory with central
charge
\begin{equation}
c=1-6(b^{-1}-b)^{2} \label{cM}%
\end{equation}
which may involve fields $\Phi_{\alpha}$ of any dimension. Continuous
parameter $\alpha$ is introduced to parameterize a continuous family of
primary fields with dimensions
\begin{equation}
\Delta_{\alpha}^{\text{(M)}}=\alpha(\alpha-q) \label{DM}%
\end{equation}
where
\begin{equation}
q=b^{-1}-b \label{q}%
\end{equation}
Also we always use the ``canonical'' CFT normalization of the primary fields
$\Phi_{\alpha}$ through the two-point functions
\begin{equation}
\left\langle \Phi_{\alpha}\Phi_{\alpha}\right\rangle _{\text{GMM}}=\left(
x\bar x\right)  ^{-2\Delta_{\alpha}} \label{Mnorm}%
\end{equation}

Degenerate fields $\Phi_{m,n}$ have dimensions$\ $%
\begin{equation}
\Delta_{m,n}^{\text{(M)}}=-q^{2}/4+\lambda_{m,-n}^{2} \label{DMatmn}%
\end{equation}
where yet another convenient notation
\begin{equation}
\lambda_{m,n}=(mb^{-1}+nb)/2 \label{lmn}%
\end{equation}
is introduced. They correspond to either $\alpha=\alpha_{m,n}$ or
$\alpha=q-\alpha_{m,n}$ with
\begin{equation}
\alpha_{m,n}=q/2+\lambda_{-m,n} \label{alphamn}%
\end{equation}
The main restrictions, which singles out this apparently loose construction,
is that the

\textbf{1. Degenerate fields} $\Phi_{1,2}$ and $\Phi_{2,1}$ (and therefore in
general the whole set $\left\{  \Phi_{m,n}\right\}  $) are in the spectrum

\textbf{2. The null-vectors} in the degenerate representations $\Phi_{m,n}$
vanish
\begin{equation}
D_{m,n}^{\text{(M)}}\Phi_{m,n}=\bar D_{m,n}^{\text{(M)}}\Phi_{m,n}=0
\label{GMM}%
\end{equation}
Here $D_{m,n}^{\text{(M)}}$ ($\bar D_{m,n}^{\text{(M)}}$) are the operators
made of the right Virasoro generators $M_{n}$ \footnote{Unusual notations
$M_{n}$ for the Virasoro generators of the matter conformal symmetry are
chosen to save $L_{n}$ for the generators of the Liouville Virasoro.}
(respectively left $\bar M_{n}$) which create the level $mn$ singular vector
in the Virasoro module of $\Phi_{m,n}$. For definiteness we normalize these
operators through the $M_{-1}^{mn}$ term as
\begin{equation}
D_{m,n}^{\text{(M)}}=M_{-1}^{mn}+d_{1}^{(m,n)}(b^{2})M_{-2}M_{-1}%
^{mn-2}+\ldots\label{DMnorm}%
\end{equation}
$\ $ First examples read explicitly
\begin{align}
D_{1,2}^{\text{(M)}}  &  =M_{-1}^{2}-b^{2}M_{-2}\label{DMmn}\\
D_{1,3}^{\text{(M)}}  &  =M_{-1}^{3}-2b^{2}(M_{-2}M_{-1}+M_{-1}M_{-2}%
)+4b^{4}M_{-3}\nonumber\\
&  \ldots\nonumber
\end{align}

\textbf{3. The identification} $\Phi_{\alpha}\equiv\Phi_{q-\alpha}$ is also
often added.

It turns out that this set of definitions imposes important restrictions on
the structure of this formal construction. The three-point function
\begin{equation}
C_{\text{M}}(\alpha_{1},\alpha_{2},\alpha_{3})=\left\langle \Phi_{\alpha_{1}%
}\Phi_{\alpha_{2}}\Phi_{\alpha_{3}}\right\rangle _{\text{GMM}} \label{CMPhi}%
\end{equation}
of the ``generic'' primary fields can be restored uniquely from the above
requirements \cite{threept}
\begin{equation}
C_{\text{M}}(\alpha_{1},\alpha_{2},\alpha_{3})=\frac{b^{b^{-2}-b^{2}%
-1}\Upsilon_{b}(2b-b^{-1}+\alpha)}{\left[  \gamma(1-b^{2})\gamma
(2-b^{-2})\right]  ^{1/2}\Upsilon_{b}(b)}\prod_{i=1}^{3}\frac{\Upsilon
_{b}(\alpha-2\alpha_{i}+b)}{\left[  \Upsilon_{b}(2\alpha_{i}+b)\Upsilon
_{b}(2\alpha_{i}+2b-b^{-1})\right]  ^{1/2}} \label{CM}%
\end{equation}
where again $\alpha=\alpha_{1}+\alpha_{2}+\alpha_{3}$ and $\Upsilon_{b}(x)$ is
the same function as in eq.(\ref{CL}). At the degenerate values of the
parameters $\alpha_{i}=\alpha_{m_{i},n_{i}}$ (and if the standard ``fusion''
relations are satisfied) the known degenerate structure constants \cite{DF}
are recovered from (\ref{CM}).

Explicit form of the OPE of $\Phi_{1,2}$ and a generic primary field
$\Phi_{\alpha}$%
\begin{equation}
\Phi_{1,2}(x)\Phi_{\alpha}(0)=C_{+}^{\text{(M)}}(\alpha)(x\bar x)^{\alpha
b}\left[  \Phi_{\alpha+b/2}\right]  +C_{-}^{\text{(M)}}(\alpha)(x\bar
x)^{1-\alpha b-b^{2}}\left[  \Phi_{\alpha-b/2}\right]  \label{Phi12}%
\end{equation}
($\left[  \Phi_{\alpha}\right]  $ stands for a primary field $\Phi_{\alpha}$
and all the tower of its conformal descendants) will be of use below. In our
normalization
\begin{equation}
C_{+}^{\text{(M)}}(\alpha)=\left[  \frac{\gamma(b^{2})\gamma(2\alpha
b+2b^{2}-1)}{\gamma(2b^{2}-1)\gamma(b^{2}+2\alpha b)}\right]  ^{1/2}%
\;;\;\;C_{-}^{\text{(M)}}(\alpha)=\left[  \frac{\gamma(b^{2})\gamma(2\alpha
b+b^{2}-1)}{\gamma(2b^{2}-1)\gamma(2\alpha b)}\right]  ^{1/2} \label{Cpm}%
\end{equation}
General form of (\ref{Phi12}) reads
\begin{equation}
\Phi_{m,n}(x)\Phi_{\alpha}(0)=\sum_{r,s}^{(m,n)}\left(  x\bar x\right)
^{\lambda_{r,s}(2\alpha+\lambda_{r,s}-q)-\Delta_{m,n}^{\text{(M)}}}%
C_{\text{M}}(\alpha_{m,n},\alpha,\alpha+\lambda_{r,s}))\left[  \Phi
_{\alpha+\lambda_{r,s}}\right]  \label{Phimn}%
\end{equation}
where $\lambda_{r,s}$ are as in eq.(\ref{lmn}) and the sign $\sum
_{r,s}^{(m,n)}$ stands for the sum over the following set of integers (we use
the notation $\{n_{1}:d:n_{1}+nd\}=\{n_{1},n_{1}+d,\ldots,n_{1}+nd\}$)
\begin{equation}
(r,s)=(\{-m+1:2:m-1\},\{-n+1:2:n-1\}) \label{rsset}%
\end{equation}

Other exact results in GMM form somewhat miscellaneous collection. What is
important for our program is the construction of the four-point function
\begin{equation}
G_{(m,n),\alpha_{1},\alpha_{2},\alpha_{3}}^{\text{(GMM)}}(x)=\left\langle
\Phi_{m,n}(x)\Phi_{\alpha_{1}}(x_{1})\Phi_{\alpha_{2}}(x_{2})\Phi_{\alpha_{3}%
}(x_{3})\right\rangle _{\text{GMM}} \label{GM4mn}%
\end{equation}
with one degenerate field $\Phi_{m,n}$ and three generic primaries
$\Phi_{\alpha}$\cite{BPZ}. The null-vector decoupling condition (\ref{GMM})
entails certain partial differential equation for the corresponding
correlations functions. In the four-point case this equation reduces to an
ordinary linear differential equation of the order $mn$, whose independent
solutions are the four-point conformal blocks (in the picture we denote
$\Delta_{i}=\Delta_{\alpha_{i}}^{\text{(M)}}$ and $\Delta_{m,n}=\Delta
_{m,n}^{\text{(M)}}$)
\begin{equation}
\mathcal{F}_{r,s}(x)=%
\raisebox{-0.569in}{\includegraphics[
height=1.2133in,
width=2.4751in
]%
{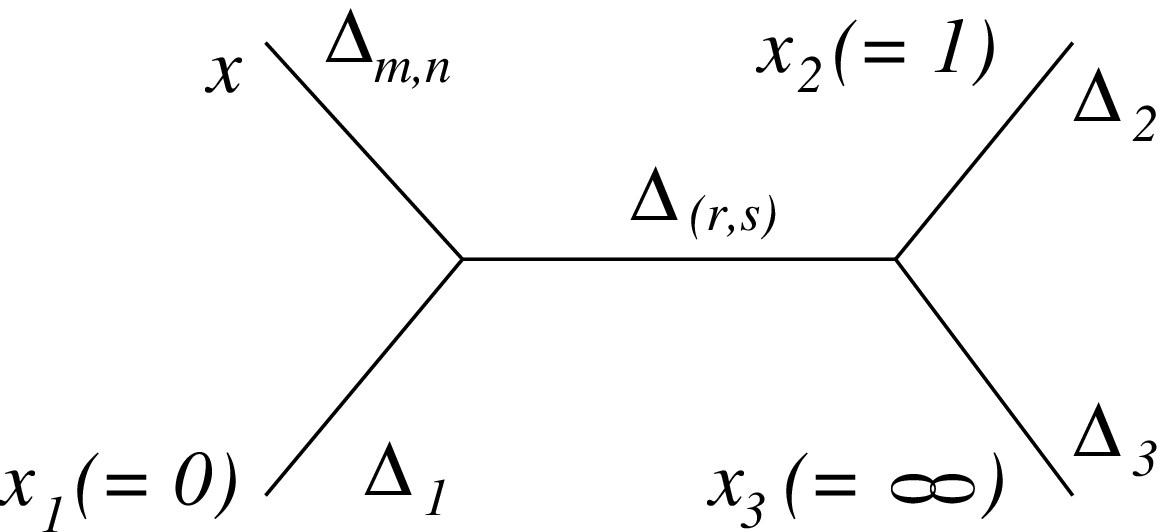}%
}%
\label{rsblock}%
\end{equation}
and $\Delta_{(r,s)}=\Delta_{\alpha_{1}}^{\text{(M)}}+\lambda_{r,-s}\left(
2\alpha_{1}-q+\lambda_{r,-s}\right)  $. The four point function is then
combined as
\begin{align}
&  G_{(m,n),\alpha_{1},\alpha_{2},\alpha_{3}}^{\text{(GMM)}}(x)=\label{G4mn}\\
&  \sum_{r,s}^{(m,n)}C_{\text{M}}(\alpha_{m,n},\alpha_{1},\alpha_{1}%
+\lambda_{r,-s})C_{\text{M}}(\alpha_{1}+\lambda_{r,-s},\alpha_{2},\alpha
_{3})\mathcal{F}_{r,s}(x)\mathcal{F}_{r,s}(\bar x)\nonumber
\end{align}
where $\sum_{r,s}^{(m,n)}$ has the same meaning as in eq.(\ref{Phimn}).

The following remark is very relevant for the subsequent developments. In the
present study, when considering the GMG, we restrict ourselves only to the
four point function with one degenerate matter field $\Phi_{m,n}$, leaving the
other three to be formal generics $\Phi_{\alpha}$. In particular, expression
(\ref{G4mn}) is the relevant construction for the matter part of the integrand
in eq.(\ref{U4}). What is important is that if one or more of the operators
$\Phi_{\alpha}$ are also degenerate\footnote{Note, that this doesn't simply
mean that the corresponding dimension, e.g., $\Delta_{1}^{\text{(M)}}=$
$\alpha_{1}(\alpha_{1}-q)$, belongs to the Kac spectrum of degenerate
dimensions. Definite relations must also hold between the other two parameters
$\alpha_{2}$ and $\alpha_{3}$ to ensure the vanishing of the corresponding
singular vector.}, the number of conformal blocks entering the correlation
function (\ref{G4mn}) might be reduced and this expression doesn't hold
literally. In this case the considerations below are not literally valid.
Important and sometimes rather subtle modifications has to be made. In the
present paper we will not study this interesting but more delicate situation
(although it is extremely actual in quantum gravity applications).

There is another interesting aspect similar to the subtlety mentioned above.
When dealing with GMM one should keep in mind that there are objects of
different nature. Some are continuous in the parameter $b^{2}$, like the
central charge, degenerate dimensions or certain correlation functions. Others
may be highly discontinuous and dependent on the arithmetic nature of the
numbers $p$ and $p^{\prime}$. Simplest example is the number of irreducible
Virasoro representations entering the theory. This warns us to be careful when
trying to reproduce the results of $\mathcal{M}_{p/p^{\prime}}$ as a naive
limit of $\mathcal{M}_{b^{2}}$ as $b^{2}\rightarrow p/p^{\prime}$ and
$\alpha\rightarrow\alpha_{m,n}$ in the formal primary fields. This is why we
stress again that the three matter fields $\Phi_{\alpha}$ in the matter
correlation function have\textit{\ generic non-degenerate values }of the
parameters $\alpha_{1}$, $\alpha_{2}$ and $\alpha_{3}$.

\section{Higher equations of motion}

Let $a_{m,n}=Q/2-\lambda_{m,n}$ with $(m,n)$ a pair of positive integers, so
that $V_{m,n}=V_{a_{m,n}}$ are the Liouville exponentials corresponding to
degenerate representations of the Liouville Virasoro algebra. Let also
$D_{m,n}^{\text{(L)}}$ be the corresponding ``singular vector creating''
operators made of the Liouville Virasoro generators $L_{n}$, similar to the
operators $D_{m,n}^{\text{(M) }}$ introduced above. In fact $D_{m,n}%
^{\text{(L)}}$ is obtained from $D_{m,n}^{\text{(M)}}$ through the
substitution $M_{n}\rightarrow L_{n}$ and $b^{2}\rightarrow-b^{2}$. Like in
GMM, in LFT the corresponding singular states vanish \cite{Poincare,
Polyakov79}
\begin{equation}
D_{m,n}^{\text{(L)}}V_{m,n}=\bar D_{m,n}^{\text{(L)}}V_{m,n}=0
\label{Ldecoupling}%
\end{equation}
Let $D_{m,n}^{\text{(L)}}$ be normalized similarly to (\ref{DMnorm}) as
\begin{equation}
D_{m,n}^{\text{(L)}}=L_{-1}^{mn}+d_{1}^{(m,n)}(-b^{2})L_{-2}L_{-1}%
^{mn-2}+\ldots\label{DLnorm}%
\end{equation}

Define also the ``logarithmic degenerate'' fields
\begin{equation}
V_{m,n}^{\prime}=\frac12\frac\partial{\partial a}V_{a}|_{a=a_{m,n}}
\label{Vmnp}%
\end{equation}
for every pair $(m,n)$ of natural numbers. These fields are not primary. Under
conformal transformations $x\rightarrow y$ they transform as
\begin{equation}
\left|  y_{x}\right|  ^{2\Delta_{m,n}}V_{m,n}^{\prime}(y)=V_{m,n}^{\prime
}(x)-\Delta_{m,n}^{\prime}V_{m,n}(x)\log\left|  y_{x}\right|  \label{Vplog}%
\end{equation}
where $y_{x}$ stands for $\partial y/\partial x$. Nevertheless, as it is shown
in \cite{higher} $D_{m,n}^{\text{(L)}}\bar D_{m,n}^{\text{(L)}}V_{m,n}%
^{\prime}$ is a primary field and, moreover, the following identity holds for
the LFT operator fields
\begin{equation}
D_{m,n}^{\text{(L)}}\bar D_{m,n}^{\text{(L)}}V_{m,n}^{\prime}=B_{m,n}\tilde
V_{m,n} \label{HEM}%
\end{equation}
where $\tilde V_{m,n}=V_{a}|_{a=a_{m,-n}}$ is the Liouville exponential of
dimension $\Delta_{m,n}^{\text{(L)}}+mn$. The numerical constant $B_{m,n}$
reads
\begin{equation}
B_{m,n}=\frac{\left(  \pi\mu\gamma(b^{2})\right)  ^{n}b^{1+2n-2m}}%
{\gamma(1-m+nb^{2})}%
{\textstyle\prod_{k,l}^{\{m,n\}}}
2\lambda_{k,l} \label{Bmn}%
\end{equation}
where $\prod_{k,l}^{\{m,n\}}$ stands for the product over
\begin{equation}
(k,l)=(\{-m+1:1:m-1\}\otimes\{-n+1:1:n-1\})\setminus(0,0) \label{klset}%
\end{equation}
It is important to observe is that in GMG the exponential $\tilde V_{m,n}$ is
naturally combined with the corresponding minimal matter field $\Phi_{m,n} $
to form the dressed $(1,1)$ form (\ref{Umn}). This fact makes HEM crucial for
the integrability of (\ref{U4}) in MG.

\section{Generalized Minimal Gravity}

Here we quote some known results in GMG. It is repeatedly observed in the
literature, that in GMG the matter GMM parameter $b$ coincides with the one of
the corresponding LFT. This is why we keep the same notation throughout this
paper. For the dressed matter fields $U_{a}=\Phi_{\alpha}V_{a}$,
eq.(\ref{Dbalance}) allows two solutions. For definiteness let's take
\begin{equation}
U_{a}=\Phi_{a-b}V_{a} \label{Ua}%
\end{equation}
The GMG problem is to evaluate the gravitational correlation functions
(\ref{UG}) with the matter part given by the GMM expressions. Thus in GMG
we're restricted to the cases where the GMM correlation function is
unambiguously determined.

The three-point function is easily calculated by multiplying $C_{\text{M}%
}(a_{1}-b,a_{2}-b,a_{3}-b)$ by the corresponding Liouville three-point
function $C_{a_{1},a_{2},a_{3}}^{\text{(L)}}$. The resulting product can be
written in the form
\begin{equation}
\left\langle W_{a_{1}}W_{a_{2}}W_{a_{2}}\right\rangle _{\text{GMG}}=\Omega
N(a_{1})N(a_{2})N(a_{3}) \label{WWW}%
\end{equation}
where $W_{a}=C\bar CU_{a}$\footnote{Later on we'll use sometimes less compact
notations $U(a)=U_{a}$ and $W(a)=W_{a}$.},
\begin{equation}
\Omega=-\left[  \pi\mu\gamma(b^{2})\right]  ^{Q/b}\left[  \gamma(b^{2}%
)\gamma(b^{-2}-1)b^{-2}\right]  ^{1/2} \label{Omega}%
\end{equation}
and the ``leg-factors'' $N(a)$ read
\begin{equation}
N(a)=\left[  \pi\mu\gamma(b^{2})\right]  ^{-a/b}\left[  \gamma(2ab-b^{2}%
)\gamma(2ab^{-1}-b^{-2})\right]  ^{1/2} \label{NA}%
\end{equation}
The two-point function $\left\langle U_{a}U_{a}\right\rangle _{\text{GMG }}$
and the partition sum $Z_{\text{L}}$ can be restored from this expression in
the form
\begin{equation}
\left\langle U_{a}U_{a}\right\rangle _{\text{GMG }}=\left[  \pi\mu\gamma
(b^{2})\right]  ^{Q/b}\frac{N^{2}(a)}{\pi(2a-Q)} \label{UU}%
\end{equation}
and
\begin{equation}
Z_{\text{L}}=\left[  \pi\mu\gamma(b^{2})\right]  ^{Q/b}\frac{1-b^{2}}{\pi
^{3}Q\gamma(b^{2})\gamma(b^{-2})} \label{ZL}%
\end{equation}

For the normalized correlation functions $\left\langle \left\langle W_{a_{1}%
}W_{a_{2}}W_{a_{2}}\right\rangle \right\rangle =Z_{\text{L}}^{-1}\left\langle
W_{a_{1}}W_{a_{2}}W_{a_{2}}\right\rangle _{\text{GMG}}$ and $\left\langle
\left\langle U_{a}U_{a}\right\rangle \right\rangle =Z_{\text{L}}%
^{-1}\left\langle U_{a}U_{a}\right\rangle _{\text{GMG}}$ it is convenient to
use slightly different leg-factors
\begin{equation}
\mathcal{N}(a)=\pi N(a)\left[  \frac{\gamma(b^{2})\gamma(b^{-2})}%
{-(1-b^{-2})^{2}}\right]  ^{1/2}=\frac\pi{(\pi\mu)^{a/b}}\left[  \frac
{\gamma(2ab-b^{2})\gamma(2ab^{-1}-b^{-2})}{\gamma^{2a/b-1}(b^{2}%
)\gamma(2-b^{-2})}\right]  \label{NNa}%
\end{equation}
where for definiteness we suppose that the branch of the square root
is chosen in the way that
\begin{equation}
\mathcal{N}(b)=\mu^{-1} \label{NNb}%
\end{equation}%
\begin{align}
\left\langle \left\langle W_{a_{1}}W_{a_{2}}W_{a_{2}}\right\rangle
\right\rangle  &  =-(1+b^{-2})b^{-2}(b^{-2}-1)\prod_{i=1}^{3}\mathcal{N}%
(a_{i})\nonumber\\
\left\langle \left\langle U_{a}U_{a}\right\rangle \right\rangle  &
=\frac{(b^{-2}+1)b^{-2}(b^{-2}-1)}{(2ab^{-1}-b^{-2}-1)}\mathcal{N}^{2}(a)
\label{twothree}%
\end{align}

At the generic values of $a$ it will prove convenient to define renormalized
fields
\begin{equation}
\mathcal{U}(a)=\mathcal{N}^{-1}(a)U_{a}\;;\;\;\;\mathcal{W}(a)=\mathcal{N}%
^{-1}(a)W_{a} \label{Wfields}%
\end{equation}
for which (\ref{twothree}) is reduced to
\begin{align}
\left\langle \left\langle \mathcal{U}(a)\mathcal{U}(a)\right\rangle
\right\rangle  &  =\frac{(g+1)g(g-1)}{(2s-g-1)}\label{UUU23}\\
\left\langle \left\langle \mathcal{W}(a_{1})\mathcal{W}(a_{2})\mathcal{W}%
(a_{3})\right\rangle \right\rangle  &  =-(g+1)g(g-1)\nonumber
\end{align}
where $s=ab^{-1}$ and $g=b^{-2}$. It is readily verified that formally
$\mathcal{W}(a)=\mathcal{W}(Q-a)$, i.e., in this normalization the dressed
matter operators are independent on the choice of the dressing. This might
seem an important advantage. The price to pay is that the leg-factors
(\ref{NA}) are sometimes singular (and in any case depend on the cosmological
constant $\mu$).

\section{Discrete states and four-point integral}

The next level of difficulty is the four-point correlation number
$\left\langle U_{a_{1}}U_{a_{2}}U_{a_{3}}U_{a_{4}}\right\rangle _{\text{GMG}}$
given by the integral (\ref{U4}). If one of the four matter operators is
degenerate, e.g., $\Phi_{\alpha_{4}}=\Phi_{m,n}$, the matter four-point
function is constructed explicitly through (\ref{G4mn}). Let the other three
fields remain generic formal primaries of GMM\footnote{As we have discussed
above, the last requirement is essential, because sometimes correlation
functions with degenerate fields are not straightforward limits of those with
generic ones with the appropriate specialization of the parameter.}. Our
purpose is to evaluate the integral
\begin{equation}
\langle U_{m,n}U_{a_{1}}U_{a_{2}}U_{a_{3}}\rangle_{\text{GMG}}=\int
\left\langle U_{m,n}(x)W_{a_{1}}(x_{1})W_{a_{2}}(x_{2})W_{a_{3}}%
(x_{3})\right\rangle d^{2}x \label{U4mn}%
\end{equation}
where $U_{m,n}$ is the dressed degenerate field $\Phi_{m,n}$ defined in
(\ref{Umn}). Denote
\begin{equation}
\Theta_{m,n}=\Phi_{m,n}V_{m,n} \label{Tmn}%
\end{equation}
the direct product of the matter and Liouville degenerate fields, and
introduce the operators
\begin{equation}
\mathcal{D}_{m,n}=D_{m,n}^{\text{(M)}}+(-)^{mn}D_{m,n}^{\text{(L)}}
\label{DDmn}%
\end{equation}
(and similarly $\mathcal{\bar D}_{m,n}$) where $D_{m,n}^{\text{(M)}}$ and
$D_{m,n}^{\text{(L)}}$ are the matter and Liouville ``singular vector
creating'' operators (\ref{DMnorm}) and (\ref{DLnorm}).

\textbf{Proposition 1: }For every pair $(m,n)$ of positive integers an
operator $H_{m,n}$ exists, made of the Virasoro generators $M_{n},$ $L_{n}$
and the ghost fields $B$ and $C$ as a graded polynomial of order $mn-1$ and
ghost number $0$, such that $H_{m,n}\Theta_{m,n}$ is closed but non-trivial.
Operator $H_{m,n}$ is unique modulo exact terms, i.e., represents a
one-dimensional cohomology class.

Statement 1 can be verified by explicit calculations on the first levels. One
finds
\begin{align}
&  H_{1,2}=M_{-1}-L_{-1}+b^{2}CB\label{Hmn}\\
&  H_{1,3}=M_{-1}^{2}-M_{-1}L_{-1}+L_{-1}^{2}-2b^{2}\left(  M_{-2}%
+L_{-2}\right)  +\nonumber\\
&  \qquad\qquad\qquad+2b^{2}\left(  M_{-1}-L_{-1}\right)  CB-4b^{4}C\partial
B\nonumber
\end{align}
For the series $(1,n)$ a proof based on the explicit expression for the
operators $D_{m,n}^{\text{(L)}}$ and $D_{m,n}^{\text{(M)}}$ \cite{StAubin}, is
given in ref.\cite{Imbimbo}. At general $(m,n)$ the statement is most
certainly also true \cite{Feigin}.

Cohomology classes $H_{m,n}\Theta_{m,n}$ were discovered in \cite{Witten,
Discstates} and are called the ``discrete states''. Although the generic form
of the operators $H_{m,n}$ is not known to us, the normalization is supposed
to be fixed as
\begin{equation}
H_{m,n}=\sum_{k=0}^{mn-1}\left(  M_{-1}\right)  ^{mn-1-k}\left(
-L_{-1}\right)  ^{k}+\ldots\label{Hmnnorm}%
\end{equation}
Apparently
\begin{equation}
\left(  \partial H_{m,n}-\mathcal{Q}R_{m,n}\right)  \Theta_{m,n}=\left(
\bar\partial\bar H_{m,n}-\mathcal{\bar Q}\bar R_{m,n}\right)  \Theta_{m,n}=0
\label{DmnTmn}%
\end{equation}
where $R_{m,n}$ is again a graded polynomial in $M_{n},$ $L_{n}$ and ghosts.

\textbf{Proposition 2:}
\begin{equation}
\mathcal{D}_{m,n}\mathcal{\bar D}_{m,n}\Theta_{m,n}^{\prime}=\left(  \partial
H_{m,n}-\mathcal{Q}R_{m,n}\right)  \left(  \bar\partial\bar H_{m,n}%
-\mathcal{\bar Q}\bar R_{m,n}\right)  \Theta_{m,n}^{\prime} \label{DmnDmn}%
\end{equation}
where
\begin{equation}
\Theta_{m,n}^{\prime}=\Phi_{m,n}V_{m,n}^{\prime} \label{Tmnp}%
\end{equation}
and $V_{m,n}^{\prime}$ is from eq.(\ref{Vmnp}).

We verified relation (\ref{DmnDmn}) directly for $(m,n)=(1,2)$ and $(1,3)$.
Thus, it is not excluded that more general case might require some
modifications. Combined with HEM (\ref{HEM}) this statement gives the precise
local form of the ``cohomological HEM'' (\ref{CHEM}). In particular, it
permits us to replace eq.(\ref{U4mn}) by (\ref{dd}) and then rewrite it as
\begin{equation}
B_{m,n}^{-1}\int_{\partial\Gamma}\partial\left\langle O_{m,n}^{\prime
}(x)W_{a_{1}}(x_{1})W_{a_{2}}(x_{2})W_{a_{3}}(x_{3})\right\rangle \frac
{dx}{2i} \label{dGamma}%
\end{equation}
where
\begin{equation}
O_{m,n}^{\prime}=H_{m,n}\bar H_{m,n}\Theta_{m,n}^{\prime} \label{Omnp}%
\end{equation}
The moduli integral is hence reduced to the boundary integral and the
so-called curvature contribution. The boundary consists of three small circles
$\partial\Gamma=\sum_{i=1}^{3}\partial\Gamma_{i}$ around the $W$-insertions
(integrated clockwise) and a big circle $\partial\Gamma_{\infty}$ near
infinity (integrated counterclockwise), leading to what is called the
curvature contribution. To evaluate the boundary terms we need to understand
better the short-distance behavior of the operator product $O_{m,n}^{\prime
}(x)W_{a}(0)$. As the first step we discuss the curvature term.

\section{Curvature term}

The curvature term comes from the fact that the operator $O_{m,n}^{\prime}$ is
not exactly a scalar (a $(0,0)$ form) but a logarithmic field. Under conformal
coordinate transformations $x\rightarrow y$ it acquires an inhomogeneous part
\begin{equation}
O_{m,n}^{\prime}(y)=O_{m,n}^{\prime}(x)-\Delta_{m,n}^{\prime}O_{m,n}%
(x)\log\left|  y_{x}\right|  \label{Olog}%
\end{equation}
where
\begin{equation}
O_{m,n}=H_{m,n}\bar H_{m,n}\Theta_{m,n} \label{ringmn}%
\end{equation}
is the ground ring element (see below) and
\begin{equation}
\Delta_{m,n}^{\prime}=\frac d{da}\Delta_{a}^{\text{(L)}}|_{a=a_{m,n}}%
=mb^{-1}+nb=2\lambda_{m,n} \label{Dmnp}%
\end{equation}
This subtlety is can be treated in two ways. First, it is easy to show that on
the sphere the transformation (\ref{Olog}) leads to the following behavior of
the correlation function with $O_{m,n}^{\prime}(x)$ at $x\rightarrow\infty$%
\begin{equation}
\left\langle O_{m,n}^{\prime}(x)W_{a_{1}}(x_{1})W_{a_{2}}(x_{2})W_{a_{3}%
}(x_{3})\right\rangle \sim-2\Delta_{m,n}^{\prime}\log(x\bar x)\left\langle
O_{m,n}W_{a_{1}}W_{a_{2}}W_{a_{3}}\right\rangle \label{Opinf}%
\end{equation}
Therefore the curvature contribution can be included as a boundary term
$\partial\Gamma_{\infty}$ at $\infty$. It is evaluated as
\begin{equation}
\frac1{2i}\int_{\partial\Gamma_{\infty}}\partial\left\langle O_{m,n}^{\prime
}(x)W_{a_{1}}(x_{1})W_{a_{2}}(x_{2})W_{a_{3}}(x_{3})\right\rangle
dx=-2\pi\lambda_{m,n}\left\langle O_{m,n}W_{a_{1}}W_{a_{2}}W_{a_{3}%
}\right\rangle \label{G4curv}%
\end{equation}
Another trick, which is easier generalized for more complicated surfaces, is
to keep a trace of the background metric $\widehat{g}_{ab}=e^{\sigma}%
\delta_{ab}$. Since the scale factor $\sigma(x)$ transforms as
\begin{equation}
\sigma(y)=\sigma(x)-2\log\left|  y_{x}\right|  \label{slog}%
\end{equation}
under conformal maps, the combination
\begin{equation}
\tilde O_{m,n}^{\prime}(x)=O_{m,n}^{\prime}(x)-\Delta_{m,n}^{\prime}%
\sigma(x)O_{m,n}(x)/2 \label{Otwiddle}%
\end{equation}
is a scalar (the dependence on the background metric is the price to pay).
Thus, in the BRST invariant environment equation (\ref{DmnDmn}) can be
rewritten as
\begin{equation}
B_{m,n}U_{m,n}=\sqrt{\widehat{g}}\left(  \frac14\widehat{\triangle}\tilde
O_{m,n}^{\prime}-\frac{\Delta_{m,n}^{\prime}}8\widehat{R}O_{m,n}\right)
\label{covHEM}%
\end{equation}
where $\widehat{\triangle}$ is the covariant Laplace operator with respect to
$\widehat{g}_{ab}$ and $\widehat{R}$ is the corresponding scalar curvature. On
a sphere the contribution of the second term apparently reduces to
(\ref{G4curv}).

At this step it is clear that a better understanding of the ground ring
structure in GMG, in particular the evaluation of the expectation value in the
right hand side of eq.(\ref{G4curv}), is of importance in the program.

\section{Ground ring in GMG}

It has been discovered in refs.\cite{Discstates,Witten} that in MG the
degenerate fields $\Phi_{m,n}$ of GMM, when combined with the degenerate
exponentials $V_{m,n}$ of the corresponding LFT, give rise to non-trivial BRST
invariant operators (\ref{ringmn}) with ghost number $0$ and conformal
dimension $(0,0)$. Some of these operators were evaluated explicitly in
\cite{Imbimbo}. The spatial derivatives $\partial O_{m,n}$ and $\bar\partial
O_{m,n}$ are exact (\ref{DmnTmn}). Therefore the correlation functions of
these discrete states in the BRST closed environment do not depend on their
positions. Moreover, as the BRST cohomology classes, they form a closed ring
under the operator product expansions, called the ground ring. This
observation led E.Witten \cite{Witten} to conclude that this object plays a
crucial role in MG and probably the complete algebraic structure of the theory
is in fact that of the ground ring. In this section we present few explicit
calculations revealing the GR properties. Cohomology properties of $O_{m,n}$
are relevant only in a $\mathcal{Q}$-invariant environment. The simplest
invariant state on a sphere is created by three operators $W_{a}$. For this
reason we perform actual calculation of the correlation function of
$\left\langle O_{m,n}W_{a_{1}}W_{a_{2}}W_{a_{3}}\right\rangle $ on a sphere
with three generic $W_{a}$ insertions. Notice, that we again suppose all the
three parameters $a_{1}$, $a_{2}$ and $a_{3}$ generic. Certain delicate
effects, which we're not going to touch here, might take place if one or more
of $W_{a}$ involve reducible representations of either matter or Liouville
Virasoro algebras.

Modulo exact forms the discrete states $O_{m,n}$ act in the space of classes
$W_{a}$. This is because their action doesn't change the ghost number and
generically all non-trivial classes are exhausted by the composite fields
$W_{a}$ with different $a$. Moreover, due to the decoupling restrictions in
the OPE of the degenerate fields $\Phi_{m,n}$ and $V_{m,n}$ with the primaries
$\Phi_{\alpha}$ and $V_{a}$ respectively, the general structure of the
operator product $O_{m,n}(x)W(a)$ is doomed to have the form
\begin{equation}
O_{m,n}W(a)=\sum_{r,s=0}^{(m,n)}A_{r,s}^{(m,n)}W(a+\lambda_{r,s})+\text{exact}
\label{OW}%
\end{equation}
with some numerical coefficients $A_{r,s}^{(m,n)}$. Our immediate goal is to
evaluate these numbers.

It is instructive to perform explicit calculations in the simplest case
$(m,n)=(1,2)$. The special OPE we need in this case are (\ref{Phi12}) and
\begin{equation}
V_{1,2}(y)V_{a}(0)=C_{+}^{\text{(L)}}(a)(y\bar y)^{ab}\left[  V_{a-b/2}%
\right]  +C_{-}^{\text{(L)}}(a)(y\bar y)^{1-ab+b^{2}}\left[  V_{a+b/2}\right]
\label{V12Va}%
\end{equation}
where
\begin{equation}
C_{+}^{\text{(L)}}(a)=1\;;\;\;\;C_{-}^{\text{(L)}}(a)=-\frac{\pi\mu}%
{\gamma(-b^{2})}\frac{\gamma(2ab-b^{2}-1)}{\gamma(2ab)} \label{CLpm}%
\end{equation}
It is easy to verify by explicit calculation (at least at the primary field
level) that in the product $U_{a}=\Phi_{a-b}V_{a}$ the action of $H_{1,2}$ and
$\bar H_{1,2}$ eliminates the ``wrong terms'' (i.e., those which include the
combinations $\Phi_{a-b/2}V_{a-b/2}$ and $\Phi_{a-3b/2}V_{a+b/2}$) and we are
left with\footnote{A different derivation of the action of $O_{1,2}$ on a generic
class $W(a)$ has been carried out in ref.\cite{Valya}. As we got to know,
V.Petkova has arrived to this form as early as the spring of 2004. See also
earlier discussion in \cite{SeibergShish}.}
\begin{equation}
O_{1,2}W(a)=A_{0,-1}^{(1,2)}W(a-b/2)+A_{0,1}^{(1,2)}W(a+b/2)+\text{exact}
\label{O12W}%
\end{equation}
Here explicitly
\begin{align}
&  A_{0,-1}^{(1,2)}=\left(  1-2ab+b^{2}\right)  ^{2}C_{-}^{\text{(M)}%
}(a-b)C_{+}^{\text{(L)}}(a)\label{Ars}\\
&  A_{0,1}^{(1,2)}=\left(  1-2ab+b^{2}\right)  ^{2}C_{+}^{\text{(M)}%
}(a-b)C_{-}^{\text{(L)}}(a)\nonumber
\end{align}
The polynomial multipliers in the coefficients are the result of the action of
$H_{1,2}\bar H_{1,2}$ on the corresponding terms in the expansion of
$\Theta_{1,2}(x)W_{a}(0)$. Similar calculation can be performed directly for
the action of every $\Theta_{m,n}$, level by level. We verified the
cancelation of the ``wrong terms'' explicitly in the case $(m,n)=(1,3)$ and
carried out the polynomials appearing due to the action of $H_{1,3}\bar
H_{1,3}$. The result is summarized as follows
\begin{equation}
N(a+\lambda_{r,s})A_{r,s}^{(m,n)}=\Lambda_{m,n}N(a) \label{Aex}%
\end{equation}
where
\begin{equation}
\pi\Lambda_{m,n}=B_{m,n}\mathcal{N}(a_{m,-n}) \label{Lambda}%
\end{equation}
and $B_{m,n}$ are the same as in eq.(\ref{Bmn}) and the factor $\mathcal{N}%
(a)$ was introduced in eq.(\ref{NNa}).

It seems tempting to simplify these relations by introducing the renormalized
fields $\mathcal{W}(a)$ as in eq.(\ref{Wfields}) and
\begin{equation}
\mathcal{O}_{m,n}=\Lambda_{m,n}^{-1}O_{m,n} \label{OOmn}%
\end{equation}
Expression (\ref{OW}) is reduced to
\begin{equation}
\mathcal{O}_{m,n}\mathcal{W}(a)=\sum_{k,l}^{(m,n)}\mathcal{W}(a+kb+lb^{-1})
\label{OOW}%
\end{equation}

This coincides with what has been figured out previously
\cite{SeibergShish} on the basis of more general arguments. Here we
arrive at the same expression through a direct calculation
\cite{BZ1} (see also \cite{Valya} for a related treatment). So far
we did it explicitly only for a restricted number of particular
examples. In particular, simple expression (\ref{Aex}) appears as a
result of mysterious interplay between different terms in the
explicit expressions for $H_{m,n}$ and the singular vectors. Simple
result of complicated calculations certainly implies a hidden
structure behind. However, the general derivation revealing this
structure, remains an open problem. Another important feature of our
treatment, which differs it from that of ref.\cite{SeibergShish}, is
that we considered the action of $O_{m,n}$ on a cohomology $W_{a}$
with generic $a$. It is natural to expect, that the relations
(\ref{OOW}) are modified when specialized to the degenerate classes
$W_{m,n}=C\bar C\Phi_{m,n}\tilde V_{m,n}$ corresponding to
irreducible representations of the matter Virasoro symmetry (i.e.,
with vanishing singular vector in the matter sector). Although the
effect might simply result in the proper truncation of the sum in
eq.(\ref{OOW}) implied by the fusion algebra of the degenerate
fields, technically the limit $a\rightarrow a_{m,-n}$ in this
expression turns out to be subtle and requires more careful
analysis. Therefore, as it has been already mentioned above, in this
article we restrict ourselves to the case of generic values of $a$,
leaving the degenerate irreducible situation for further study. The
present result is sufficient for our subsequent treatment of the
integral (\ref{U4mn}) with three generic non-degenerate values of
$a_{1} $, $a_{2}$ and $a_{3}$.

The simple form (\ref{OOW}) naturally implies the following structure of the
ground ring algebra
\begin{equation}
\mathcal{O}_{m,n}\mathcal{O}_{m^{\prime},n^{\prime}}=\sum_{l}^{[m,m^{\prime}%
]}\sum_{k}^{[n,n^{\prime}]}\mathcal{O}_{l,k} \label{rings}%
\end{equation}
where the symbol $\sum_{k}^{[n,n^{\prime}]}$ stands for the sum over
$k=\left\{  \left|  n-n^{\prime}\right|  +1:2:n+n^{\prime}-1\right\}  $. Or,
following \cite{SeibergShish} one can introduce the generating elements
$X=\mathcal{O}_{1,2}/2$ and $Y=\mathcal{O}_{2,1}/2$ and rewrite (\ref{rings})
in the form
\begin{equation}
\mathcal{O}_{m,n}=U_{m-1}\left(  Y\right)  U_{n-1}\left(  X\right)
\label{Chebyshev}%
\end{equation}
where $U_{n}(x)$ are the Chebyshev polynomials of the second kind.

\section{Boundary terms}

The three boundary integrals in (\ref{dGamma})
\begin{equation}
g_{i}=\int_{\partial\Gamma_{i}}\partial\left\langle O_{m,n}^{\prime
}(x)W_{a_{1}}(x_{1})W_{a_{2}}(x_{2})W_{a_{3}}(x_{3})\right\rangle \frac
{dx}{2i} \label{dGammai}%
\end{equation}
are controlled by the OPE of the ``logarithmic primitive'' $O_{m,n}^{\prime} $
and the (generic in our case) states $W(a_{i})$. A straightforward way to
evaluate this expansion is to carry out first such expansion for the
``primitive'' product $\Theta_{m,n}^{\prime}(x)=\Phi_{m,n}(x)V_{m,n}^{\prime
}(x)$ and then decorate it with $C\bar C(0)$ and apply $H_{m,n}\bar H_{m,n}$.
Again the last OPE is a product of two independent ones, those for $\Phi
_{m,n}(x)\Phi_{\alpha}(0)$ and for $V_{m,n}^{\prime}(x)V_{a}(0)$. While the
first is the same discrete degenerate OPE (\ref{Phimn}) as in the case of
ground ring calculations, the second is more complicated and requires a
separate analysis.

\begin{figure}
[tbh]
\begin{center}
\includegraphics[
trim=0.000000in 0.000000in -0.010918in 0.007004in,
height=2.9092in,
width=5.079in
]%
{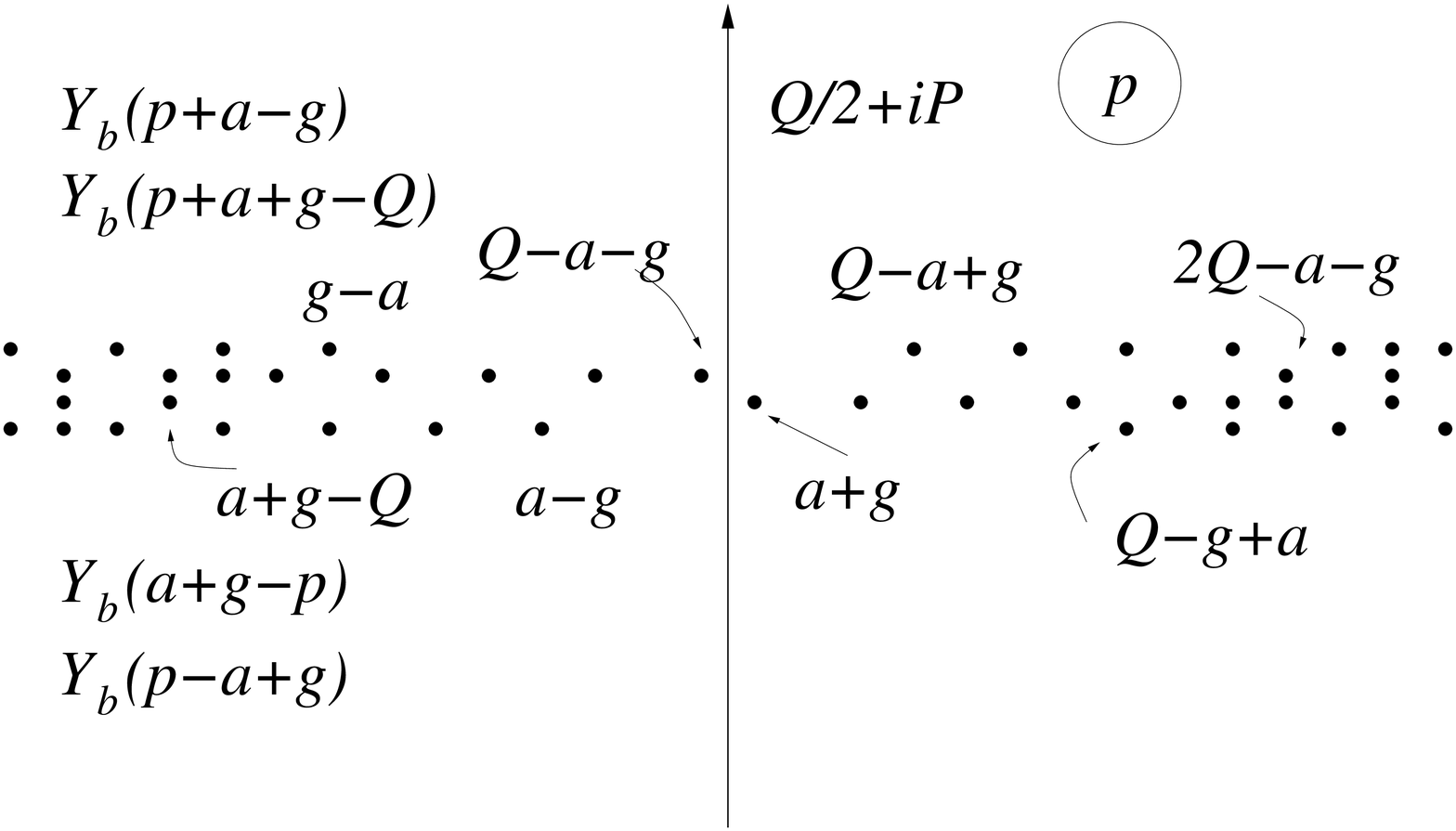}%
\caption{Location of the poles of the sturcture constant while $a+g>Q/2.$}%
\label{c1}%
\end{center}
\end{figure}

\begin{figure}
[tbhtbh]
\begin{center}
\includegraphics[
trim=0.000000in 0.000000in -0.013647in -0.001531in,
height=2.9092in,
width=5.1491in
]%
{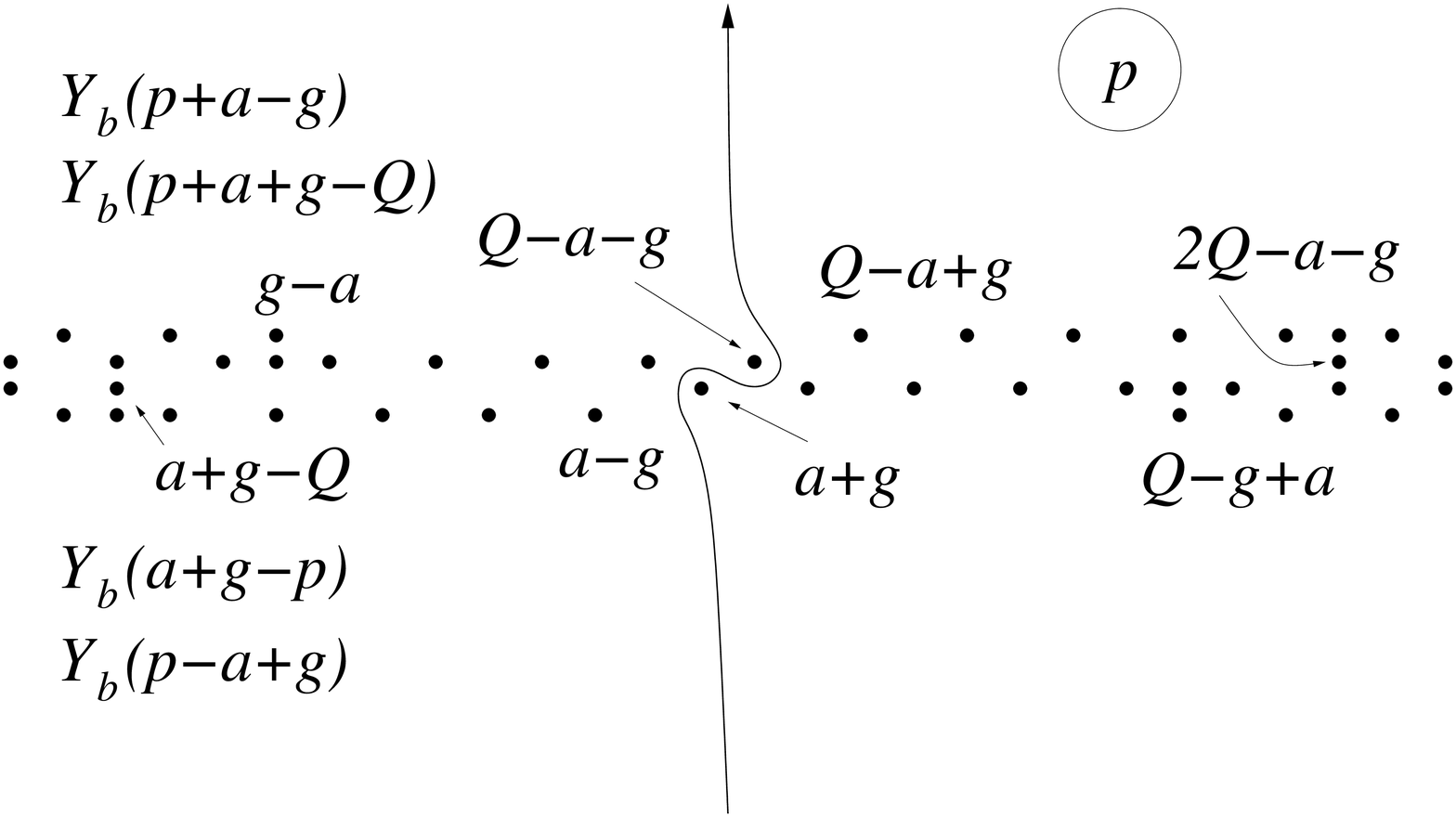}%
\caption{The contour deformation due to the analytic continuation of the OPE
(\ref{LOPEp}) away from the basic domain.}%
\label{c2}%
\end{center}
\end{figure}

The most direct way is to start with the general ``continuous'' OPE
(\ref{LOPE}), which we rewrite here as\footnote{In this section the letter $g$
does not stand for $b^{-2}$ as in sect.4 and below in sect.11.}
\begin{equation}
V_{g}(x)V_{a}(0)=\int_{\uparrow}^{\prime}\frac{dp}{4\pi i}C_{g,a}%
^{\text{(L)}p}(x\bar{x})^{\Delta_{p}^{\text{(L)}}-\Delta_{g}^{\text{(L)}%
}-\Delta_{a}^{\text{(L)}}}\left[  V_{p}(0)\right]  \label{LOPEp}%
\end{equation}
where $\uparrow$ passes through $Q/2$ along the imaginary axis and the prime
indicates the deformations necessary for the analytic continuation from the
``basic domain'' (\ref{basic}). The singularities of the structure constant
\begin{equation}
C_{g,a}^{\text{(L)}p}=\frac{(\pi\mu\gamma(b^{2})b^{2-2b^{2}})^{(p-a-g)}%
\Upsilon_{b}(b)\Upsilon_{b}(2g)\Upsilon_{b}(2a)\Upsilon_{b}(2Q-2p)}%
{\Upsilon_{b}(p+a-g)\Upsilon_{b}(a+g+p-Q)\Upsilon_{b}(a+g-p)\Upsilon
_{b}(p+g-a)}\label{Lstr}%
\end{equation}
are determined by zeros of the four $\Upsilon_{b}$-functions in the
denominator. An example of their location is shown in fig.\ref{c1}, where we
have chosen both $a$ and $g$ real, positive and less then $Q/2$. The pattern
in this figure corresponds to the ``basic domain'', i.e., $a+g>Q/2$. The
``right'' zeros of all the four multipliers in the denominator are to the
right and all ``left'' ones are to the left from the integration contour
$\uparrow$, which in this case remains a straight line going vertically
through $Q/2$. The strings of zeros are shifted slightly from the real axis to
better distinguish zeros coming from different factors. The uppermost and the
next string from the above are due to the factors $\Upsilon_{b}(p+a-g)$ and
$\Upsilon_{b}(a+g+p-Q)$ respectively. Then lie the zeros of $\Upsilon
_{b}(a+g-p)$ and the lowest string belongs to the multiplier $\Upsilon
_{b}(p+g-a)$.
Then if e.g. the parameter $g$ decreases and $a+g$ becomes less than $Q/2$ the
two poles at $a+g$ and $Q-a-g$ cross the vertical line $\operatorname*{Re}%
p=Q/2$ (called often the Seiberg bound \cite{Seiberg}). Analyticity requires
the integration contour to be deformed accordingly (fig.\ref{c2}).
\begin{figure}
[tbhtbhtbh]
\begin{center}
\includegraphics[
trim=0.000000in 0.000000in -0.006253in -0.010642in,
height=2.8694in,
width=5.2788in
]%
{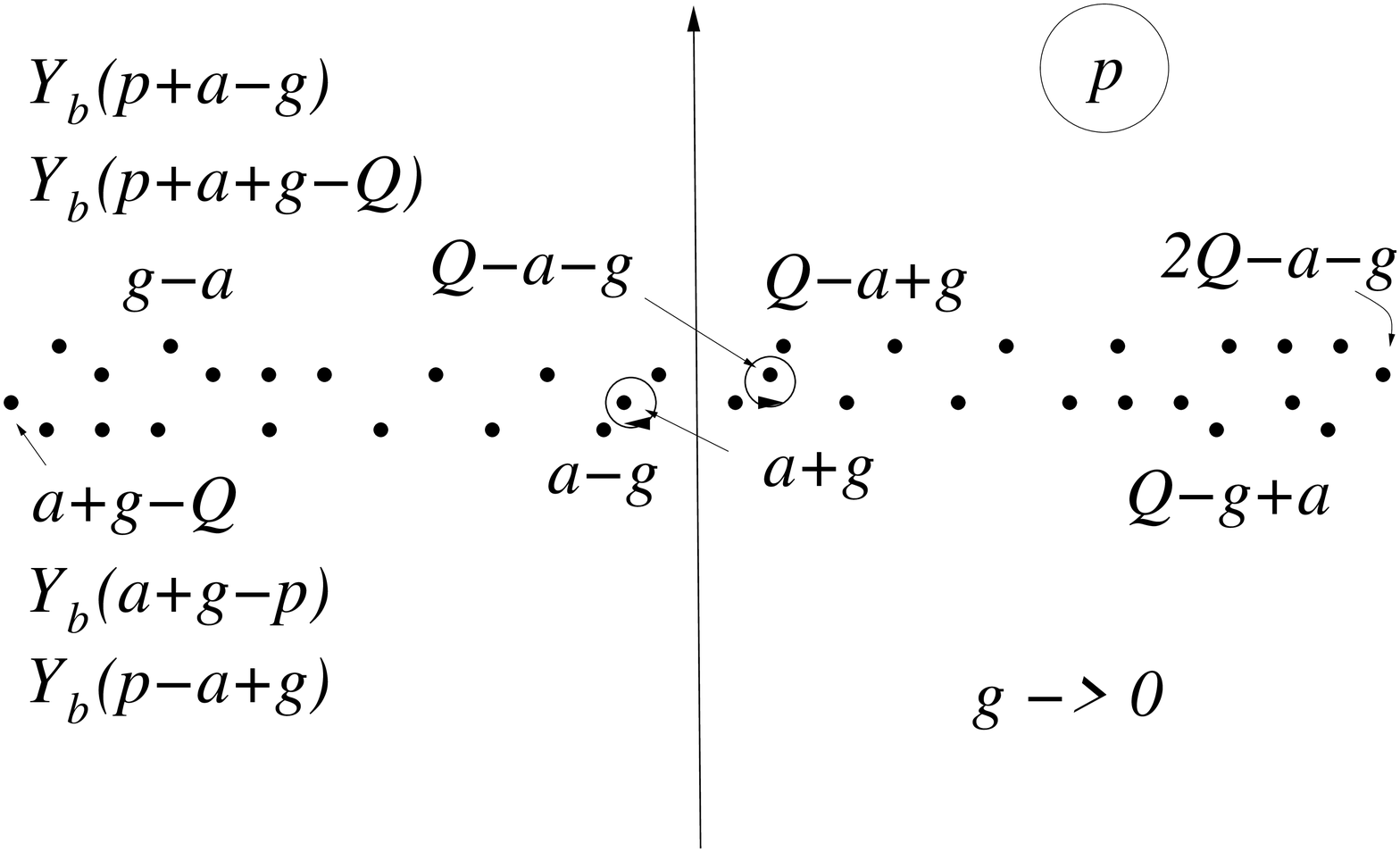}%
\caption{``Discrete terms'' due to the poles at $p=a+g$ and $p=Q-a-g$ are
picked up explicitly. These contributions are singular at $g\rightarrow0$ due
to close poles at $p=a-g$ (resp. at $p=Q-a+g$). Notice that we pick up as the
discreted term the pole located close to the vertical line $\operatorname*{Re}%
p=Q/2$.}%
\label{c3}%
\end{center}
\end{figure}
The effect of this deformation can be separated as the so called discrete
terms
\begin{align}
V_{g}(x)V_{a}(0) &  =\frac{1}{2}(x\bar{x})^{-2ag}[V_{a+g}(0)]+\frac{1}%
{2}(x\bar{x})^{-2ag}R_{\text{L}}(a+g)[V_{Q-a-g}(0)]\label{disc1}\\
&  +\int_{\uparrow}\frac{dp}{4\pi i}C_{g,a}^{\text{(L)}p}(x\bar{x}%
)^{\Delta_{p}^{\text{(L)}}-\Delta_{g}^{\text{(L)}}-\Delta_{a}^{\text{(L)}}%
}\left[  V_{p}(0)\right]  \nonumber
\end{align}
like it is shown in fig.\ref{c3}, where the two poles $a+g$ and $Q-a-g$ are
picked up explicitly and the corresponding residues are evaluated. Notice,
that the two discrete terms in (\ref{disc1}) are in fact identical due to the
reflection relation (\ref{Lrefl}). This is in fact a consequence of the
complete symmetry of the integral (\ref{LOPEp}) under the reflection
$p\rightarrow Q-p$ and therefore holds for all ``mirror images'' w.r.t. this
symmetry. Below we'll use this feature to keep only one of each pair of
images, say that with $\operatorname*{Re}p<Q/2$ and then supply the answer
with the factor of $2$. Further change of the parameters may force more poles
to cross the contour and there will be more discrete terms in the right hand
side of (\ref{disc1}).

Another important remark is in order here. In the derivation of (\ref{disc1})
above we implied that $\operatorname*{Re}(a+g)<Q/2$. A quick reconsideration
of the opposite case $\operatorname*{Re}(a+g)>Q/2$ shows that we have to pick
up instead the poles $p=Q-a+g$ and $p=a-g$. This replaces eq.(\ref{disc1}) by
\begin{equation}
V_{g}(x)V_{a}(0)=(x\bar x)^{-2(Q-a)g}R_{\text{L}}(a)[V_{Q-a+g}(0)]+\int
_{\uparrow}\frac{dp}{4\pi i}C_{g,a}^{\text{(L)}p}(x\bar x)^{\Delta
_{p}^{\text{(L)}}-\Delta_{g}^{\text{(L)}}-\Delta_{a}^{\text{(L)}}}\left[
V_{p}(0)\right]  \label{dr1}%
\end{equation}
In general, if two poles of the integrand pinch the integration contour,
similarly to what we have just observed in a simple example, it is the correct
choice to pick up explicitly the residue of the pole which is closer to the
bound $\operatorname*{Re}p=Q/2$ (and finally to put remaining integration
contour to its original position $\operatorname*{Re}p=Q/2$). Opposite choice
is misleading since in this case the residual integral part contains more
important terms than those taken into account. Finally it is convenient to use
the reflection relations and put all the discrete terms $V_{a}$ to the half
plane $\operatorname*{Re}a<Q/2$.

Our purpose is to study (\ref{LOPEp}) at $g$ close to certain degenerate value
$g\rightarrow a_{m,n}=Q/2-\lambda_{m,n}$. It is seen immediately that the
structure constant (\ref{Lstr}) contains an overall multiplier $\Upsilon
_{b}(2g)$ vanishing in this limit. Hence, the singularities arising from the
divergencies of the integral are very important. To give an idea of what
happens in general we consider first the simplest possible case $g\rightarrow
a_{1,1}=0$. The corresponding degenerate field $V_{1,1}$ is just the identity
operator while the logarithmic primary $V_{1,1}^{\prime}$ coincides with the
basic Liouville field $\phi$. In the limit $g\rightarrow0$ the integral term
in both equations (\ref{disc1}) and (\ref{dr1}) disappears and we arrive at
pure $V_{a}(0)$ (as it of course should be for the identity operator at the
place of $V_{g}$ at the left hand side). This is the simplest, trivial case of
the discrete degenerate OPE (similarly to (\ref{Phimn}) in GMM)
\begin{equation}
V_{m,n}(x)V_{a}(0)=\sum_{r,s}^{(m,n)}\left(  x\bar x\right)  ^{\lambda
_{r,s}(Q-2a-\lambda_{r,s})-\Delta_{m,n}^{\text{(L)}}}C_{r,s}^{\text{(L)}%
}(a)\left[  V_{a+\lambda_{r,s}}\right]  \label{VmnV}%
\end{equation}
which hold for the fields $V_{m,n}$ due to the decoupling (\ref{Ldecoupling})
of the singular vectors. Further, the linear in $g$ term in (\ref{disc1})
gives
\begin{equation}
\phi(x)V_{a}(0)=-a\log(x\bar x)V_{a}(0)+\text{less singular terms}
\label{phifree}%
\end{equation}
This logarithmic OPE, which holds at $\operatorname*{Re}a<Q/2$, apparently
simulates the similar expansion in the theory of free scalar field. However,
at $\operatorname*{Re}a>Q/2$ we have to differentiate in $g$ eq.(\ref{dr1})
instead. The net result
\begin{equation}
\phi(x)V_{a}(0)=\left(  \left|  Q/2-a\right|  _{\operatorname*{Re}%
}-Q/2\right)  \log(x\bar x)V_{a}(0)+\ldots\label{phiVg}%
\end{equation}
is easier formulated in terms of the symbol
\begin{equation}
\left|  x\right|  _{\operatorname*{Re}}=%
\genfrac{\{}{.}{0pt}{}{\;\;x\;\;\;\text{if }\operatorname*{Re}%
x>0}{-x\;\;\;\text{if }\operatorname*{Re}x<0}%
\label{are}%
\end{equation}
which will be repeatedly used throughout what follows. Notice, that for the
logarithmic degenerate field the integral term in the right hand side of
(\ref{dr1}) doesn't vanish, as it had place in the case of an authentic
degenerate field. The OPE (\ref{phiVg}) remains continuous, although the
integral term is less singular than the logarithmic one\footnote{This is
because of our prescription to always pick up the pole closest to the line
$\operatorname*{Re}p=Q/2$.}. We will see before long that the mechanism
leading to the non-analytic structure (\ref{are}) in this simple case is
general and gives rise to all non-analyticities in the boundary terms.

\begin{figure}
[tbh]
\begin{center}
\includegraphics[
trim=0.000000in 0.000000in 0.003060in 0.000596in,
height=2.9784in,
width=5.0998in
]%
{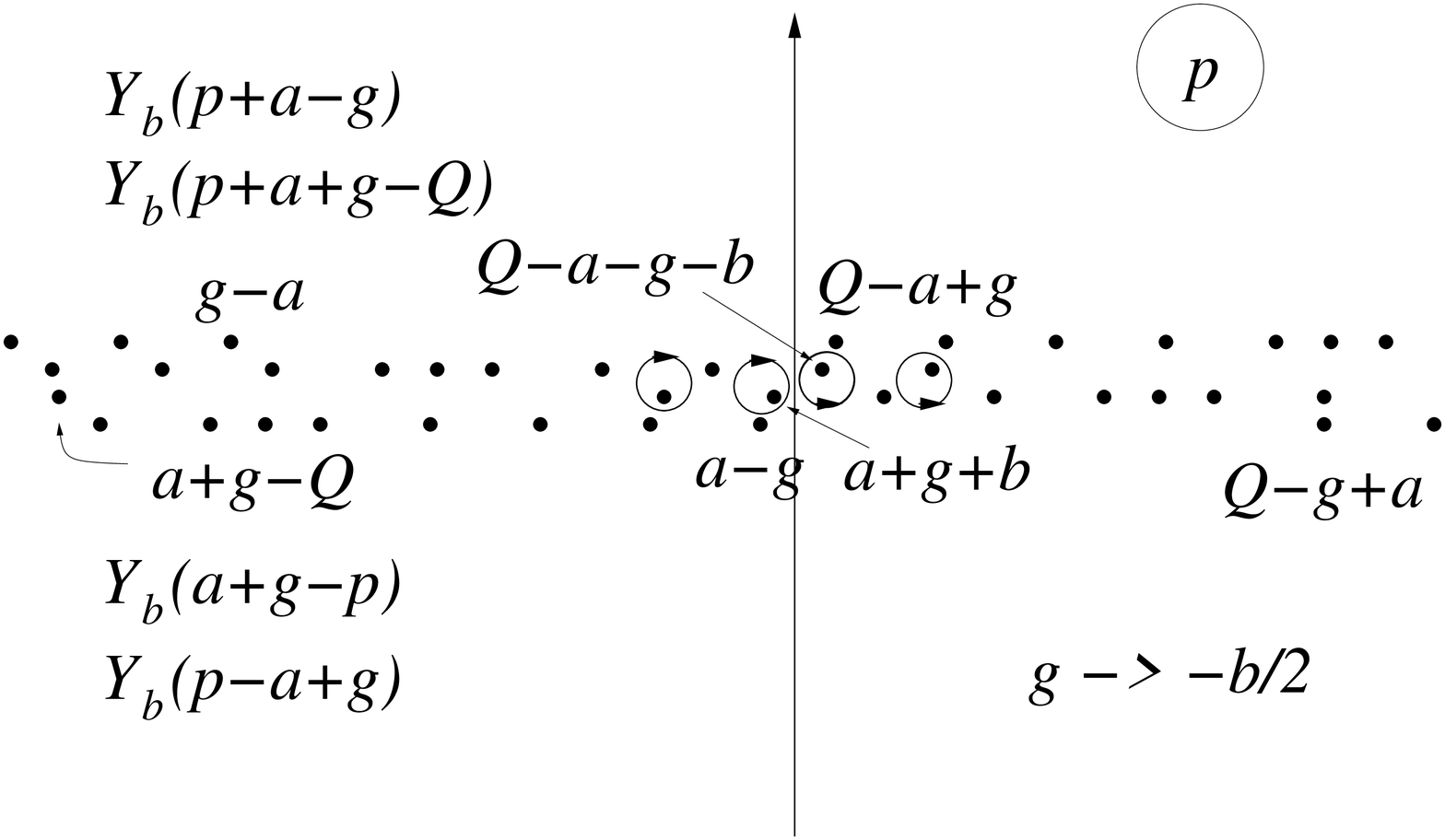}%
\caption{Singular terms in the integral (\ref{LOPEp}) in the limit
$g\rightarrow-b$. The discrete terms are picked up explicitly. This picture
corresponds to the case $\operatorname*{Re}(a+g)<Q/2-b$.}%
\label{c4}%
\end{center}
\end{figure}

After this simple warm up we consider more complicated case $g\rightarrow-b$,
which corresponds to $V_{1,2}$ and, in the logarithmic case, to $V_{1,2}%
^{\prime}$. A sample of poles of the structure constant is demonstrated in
fig.\ref{c4}. The string of right zeros of $\Upsilon_{b}(a+g-p)$ penetrates
further to the half-plane $\operatorname*{Re}p<Q/2$ and the first and second
zeros at $p=a+g$ and $p=a+g+b$ hit respectively the second and the first ones
from the ``left'' string of the factor $\Upsilon_{b}(p-a+g) $ at $p=a-g-b$ and
$p=a-g$ (as usual there are symmetric under $p\rightarrow Q-p$ pinches in the
left half-plane $\operatorname*{Re}p>Q/2$ which give identical contributions
and therefore are not discussed separately).
In the limit $g\rightarrow-b$ these pinches produce singularities, which
neutralize the overall zero in the factor $\Upsilon_{b}(2g)$ and give just the
two-term discrete OPE (\ref{V12Va}). Of course, this discrete degenerate OPE,
as well as the more general one (\ref{VmnV}), are easier figured out from the
null-vector decoupling and self-consistency conditions (the bootstrap). Here
we reproduce it more systematically from the generic Liouville OPE
(\ref{LOPE}) mainly to show the mechanism leading to the singular discrete
terms and to vanishing of the continuous integral part. Moreover, this
approach offers a way to carry out important terms at the next order in the
$g+b$ expansion, i.e., in the OPE $V_{m,n}^{\prime}(x)V_{a}(0)=\ldots$.

First of all, it is clear that the terms of the form
\begin{equation}
O_{m,n}^{\prime}(x)W_{a}(0)=\ldots+\log(x\bar x)R_{m,n}^{a}(0)+\ldots
\label{Oplog}%
\end{equation}
(where $R_{m,n}^{a}$ is some local operator to be discussed below) are of the
most interest in the calculation of the boundary terms (\ref{dGammai}). This
is because

\begin{enumerate}
\item  Such terms give finite contribution to (\ref{dGammai}).

\item  Less singular terms are not important in the integral (\ref{dGammai}).

\item  Contributions of more singular terms (if there are any\footnote{As we
have argued above, the residual integral term is less singular then the
discrete ones, separated according to our prescription. Even stronger short
distance singularities may appear as the additional discrete terms, which
vanish in the degenerate expansion but survive in the logarithmic one.})
depend singularly on the radius of the circle $\partial\Gamma_{i}$. In field
theory we attribute such divergencies to certain singular renormalizations and
therefore do not count them in the definition of the integral (\ref{U4mn}).
\end{enumerate}

Terms of this type can come only from those contributions to $V_{m,n}^{\prime
}(x)V_{a}(0)$ where the derivative with respect to $g$ in eq.(\ref{LOPEp}) (or
w.r.t. $a$ in the definition (\ref{Vmnp}))\ acts on the exponent of the
$(x\bar x)$-dependence. Moreover, such terms appear only in the discrete
terms, where the vanishing $\Upsilon_{b}(2g)$ is compensated by a singularity
of the integral (in particular, they are never present in the residual
``continuous'' terms). A short meditation makes it evident that the terms of
interest are precisely those appearing in the discrete OPE (\ref{V12Va}),
\begin{align}
&  V_{1,2}^{\prime}(x)V_{a}(0)=\label{V12pVa}\\
&  \log(x\bar x)\left(  q_{0,1}^{(1,2)}(a)(x\bar x)^{ab}C_{+}^{\text{(L)}%
}(a)V_{a-b/2}(0)+q_{0,-1}^{(1,2)}(a)(x\bar x)^{1-ab+b^{2}}C_{-}^{\text{(L)}%
}(a)V_{a+b/2}(0)\right)  +\ldots\nonumber
\end{align}
decorated however by certain multipliers
\begin{equation}
q_{0,s}^{(1,2)}(a)=\left|  a-bs/2-Q/2\right|  _{\operatorname*{Re}}%
-\lambda_{1,2} \label{q12}%
\end{equation}
These multipliers are traced back to the derivative in $g$ of the $(x\bar
x)$-exponent, the non-analyticity being attributed to the fact that in
different domains of the parameter $a$ different poles have to be taken as the
discrete terms (we remind again that in the colliding pairs of poles we always
pick up explicitly the one which is $p$ closer to the Seiberg bound
$\operatorname*{Re}p=Q/2$). This is precisely the same mechanism of
non-analyticity as we have seen in eq.(\ref{phiVg}).

Once the relevant terms in the logarithmic Liouville OPE $V_{1,2}^{\prime
}(x)V_{a}(0)$ are established, all further calculations repeat literally those
in the derivation (\ref{O12W}). Skipping the straightforward calculations
(which show that the ``wrong'' cross terms disappear in the product of
(\ref{V12pVa}) and (\ref{Phi12}) after $H_{1,2}\bar H_{1,2}$ is applied and
give again the familiar polynomials in the ``good'' terms) let us quote the
final result, which has a better look if we again renormalize the fields
$W_{a}$ as in eq.(\ref{Wfields})
\begin{equation}
O_{1,2}^{\prime}(x)\mathcal{W}_{a}=\Lambda_{1,2}\log(x\bar x)\left(
q_{0,1}^{(1,2)}(a)\mathcal{W}_{a-b/2}+q_{0,-1}^{(1,2)}(a)\mathcal{W}%
_{a+b/2}\right)  +\text{ non-logarithmic terms} \label{O12log}%
\end{equation}
Here $\Lambda_{1,2}$ is from eq.(\ref{Lambda}).

Two examples considered above make the road smooth enough to roll along in the
general case. The relevant logarithmic terms in $V_{m,n}^{\prime}(x)V_{a}(0)$
are
\begin{equation}
V_{m,n}^{\prime}(x)V_{a}(0)=\sum_{r,s}^{(m,n)}\left(  x\bar x\right)
^{\lambda_{r,s}(Q-2a-\lambda_{r,s})-\Delta_{m,n}^{\text{(L)}}}q_{r,s}%
^{(m,n)}(a)C_{r,s}^{\text{(L)}}(a)\left[  V_{a+\lambda_{r,s}}\right]
\label{VmnpVa}%
\end{equation}
where (\ref{q12}) is generalized to
\begin{equation}
q_{r,s}^{(m,n)}(a)=\left|  a-\lambda_{r,s}-Q/2\right|  _{\operatorname*{Re}%
}-\lambda_{m,n} \label{qrs}%
\end{equation}
and the sum is again over the standard set (\ref{rsset}). The general version
of (\ref{O12log}) can be directly borrowed from the expression (\ref{Aex}) in
sect.7. In the ground ring calculation this expression was never proved, just
guessed on the basis of explicit results for $(m,n)=(1,2)$ and $(m,n)=(1,3)$.
In the present context there is no need to repeat the $(1,3)$ calculation
because, as we have seen above, for the logarithmic terms it is literally the
same as that of sect.7. Hence the guess (\ref{Aex}) implies
\begin{equation}
\mathcal{O}_{m,n}^{\prime}(x)\mathcal{W}_{a}=\log(x\bar x)\sum_{r,s}%
^{(m,n)}q_{r,s}^{(m,n)}(a)\mathcal{W}_{a-\lambda_{r,s}} \label{Omnlog}%
\end{equation}
where we have found it convenient to get rid of $\Lambda_{m,n}$ renormalizing
$O_{m,n}^{\prime}$ similarly to (\ref{OOmn})
\begin{equation}
\mathcal{O}_{m,n}^{\prime}(x)=\Lambda_{m,n}^{-1}O_{m,n}^{\prime} \label{OOmnp}%
\end{equation}

\section{Four point correlation number}

Now we are in the position to write down the main result of this paper, i.e.,
the GMG four-point function with one degenerate and three generic matter
fields. Summing up the boundary contributions (\ref{Omnlog}) and the curvature
term (\ref{G4curv}) we find for the normalized correlation number the
following expression
\begin{equation}
Z_{\text{L}}^{-1}\int\left\langle \mathcal{U}_{m,n}(x)\mathcal{W}_{a_{1}%
}\mathcal{W}_{a_{2}}\mathcal{W}_{a_{3}}\right\rangle d^{2}x=-(b^{-2}%
+1)b^{-3}(b^{-2}-1)\Sigma_{m,n}(a_{1},a_{2},a_{3}) \label{main}%
\end{equation}
where
\begin{equation}
\Sigma_{m,n}(a_{1},a_{2},a_{3})=-2mn\lambda_{m,n}+\sum_{i=1}^{3}\sum
_{r,s}^{(m,n)}\left(  \lambda_{m,n}-\left|  a_{i}-\lambda_{r,s}-Q/2\right|
_{\operatorname*{Re}}\right)  \label{Sigma}%
\end{equation}
This expression seems to gain somewhat in the transparency after a simple
resummation of the $\lambda_{m,n}$ terms
\begin{equation}
\Sigma_{m,n}(a_{1},a_{2},a_{3})=mn\lambda_{m,n}-\sum_{i=1}^{3}\sum
_{r,s}^{(m,n)}\left|  \lambda_{i}-\lambda_{r,s}\right|  _{\operatorname*{Re}}
\label{Sigma1}%
\end{equation}
Here convenient ``momentum'' parameters $\lambda_{i}=Q/2-a_{i}$ are introduced
for the generic matter insertions. This parametrization makes the symmetry of
(\ref{Sigma}) w.r.t. $a_{i}\rightarrow Q-a_{i}$ (i.e., the independence on the
choice of the ``Liouville dressing'' of the matter fields) apparent and
suggests the following diagrammatic representation
\begin{align}
&  \Sigma_{m,n}(a_{1},a_{2},a_{3})=\label{Sigmapict}\\
&  mn\lambda_{m,n}-\sum_{r,s}^{(m,n)}%
\raisebox{-0.2776in}{\includegraphics[
height=0.6841in,
width=1.1908in
]%
{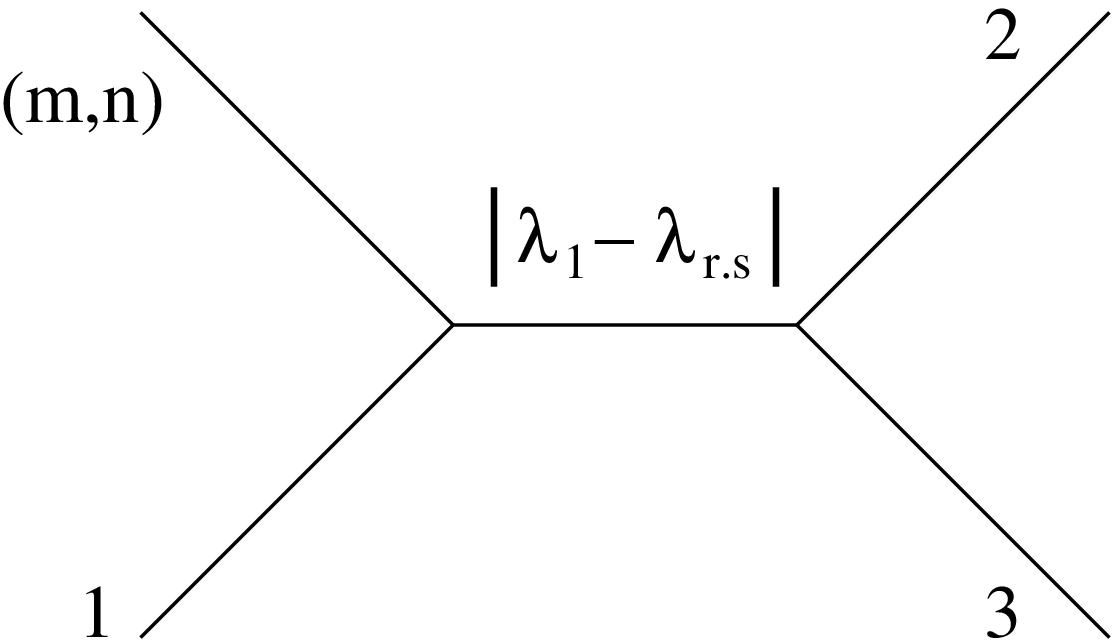}%
}%
-\sum_{r,s}^{(m,n)}%
\raisebox{-0.4514in}{\includegraphics[
height=1.0352in,
width=0.857in
]%
{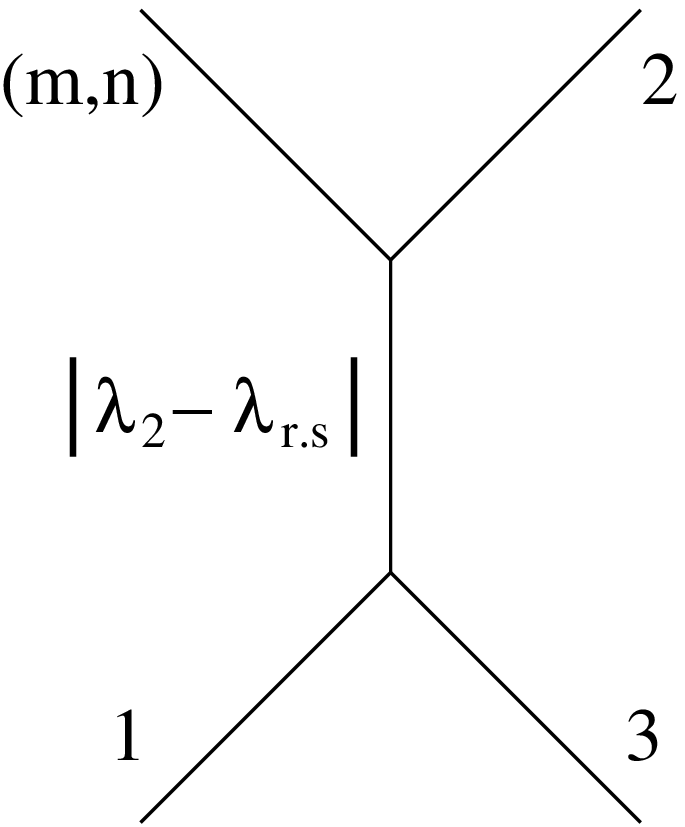}%
}%
-\sum_{r,s}^{(m,n)}%
\raisebox{-0.4791in}{\includegraphics[
height=1.0784in,
width=0.8916in
]%
{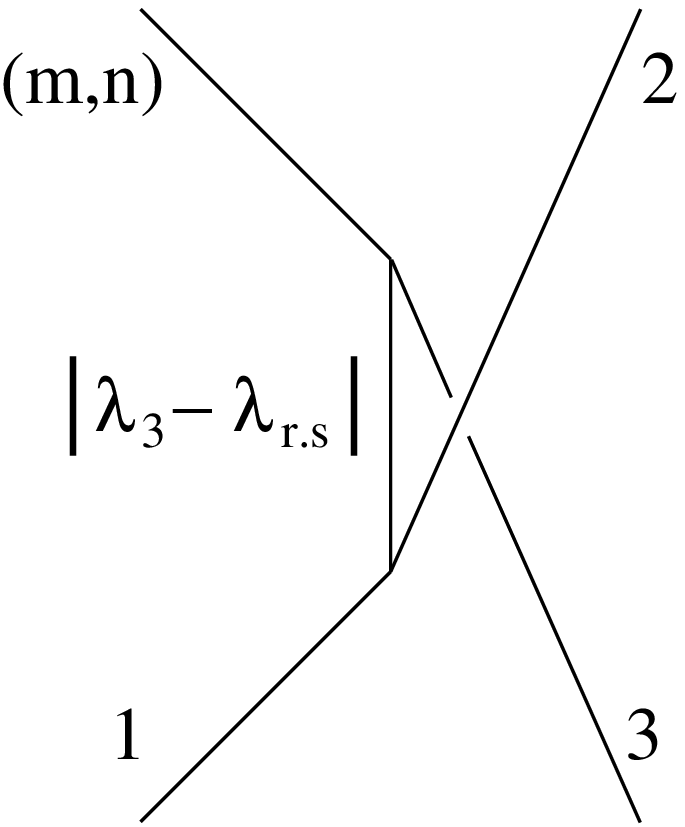}%
}%
\nonumber
\end{align}

This expression looks extremely simple, especially compared with the
complicated root leading to it in our treatment. Most probably this means that
we miss a much simpler and therefore more profound look at the physics behind
the problem. Nevertheless, we believe that our expression is correct, at least
if the generic non-degenerate fields are involved together with the degenerate
composite $U_{m,n}$. Few comparisons with the direct numerical integration in
eq.(\ref{U4mn}), as well as with the numbers coming from the matrix model
approach, are presented in the subsequent two sections. They both support our
expression (\ref{main}).

However a little closer inspection of (\ref{Sigma1}) reveals striking
problems. There is apparent inconsistency if one (or more) of the fields
$W_{i} $ in our correlation function corresponds to a degenerate matter
primary $\Phi_{m^{\prime},n^{\prime}}$. This problem comes out immediately if
one inserts the matter identity $I=\Phi_{1,1}$, e.g., $W_{3}=W_{b}=V_{b}$
instead of one of the generic $W$-insertions. Our formula gives in this case
\begin{equation}
\Sigma_{m,n}(a_{1},a_{2},b)=mn\lambda_{m,n}-\sum_{r,s}^{(m,n)}\left|
\lambda_{1,-1}-\lambda_{r,s}\right|  -\sum_{i=1}^{2}\sum_{r,s}^{(m,n)}\left|
\lambda_{i}-\lambda_{r,s}\right|  _{\operatorname*{Re}} \label{Sigmab}%
\end{equation}
while the usual interpretation of $V_{b}$ as the area element implies (or,
equivalently, eq.(\ref{phiVg}))
\begin{equation}
\Sigma_{m,n}(a_{1},a_{2},b)=b-\sum_{i=1}^{2}\left|  \lambda_{i}\right|
_{\operatorname*{Re}} \label{Sigmad}%
\end{equation}
and evidently contradicts (\ref{Sigmab}) if $(m,n)\neq(1,1)$. Similar problem
arises always if the degeneracy of a matter insertion in one (or more) $W_{i}$
entails a reduction of the number of the conformal blocks involved in the
matter four-point function (\ref{GM4mn}). We have already mentioned this
subtlety as the reason to consider only all three generic non-degenerate
$W_{i}$ insertions in our calculations. From the analytic point of view this
effect is certainly due to the delicate interplay between two limits, e.g.,
$\alpha_{1}\rightarrow\alpha_{m^{\prime},n^{\prime}}$ in $W_{1}$ and
$\lambda_{2}\pm\lambda_{3}\rightarrow\lambda_{r,s}$ (with $(r^{\prime
},s^{\prime})$ from the set prescribed by the standard fusion rules) in
$W_{2}$ and $W_{3}$, required to turn $\Phi_{1}$ to a degenerate $(m^{\prime
},n^{\prime})$ GMM field.

For the moment we are not certain about the resolution of this important
difficulty. We believe, however, that our formula (\ref{Sigma}) gives correct
answer as far the number of matter conformal blocks is equal to the degeneracy
level $mn$ of the field $U_{m,n}$ chosen in (\ref{U4mn}) as the integrated
insertion. If not, sometimes we can use the freedom in the choice of the
integration point in (\ref{U4}) and take another degenerate field
$U_{m^{\prime},n^{\prime}}$ as the starting point, such that $m^{\prime
}n^{\prime}$ gives the actual number of blocks. For example, this always can
be done if all the degenerate fields belong to the $(1,n)$ (or $(m,1)$)
subset. In this case it is sufficient to begin with the field with the least
value of $n$.

In the case of more general degenerate insertions this often cannot be always
done. A simple example is the OPE$\;\Phi_{1,2}\Phi_{2,1}$ which results in a
single representation $\left[  \Phi_{2,2}\right]  $ and therefore the four
point function with these two matter operators contains only one conformal
block. In connection with the problems indicated above we find it instructive
to consider an example of the four-point function $\left\langle U_{1,2}%
U_{2,1}U_{a_{1}}U_{a_{2}}\right\rangle _{\text{GMG}}$ which involves this pair
of matter operators. Of course the decoupling of the matter singular vectors
(\ref{GMM}) require the parameters $a_{1}$ and $a_{2}$ to be related according
to the fusion rules with the ``intermediate'' representation $\left[
\Phi_{2,2}\right]  $. We consider here the choice $a_{1}+a_{2}=Q-\lambda
_{1,1}$, because it illustrates an interesting and important effect. Namely,
with this choice the Liouville four-point function is resonant. This is
apparent under the following choice of the Liouville ``dressings'' of the
fields $\Phi_{1,2}$ and $\Phi_{2,1}$
\begin{align}
\left\langle U_{1,2}U_{2,1}U_{a}U_{Q/2-a}\right\rangle _{\text{GMG}}  &
=\int\left\langle \Phi_{1,2}(x)\Phi_{2,1}(0)\Phi_{a-b}(1)\Phi_{q/2-a}%
(\infty)\right\rangle _{\text{GMM}}\label{U1221}\\
&  \times\left\langle V_{b^{-1}-b/2}(x)V_{b-b^{-1}/2}(0)V_{a}(1)V_{Q/2-a}%
(\infty)\right\rangle _{\text{L}}d^{2}x\nonumber
\end{align}
and therefore in the Liouville part $\sum_{i}a_{i}=Q$. In LG such poles are
interpreted as certain $\log\mu$ decoration of the standard power-like $\mu
$-dependence of the correlation function. On the other hand, everyone familiar
with the matrix model machinery (whose net result is quite similar to the mean
field picture) would have a problem to imagine how logarithms can appear those
context. Also our result (\ref{Sigma1})) is never singular in the parameters.
Both confrontations suggest that even if the Liouville correlation function
has a pole as a function of $a_{i}$, the moduli integral vanishes to cancel
the singularity. This is what we are want to demonstrate now for the case
(\ref{U1221}), although we're not yet in the position to resolve the
singularity and give a prescription for the finite part of this integral.

The resonant Liouville correlation function is
\begin{equation}
\left\langle V_{b^{-1}-b/2}(x)V_{b-b^{-1}/2}(0)V_{a}(1)V_{Q/2-a}%
(\infty)\right\rangle _{\text{L}}=-(x\bar x)^{b^{-2}+b^{2}-5/2}\left[
(1-x)(1-\bar x)\right]  ^{ab-2ab^{-1}}\log\mu\label{V1221}%
\end{equation}
The matter four point function is also quite simple in this case
\begin{equation}
\left\langle \Phi_{1,2}(x)\Phi_{2,1}(0)\Phi_{a-b}(1)\Phi_{q/2-a}%
(\infty)\right\rangle _{\text{GMM}}=\frac{(1-b^{2}-(1-2ab)x)(1-b^{2}%
-(1-2ab)\bar x)}{(1-b^{2})^{2}(x\bar x)^{1/2}\left[  (1-x)(1-\bar x)\right]
^{b^{2}-ab}} \label{Phi1221}%
\end{equation}
Thus
\begin{equation}
\left\langle U_{1,2}U_{2,1}U_{a}U_{Q/2-a}\right\rangle _{\text{GMG}}=-\log
\mu\int\frac{(1-b^{2}-(1-2ab)x)(1-b^{2}-(1-2ab)\bar x)d^{2}x}{(1-b^{2}%
)^{2}(x\bar x)^{3-b^{-2}-b^{2}}\left[  (1-x)(1-\bar x)\right]  ^{b^{2}%
-2ab+2ab^{-1}}} \label{Uexp}%
\end{equation}
The integral is carried out explicitly with the use of the general integration
formula\footnote{It is implied in this formula that $\mu-\bar\mu$ and
$\nu-\bar\nu$ are integer numbers, otherwise the integral has no sense.}
\begin{equation}
\int x^{\mu-1}\bar x^{\bar\mu-1}(1-x)^{\nu-1}(1-\bar x)^{\bar\nu-1}%
d^{2}x=\frac{\pi\Gamma(\mu)\Gamma(\nu)\Gamma(1-\bar\mu-\bar\nu)}{\Gamma
(1-\bar\mu)\Gamma(1-\bar\nu)\Gamma(\mu+\nu)} \label{int}%
\end{equation}
and turns out to vanish, as we expected on the basis of the general arguments.
This finite part of this integral requires more delicate analysis. We hope to
clarify this important question shortly.

\section{Numerical check: direct integration}

In this section we perform a numerical check of our analytic expression
(\ref{Sigma}) for the integral (\ref{U4mn}) through the direct numerical
integration over the moduli space. Of course, the space of three generic
parameters $a_{1}$, $a_{2}$ and $a_{3}$, together with the central charge
parameter $b$, is too big to investigate it to any extent of
comprehensiveness. Here we restrict ourselves to a very preliminary study,
taking a simple example of the four point function $\left\langle U_{1,2}%
^{4}\right\rangle _{\text{GMG}}$ with four identical $U_{1,2}$
insertions\footnote{We postpone a more comprehensive analysis for the further
work.}. Although it is not precisely the case of three generic non-degenerate
matter primaries, the OPE $\Phi_{1,2}\Phi_{1,2}$ always contains two
representations, $\left[  I\right]  $ and $\left[  \Phi_{1,3}\right]  $ and
therefore there always two conformal blocks in the matter correlation
function. This case therefore satisfies our above criteria and is supposed to
be given formally by the expressions (\ref{main},\ref{Sigma}).

On the other hand, this particular example has some important advantages for
the numerical work. The matter structure constants (and to some extent the
Liouville ones) are simplified. The matter conformal blocks are expressed
explicitly in terms of the hypergeometric functions. And finally, all the four
insertions are identical and therefore the integrated 2-form in (\ref{U4mn})
enjoys the complete modular symmetry, i.e., is invariant under the
transformations
\begin{align}
R:x  &  \rightarrow1-x\;;\;\;\;\;\;\;\;\;T:x\rightarrow1/x\nonumber\\
RT:x  &  \rightarrow1/(1-x)\;;\;\;\;TR:x\rightarrow1-1/x\label{modular}\\
\;\;TRT  &  =RTR:x\rightarrow x/(x-1)\nonumber
\end{align}
This allows to reduce the integration region in the integral (\ref{U4mn}) from
the whole complex plane to a fundamental domain, e.g., the segment
$\mathbf{F=}\{{\operatorname*{Re}}x<1/2;\;\left|  1-x\right|  <1\}$.

\begin{figure}
[tbh]
\begin{center}
\includegraphics[
trim=0.000000in 0.000000in 0.001336in -0.001578in,
height=4.0093in,
width=5.079in
]%
{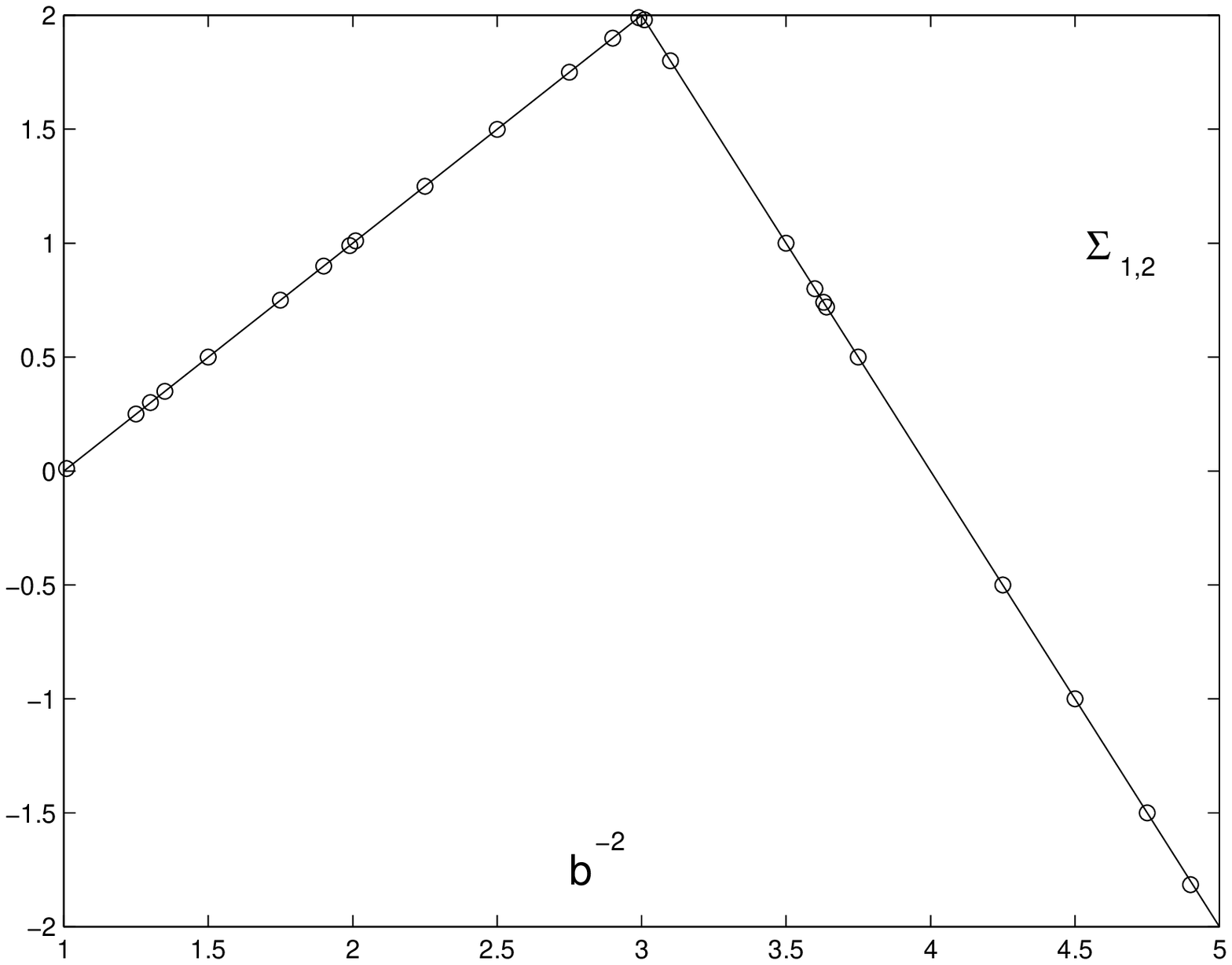}%
\caption{ Direct numerical evaluation of the integral (\ref{U412}) (circles)
versus the exact result (continuous straight line) }%
\label{plot}%
\end{center}
\end{figure}

Our general result (\ref{main}) for $\left\langle U_{1,2}^{4}\right\rangle
_{\text{GMG}}$ can be written as
\begin{equation}
Z_{\text{L}}^{-1}\left\langle U_{1,2}^{4}\right\rangle _{\text{GMG}}%
=-(2\pi)^{4}(b^{-2}+1)b^{-3}(b^{-2}-1)\Sigma_{1,2}(b^{-2})\mathcal{L}%
^{4}(g)\label{main12}%
\end{equation}
where $Z_{\text{L}}$ is as in (\ref{ZL}) and the ``leg-factor'' $\mathcal{L}%
(g)$ reads
\begin{equation}
\mathcal{L}(g)=\left[  \left|  \frac{\gamma(2gb-b^{2})\gamma(2gb^{-1}-b^{-2}%
)}{4\gamma^{2g/b-1}(b^{2})\gamma(2-b^{-2})}\right|  \right]  ^{1/2}\label{Leg}%
\end{equation}
Here $g$ is the solution of the ``dressing condition'' (\ref{Dbalance}%
)\footnote{Again in this section we use the letter $g$ in different meaning
then $b^{-2}$.}
\begin{equation}
g=Q/2-\sqrt{(b-b^{-1}/2)^{2}}=b^{-1}/2+b/2-\left|  b-b^{-1}/2\right|
\label{g12}%
\end{equation}
Factor (\ref{Sigma1}) then becomes
\begin{equation}
b^{-1}\Sigma_{1,2}(b^{-2})=-\frac{1}{2}b^{-2}+\frac{7}{2}-\frac{3}{2}\left|
b^{-2}-3\right|  \label{Sigma12}%
\end{equation}
It is plotted in fig.\ref{plot} as the function of $b^{-2}$ as a broken
straight line.%

The original four-point integral (\ref{U4mn}) turns in this case to
\begin{equation}
\left\langle U_{1,2}^{4}\right\rangle _{\text{GMG}}=6\int_{\mathbf{F}%
}G_{\text{M}}(x,\bar x)G_{\text{L}}(x,\bar x)d^{2}x \label{U412}%
\end{equation}
where the GMM $G_{\text{M}}(x,\bar x)=\left\langle \Phi_{1,2}(x)\Phi
_{1,2}(0)\Phi_{1,2}(1)\Phi_{1,2}(\infty)\right\rangle _{\text{GMM}}$ reads
explicitly
\begin{equation}
G_{\text{M}}(x,\bar x)=\mathcal{F}_{1,1}(x)\mathcal{F}_{1,1}(\bar
x)-\kappa^{2}\mathcal{F}_{1,3}(x)\mathcal{F}_{1,3}(\bar x) \label{GM412}%
\end{equation}
where
\begin{equation}
\kappa^{2}=\dfrac{(1-2b^{2})^{2}\gamma(b^{2})}{\gamma^{2}(2b^{2}%
)\gamma(2-3b^{2})} \label{kappa}%
\end{equation}
and the conformal blocks are
\begin{align}
\mathcal{F}_{1,1}(x)  &  =x^{1-3b^{2}/2}(1-x)^{1-3b^{2}/2}{}_{2}F_{1}%
(2-3b^{2},1-b^{2},2-2b^{2},x)\label{block412}\\
\mathcal{F}_{1,3}(x)  &  =x^{b^{2}/2}(1-x)^{b^{2}/2}{}_{2}F_{1}(-1+3b^{2}%
,b^{2},2b^{2},x)\nonumber
\end{align}
The Liouville correlation function $G_{\text{L}}(x,\bar x)=\left\langle
V_{g}(x)V_{g}(0)V_{g}(1)V_{g}(\infty)\right\rangle _{\text{L}}$ is represented
in the form
\begin{equation}
G_{\text{L}}(x,\bar x)=\mathcal{R}_{g}\int^{\prime}\frac{dP}{4\pi}%
r_{g}(P)\mathcal{F}_{P}\left(  \left.
\begin{array}
[c]{cc}%
\Delta_{g} & \Delta_{g}\\
\Delta_{g} & \Delta_{g}%
\end{array}
\right|  x\right)  \mathcal{F}_{P}\left(  \left.
\begin{array}
[c]{cc}%
\Delta_{g} & \Delta_{g}\\
\Delta_{g} & \Delta_{g}%
\end{array}
\right|  \bar x\right)  \label{GL412}%
\end{equation}
where the complicated expressions for the Liouville structure constants are
reduced to
\begin{equation}
\mathcal{R}_{g}=\left(  \gamma(b^{2})b^{2-2b^{2}}\right)  ^{(Q-4g)/b}%
\frac{\Upsilon_{b}^{4}(b)\Upsilon_{b}^{4}(2g)}{\pi^{2}\Upsilon_{b}%
^{4}(2g-Q/2)} \label{Rg}%
\end{equation}
and
\begin{align}
r_{g}(P)  &  =\frac{\pi^{2}\Upsilon_{b}(2iP)\Upsilon_{b}(-2iP)\Upsilon_{b}%
^{4}(2g-Q/2)}{\Upsilon_{b}^{2}(b)\Upsilon_{b}^{2}(2g-Q/2-iP)\Upsilon_{b}%
^{2}(2g-Q/2+iP)\Upsilon_{b}^{4}(Q/2-iP)}\label{rg}\\
&  =\sinh2\pi b^{-1}P\sinh2\pi bP\exp\left(  -8\int_{0}^{\infty}\frac
{dt}t\frac{\sin^{2}Pt\left(  \cosh^{2}(Q-2g)t-e^{-Qt}\cos^{2}Pt\right)
}{\sinh bt\sinh b^{-1}t}\right) \nonumber
\end{align}
The general symmetric conformal block $\mathcal{F}_{P}\left(  \left.
\begin{array}
[c]{cc}%
\Delta_{g} & \Delta_{g}\\
\Delta_{g} & \Delta_{g}%
\end{array}
\right|  x\right)  $ is evaluated numerically through the recursive procedure
introduced in (\cite{block}) (a summary can be found in the more accessible
paper \cite{AAl}). As usual, the prime near the integral sign indicates
possible discrete terms. In this study we consider only the region
$1<b^{-2}<5$ where such extra terms don't appear and the integral in
(\ref{GL412}) can be understood literally. In fig.\ref{plot} the results of
the numerical evaluation of the integral (\ref{U412}) are shown as circles.
Notice, that near the break at $b^{-2}=3$ this integral converges very slowly
near $x=0$, $1$ and $\infty$. Numbers in fig.\ref{plot} in this region require
proper modification of the integration algorithm. A solution to this problem,
as well as other interesting details related to the numerical evaluation of
(\ref{U412}) will be reported as a separate publication.

\section{Comparing with martix models}

It is of course very interesting to compare our expression (\ref{main}) with
the correlation numbers arising in the matrix model context. Unfortunately
this is not straightforward. The standard formulations of the matrix models
are interpreted mostly in terms of genuine rational minimal models with
$b^{2}=p/p^{\prime}$ and involve only degenerate CFT matter fields. Moreover,
the main bulk of the matrix model results contains the field theory related
information in rather ciphered form. It still takes a considerable effort to
disentangle the relevant correlation functions and interpret them in terms of
the minimal gravity.

There is a matrix model example where the continuous interpretation is
unambiguous and in addition the matter central charge varies continuously, so
that our treatment of the generalized minimal gravity seems to be relevant.
Recently I.Kostov \cite{Kostov} worked out a new exciting result about the
so-called gravitational $O(n)$ model. This model is a random lattice covered
by self-avoiding polymer loops, each component bringing up the weight factor
$n$. The critical thermodynamics is controlled by two parameters, the
``cosmological constant'' $x$ coupled to the size of the lattice and the
``mass'' parameter $t$, which regulates the length of the polymers. Both
parameters are chosen as the deviations of the corresponding absolute
activities from the double critical point where the lattice size and loop
length ``blow up'' simultaneously. The critical singularity in the
neighborhood of this point is the subject of continuous field theory.

According to ref.\cite{Kostov} the singular part $Z(t,x)$ of the genus $0$
partition function admits the following simple description. Introduce the
(standard in the $O(n)$ model) parameterization of the loop weight $-2<n<2$
\begin{equation}
n=-2\cos(\pi g) \label{Ong}%
\end{equation}
in terms of the variable $1<g<2$. Also let $Z(x,t)$ be the singular part of
the genus $0$ partition function and
\begin{equation}
u=-(g-1)Z_{xx} \label{ugZ}%
\end{equation}
its second derivative in $x$. Then $u$ is a solution to the following simple
transcendental equation
\begin{equation}
u^{p}+tu^{p-1}=x \label{ieq}%
\end{equation}
where $p=(g-1)^{-1}$. Equations (\ref{ieq}) and (\ref{ugZ}) result in the
following expansion
\begin{align}
Z  &  =tx^{2}+x^{g+1}\sum_{\substack{n=0 \\n\neq1 }}^{\infty}\frac
{\Gamma(g(n-1)-n-1)}{n!\Gamma(g(n-1)-2n+2)}\left(  tx^{1-g}\right)
^{n}\label{Zt}\\
&  =x^{g+1}\left(  \frac{-1}{(g-1)g(g+1)}+tx^{1-g}+\frac{\left(
tx^{1-g}\right)  ^{2}}{2(g-3)}+\frac{\left(  tx^{1-g}\right)  ^{3}}%
6+\frac{(g-2)\left(  tx^{1-g}\right)  ^{4}}8+\ldots\right) \nonumber
\end{align}

On the other hand, the critical dilute polymers on the random lattice admit
the standard continuous interpretation in terms of GMG with\cite{ON}
\begin{equation}
c_{\text{M}}=13-6(g+g^{-1}) \label{cn}%
\end{equation}
This is one of the occasions where the GMM $\mathcal{M}_{b^{2}}$ is relevant
as the matter field theory, the parameters being related as
\begin{equation}
g=b^{-2} \label{gb}%
\end{equation}
Moreover, the GMM operator coupled to the off-critical ``mass'' of the polymer
loop is most likely the degenerate $\Phi_{1,3}$ field and we're dealing with
the GMG perturbed by the composite field
\begin{equation}
U_{1,3}=\Phi_{1,3}V_{b^{-1}-b} \label{dr13}%
\end{equation}
Notice the particular choice of the Liouville ``dressing'', which turns out to
be relevant in the gravity context at $1<b^{-2}<3$\cite{ON, GK}. Therefore,
the coefficients in the expansion (\ref{Zt}) are interpreted as the multipoint
correlation functions of this field in the corresponding GMG. Since the
overall normalization of the partition function depends on the scale and
cannot be fixed in the universal way, it is natural to relate the normalized
correlation functions
\begin{equation}
Z_{\text{L}}^{-1}\left\langle U_{1,3}^{n}\right\rangle _{\text{GMG}}%
=\frac{\Gamma(b^{-2}(n-1)-n-1)\Gamma(2-b^{-2})}{\Gamma(b^{-2}(n-1)-2n+2)\Gamma
(-1-b^{-2})}(2\pi L_{\text{eg}})^{n} \label{n13}%
\end{equation}
The rescaling factor $L_{\text{eg}}$ is introduced here because the
normalization of the dimensional cosmological constant $\mu$ is different from
that of $x$ and also the field $U_{1,3}^{\text{(mat)}}$ coupled to the
parameter $t$ is normalized differently from our $U_{1,3}$. The rescaling
factor relates the dimensionless combinations
\begin{equation}
U_{1,3}^{\text{(mat)}}x^{g-1}=\left(  2\pi L_{\text{eg}}\right)  ^{-1}U_{1,3}
\label{xnorm}%
\end{equation}
It is easy to find this factor explicitly, comparing the GMG two- and
three-point functions (\ref{twothree}) with the corresponding terms in
(\ref{n13})\cite{threept}
\begin{equation}
L_{\text{eg}}=\frac{\gamma(b^{-2}-1)\left[  -\gamma(b^{2})\gamma
(2-3b^{2})\right]  ^{1/2}}{2(2-b^{-2})(\pi\mu\gamma(b^{2})^{b^{-2}-1})}
\label{Legg}%
\end{equation}

The four-point case of (\ref{n13}) reads
\begin{equation}
Z_{\text{L}}^{-1}\left\langle U_{1,3}^{4}\right\rangle _{\text{GMG}%
}=-3(g-2)(g-1)g(g+1)(2\pi L_{\text{eg}})^{4} \label{U134}%
\end{equation}
At the same time, our expression (\ref{main}) gives for this case
\begin{equation}
\left\langle \left\langle \mathcal{U}_{1,3}^{4}\right\rangle \right\rangle
=-(g+1)g(g-1)b^{-1}\Sigma_{1,3} \label{U413}%
\end{equation}
where
\begin{equation}
b^{-1}\Sigma_{1,3}=\frac32\left(  g+3-\left|  g-1\right|  -\left|  g-3\right|
-\left|  g-5\right|  \right)  \label{Sigma13}%
\end{equation}
and at $1<g<3$%
\begin{equation}
b^{-1}\Sigma_{1,3}=3(g-2) \label{consent}%
\end{equation}
in agreeable consent with (\ref{U134}).

\textbf{Acknowledgments}

Authors are grateful to B.Feigin, I.Kostov, A.Litvinov, V.Petkova and
A.Zamolodchikov for useful discussions. Our collaboration was supported by
INTAS under the grant \#INTAS-OPEN-03-51-3350. A.B was supported by Russian
Foundation for Basic Research under the grant \#RBRF 04-02-16027, by the RAS
program ''Elementary particles and the fundamental nuclear physics'' and the
Scienific School grant 2044.2003.2. As usual, Al.Z's creativity was mostly
inspired by Galina Gritsenko. Besides that Al.Z acknowledges the hospitality
and stimulating scientific atmosphere of the Theoretical Physics Laboratory at
RIKEN, where the main part of the article has been written. His efforts were
supported also by the European Committee under contract EUCLID
HRPN-CT-2002-00325. The organizers of the LAPP conference ``RAQIS 05'' gave us
a rare occasion to work for a while in a personal contact and coordinate in
this way the final version the manuscript.

\setcounter{figure}{0}
\setcounter{section}{0}
\setcounter{footnote}{0}

\newpage

\begin{center}
{\Large \textbf{Massive Majorana Fermion Coupled to 2D Gravity and
Random Lattice Ising Model}}

\vspace{1.0cm}

{\large Y.Ishimoto}

\vspace{0.2cm}

Kawai Theoretical Laboratory, RIKEN,

Saitama 351-0198, Japan

\vspace{0.2cm}

and

\vspace{0.2cm}

{\large Al.Zamolodchikov}\footnote{On leave of absence from
Institute of Theoretical and Experimental Physics,
B.Cheremushkinskaya 25, 117259 Moscow, Russia.}

\vspace{0.2cm}

Laboratoire de Physique Th\'eorique et
Astroparticules\footnote{UMR-5207 CNRS UM-2}

Universit\'e Montpellier II

Pl.E.Bataillon, 34095 Montpellier, France
\end{center}

\vspace{1.0cm}

\textbf{Abstract}

We consider the partition function of the 2D free massive Majorana
fermion coupled to the quantized metric of the spherical topology.
By adding an arbitrary conformal ``spectator'' matter we get a
control over the total matter central charge. This gives an
interesting continuous family of criticalities and also enables us
to make a connection with the semicalssical limit. We use the
Liouville field theory as the effective description of the quantized
gravity. The spherical scaling function is calculated approximately
but, to our belief, to a good numerical precision in almost the
whole region of the spectator parameter. Impressive comparison with
the predictions of the solvable matrix model gives rise to a more
general model of random lattice statistics, which is most probably
not solvable by the matrix model technique but reveals a more
general pattern of critical behavior. We hope that numerical
simulations or series extrapolation will be able to reveal our
family of scaling functions.

\section{Introduction}

It was demonstrated in recent developments that the Liouville field
theory (LFT) is an effective tool to study relevant problems of 2D
quantum gravity \cite{LYgrav}. The field theory of two-dimensional
gravity, based on LFT, is currently believed to give an effective
description of the universal critical behavior in certain models of
statistical physics called the random lattice spin systems. These
systems were in fact first introduced rather as the models of the
continuous 2D gravity itself \cite{Kazakov, FDavid, KMK, BKKM} then
from the internal needs of the statistical mechanics, basic
motivation coming from the Polyakov's formulation of the string
theory \cite{Polyakov13}. The random lattice models (RLM) can be
thought of as the ordinary lattice systems but defined on a general
graph of fixed topology, the graph itself being a dynamical degree
of freedom and subject to the averaging in the partition sum. From
this point of view the name ``dynamical lattice'' or ``dynamical
triangulations'' \cite{BKKM} seems more correct then the traditional
reference RLM. It has been argued long ago that such ``naive''
version of the Regge calculus \cite{Regge} at large lattice size
reveals interesting critical behavior (see e.g. \cite{KMK}).
Moreover, the critical exponents occur to coincide with those in the
continuous description through the field theoretic path integral
over all metrics on a surface \cite{KPZ}. It has been realized that
the Liouville field theory might be a convenient instrument to treat
such problems, especially in the DDK (David-Distler-Kawai)
formulation \cite{David, Distler}.

Further development of the field theoretic approach to the 2D
gravity required essentially a better understanding of LFT as a
conformal field theory. Despite some attempts based on the
``analytic continuation in the number of integrations''
\cite{GoulianLi, Dotsenko}, the progress was slow. It was tempered
probably by the superficial but persistent idea that the structures
studied earlier in the rational conformal field theories (CFT)
feature to some extent the general pattern of CFT\footnote{This idea
is still popular, especially among younger theorists. This fact
should certainly be attributed to the drawbacks in the existing CFT
textbooks.}. Finally, in 1992 H.Dorn and H.Otto \cite{DO3} supported
the belief that LFT is a solvable CFT and gave an explicit
expression for the three-point correlation function. Unfortunately
up to this time the topic was completely out of fashion and further
development went slowly \cite{AAl3, Teschner3}. This is because,
with the exception of certain boundary versions \cite{FZZ, ZZ}, the
application of new results in the string theory remains problematic.

In this article we would like to put forward another, opposite side
of LFT (and general continuous 2D gravity), its role as the
effective field theory of the RLM critical behavior. Many
interesting models of dynamical lattice are exactly solvable by the
matrix model technique \cite{Brezinetal}. First 2D gravity
applications appeared already in \cite{Kazakov}. This gave rise to
another popular idea that the matrix model patterns of critical
behavior cover completely all universality classes in RLM and
therefore all possible continuous 2D
gravities\footnote{Incidentally, the terms ``matrix models'' and
``2D gravity'' are often used as synonyms, notably in the string
theretic circles.}. However we find this idea doubtful and one of
the main motivations of this study is to justify this concern. To
this order we study the theory of 2D massive Majorana fermion
coupled to the quantized gravity. This field theory is supposed to
be a continuous description of the random lattice Ising model
(RLIM), as first formulated and solved by V.Kazakov \cite{KIsing}
(for more detailed study see \cite{Boulatov}). And indeed, the
critical exponents of RLIM are known \cite{KPZ} to be reproduced
correctly in the LFT description. Here we go beyond the critical
exponents and try to evaluate a scaling function, the one related to
the spherical partition function of the massive (perturbed) matter.
For the time being LFT approach fails to give a comprehensive answer
for such scaling functions. Instead it offers some critical
correlation functions (see e.g., \cite{GoulianLi, DiFrancescoKut,
Dotsenko} or \cite{three, four} for more recent studies) and thus
few first terms of the conformal perturbative expansion. However, in
\cite{LYgrav} it was demonstrated that even this restricted
information turns out rather useful if certain conjectures are made
about the analytic structure of the scaling function. In fact these
conjectures, although they are not universal and some explicit
counterexamples are known, seem rather natural and are justified by
a number of particular examples. Under these analytic assumptions,
the perturbative information can be treated through a combined
analytic-numeric procedure \cite{LYgrav} (efficiency of this
procedure has been earlier observed in the case of classical gravity
in \cite{sphere}) and results in approximate, but sometimes
impressively accurate numerical description of the scaling function.

Below we perform the same program for the Majorana fermion CFT
perturbed by the mass term. The spherical scaling function obtained
in this way can be compared with the exact one evaluated in the
matrix model framework. In fact we consider a more general model
where an additional conformal matter is added, which remains always
critical but allows us an access over the LFT parameter which can
now vary continuously. We propose to attribute the corresponding
scaling functions to a kind of generalization of the standard RLIM,
a result of the ``naive'' decoration of the statistical weights by a
``determinant'' factor. This generalized RLIM is hardly exactly
solvable in general but likely can be studied numerically through
the extrapolations of the finite $N$ data \cite{David1} or through
the MC simulations \cite{MC, BKKM}. If our conjecture is correct,
such studies should reveal new types of critical behavior, never
seen before in the matrix model studies.

\section{Massive fermion coupled to Liouville}

Our model is defined through the following local Lagrangian
\begin{equation}
\mathcal{L}_{m}=\mathcal{L}_{\text{L}}(b)+\mathcal{L}_{\text{ising}%
}+\mathcal{L}_{\text{sp}}+\frac m{2\pi}\varepsilon e^{2a\phi}\label{Lm}%
\end{equation}
where $\mathcal{L}_{\text{L}}$ is the usual Liouville action with
cosmological
coupling constant $\mu$ and parameter $b$%
\begin{equation}
\mathcal{L}_{\text{L}}(b)=\frac1{4\pi}(\partial_{a}\phi)^{2}+\mu
e^{2b\phi
}\label{LL}%
\end{equation}
For what is concerned LFT, we follow the notations of \cite{AAl3},
in particular $V_{a}$ denotes the quantum version of the exponential
Liouville field $\exp(2a\phi)$, which is a Liouville primary of
dimension $\Delta
_{a}^{\text{(L)}}=a(Q-a)$. The Liouville central charge (as usual $Q=b^{-1}%
+b$)
\begin{equation}
c_{\text{L}}=1+6Q^{2}\label{cL3}%
\end{equation}
Then, $\mathcal{L}_{\text{ising}}$ is the usual $c=1/2$ free field
theory of the Majorana massless fermion
\begin{equation}
\mathcal{L}_{\text{ising}}=\frac1{2\pi}\left(
\psi\bar\partial\psi+\bar
\psi\partial\bar\psi\right) \label{Lising}%
\end{equation}
This normalization corresponds to the following massless free
fermion propagators
\begin{equation}
\left\langle \psi(z)\psi(0)\right\rangle =\frac1z;\;\;\;\left\langle
\bar
\psi(\bar z)\bar\psi(0)\right\rangle =\frac1{\bar z}\label{psinorm}%
\end{equation}
The energy density operator $\varepsilon=i\bar\psi\psi$, which plays
the role of a perturbation in (\ref{Lm}), has dimension $(1/2,1/2)$.
It is normalized in the standard in CFT way
\begin{equation}
\left\langle \varepsilon(x)\varepsilon(0)\right\rangle =\frac1{x\bar
x}\label{epsnorm}%
\end{equation}
Finally, as in ref.\cite{LYgrav} we added a ``spectator'' CFT of
central charge $c_{\text{sp}}$ to have a control over the parameters
of the Liouville gravity. This matter is not coupled to the
perturbation term and therefore remains conformal, being in our
formalism completely decoupled from the other degrees of
freedom\footnote{We remind that this decoupling is specific for the
sphere. In a less trivial topology there is a residual coupling
through the
conformal moduli.}. Its presence is indicated by the term $\mathcal{L}%
_{\text{sp}}$ in the total Lagrangian. We call the resulting family
of 2D gravity theories the gravitational Ising model (GIM).

For a semiclassical analysis the interaction term in the Lagrangian
is better written as
\begin{equation}
\mathcal{L}_{\text{int}}=\frac m{2\pi}\left(  \varepsilon
e^{2a\phi}+\frac
m{16a^{2}}e^{4a\phi}\right) \label{Lint}%
\end{equation}
The second term in the brackets is the ``trailing'' counterterm
needed to give the perturbing field definite dimension.

As usual, the following balance equations hold
\begin{align}
c_{\text{sp}}+\frac32+6Q^{2} &  =26\label{balance}\\
a(Q-a) &  =\frac12\nonumber
\end{align}
The first equation is recombined as
\begin{equation}
6Q^{2}=\frac{49}2-c_{\text{sp}}\label{Qcp}%
\end{equation}
This is the relation which determines the parameter $b$ of the model
in terms of the central charge of the background matter. In
particular the
semiclassical regime $b\rightarrow0$ is achieved when $c_{\text{bg}%
}\rightarrow-\infty$ and ``pure Ising'' case $c_{\text{bg}}=0$
corresponds to $b^{2}=3/4$.

According to the conventional wisdom (learned in fact from the
matrix model experience and from the semiclassical intuition) in the
second equation from the two solutions we choose the least one
\begin{equation}
a=\frac Q2-\sqrt{\frac{Q^{2}-2}4}\label{a3}%
\end{equation}
For the pure Ising model we have
\begin{align}
a=b/3\label{ab}%
\end{align}
while in the semiclassical regime
\begin{equation}
a=b/2-b^{3}/4+O(b^{7})\label{acl}%
\end{equation}

Self-consistency of LFT requires the parameter $b$ to remain real.
This restricts the variation of $c_{\text{sp}}$ to the region
$-\infty <c_{\text{sp}}\leq1/2$. Inside this interval we always
choose $b$ to be the smaller solution of (\ref{balance}) so that
$0<b^{2}\leq1$.

\section{Partition function}

The spherical (genus $0$) partition function $Z(m,\mu)$ scales as
\begin{equation}
Z(m,\mu)=\mu^{Q/b}G_{\text{sphere}}\left(  \frac m{\mu^{s}}\right)
\label{Fsphere}%
\end{equation}
where the scaling function $G_{\text{sphere}}$ depends on the scale
invariant combination of the coupling constants. The exponent $s$ is
easily figured out from the scale dependence of the individual terms
in (\ref{Lm})
\begin{equation}
s=ab^{-1}\label{s}%
\end{equation}
This parameter varies from $s=1/2-b^{2}/4+\ldots$ in the
semiclassical limit $b\rightarrow0$ to
$s=1-\sqrt{1/2}=0.29289\ldots$ at $b^{2}=1$. In the pure Ising model
$s=1/3$.

Formal perturbative expansion in $\mathcal{L}_{\text{int}}$ gives
\begin{equation}
Z(m,\mu)=\sum_{n=0}^{\infty}\frac{(-m)^{n}}{(2\pi)^{n}n!}\left\langle
\varepsilon^{n}\right\rangle _{\text{GIM}}\label{Zexp}%
\end{equation}
where $\left\langle ...\right\rangle _{\text{GIM}}$ means
non-normalized correlation functions in GIM. In the Liouville
gravity such $n$-point functions are given by the integrals over the
moduli of our sphere with $n-3$ punctures
\begin{equation}
\left\langle \varepsilon^{n}\right\rangle
_{\text{GIM}}=\int\left\langle C\bar C(x_{1})C\bar C(x_{2})C\bar
C(x_{3})\right\rangle \left\langle \varepsilon
(x_{1})\ldots\varepsilon(x_{n})\right\rangle
_{\text{ff}}\left\langle V_{a}(x_{1})\ldots
V_{a}(x_{n})\right\rangle _{\text{L}}d^{2}x_{3}\ldots
d^{2}x_{n}\label{epsn}%
\end{equation}
where $\left\langle \ldots\right\rangle _{\text{ff}}$ and
$\left\langle \ldots\right\rangle _{\text{L}}$ are the correlation
functions respectively in the theory of free Majorana fermion and
LFT while $C$ and $\bar C$ are familiar in 2D gravity and string
theory gauge fixing ghosts. Expression (\ref{epsn}) doesn't need any
comments for a reader familiar with the string theory, otherwise
some details and explanations are placed in ref.\cite{LYgrav}. There
one can also find how to treat the case $n<3$, which does not fit in
eq.(\ref{epsn}) and therefore requires special consideration.

Again, a simple dimensional analysis shows that $\left\langle
\varepsilon ^{n}\right\rangle _{\text{GIM}}\sim\mu^{(Q-na)b}$ and
therefore (\ref{Zexp}) is nothing but the power series for the
scaling function $G_{\text{sphere}}$. Recall also that in the theory
of fermions only the $\varepsilon$-correlations are non-zero with an
even number of insertions $n=2k$. The expansion can be rewritten in
the form
\begin{equation}
\frac{Z(m,\mu)}{Z(0,\mu)}=\sum_{k=0}^{\infty}\frac{a_{2k}}{(2\pi)^{2k}%
(2k)!}\left(  \frac{m^{2}}{\mu^{2s}}\right)  ^{k}\label{Znorm}%
\end{equation}
with
\begin{equation}
a_{n}=Z_{\text{L}}^{-1}\mu^{ns}\left\langle
\varepsilon^{n}\right\rangle
_{\text{GIM}}\label{an}%
\end{equation}
Here $Z_{\text{L}}$ denotes the Liouville partition function of the
sphere $Z_{\text{L}}=\left\langle I\right\rangle _{\text{L}}$ (we
imply that the fermionic correlation function are normalized in the
standard way so that $\left\langle I\right\rangle _{\text{ff}}=1$).

As in \cite{LYgrav} we find it convenient to introduce the fixed
area
partition function $Z_{A}(m)$%
\begin{equation}
Z_{A}(m)=A\int_{\uparrow}Z(m,\mu)e^{\mu A}\frac{d\mu}{2\pi i}\label{ZA}%
\end{equation}
the integration contour $\uparrow$ going along the imaginary axis of
$\mu$ to the right from all singularities of the integrand. This
partition function expands as follows
\begin{equation}
Z_{A}(m)=Z_{A}(0)\sum_{k=0}^{\infty}z_{2k}\eta^{2k}\label{ZAtn}%
\end{equation}
where the coefficients $z_{n}$ are easily related to those in
(\ref{Znorm})
\begin{equation}
z_{n}=\frac{\pi^{ns}a_{n}\Gamma(-b^{-2}-1)}{(2\pi)^{n}n!\Gamma(-b^{-2}%
-1+ns)}\label{zn}%
\end{equation}
and the scaling parameter $\eta$ reads
\begin{equation}
\eta=m\left(  \frac A\pi\right)  ^{s}\label{eta}%
\end{equation}
Due to the extra $\Gamma$ function in the denominator of (\ref{zn})
convergence of the series (\ref{ZAtn}) is much better then that of
(\ref{Znorm}). In fact it is \textit{absolutely convergent} and
defines an
entire scaling function of $t=\eta^{2}$%
\begin{equation}
z(t)=\sum_{k=0}^{\infty}z_{2k}t^{k}\label{zt}%
\end{equation}

\section{Perturbative terms}

As usual (see e.g., \cite{DO3, LYgrav}) the Liouville partition
function is restored from the three-point function in LFT \cite{DO3,
AAl3}
\begin{equation}
Z_{\text{L}}=\left[  \pi\mu\gamma(b^{2})\right]
^{Q/b}\frac{1-b^{2}}{\pi
^{3}Q\gamma(b^{2})\gamma(b^{-2})}\label{ZL3}%
\end{equation}
In the same way, the normalization (\ref{epsnorm}) corresponds to
the following (unnormalized) two-point function \cite{LYgrav}
\begin{equation}
\left\langle \varepsilon\varepsilon\right\rangle
_{\text{GIM}}=-\left(  \pi \mu\gamma(b^{2})\right)
^{(Q-2a)/b}\frac{\gamma(2ab-b^{2})\gamma
(2ab^{-1}-b^{-2})}{\pi(Q-2a)}\label{eps2}%
\end{equation}
For the first two coefficients in (\ref{zt}) we obtain
\begin{align}
z_{0} &  =1\label{z0z2}\\
z_{2} &  =\frac{\gamma(b^{2}(2s-1))\Gamma(b^{-2}-1)\gamma^{1-2s}(b^{2}%
)}{8\Gamma(1+b^{-2}-2s)}\nonumber
\end{align}

Next perturbative coefficient $z_{4}$ requires more involved
calculations. Basically it is reduced to the following integral
\begin{equation}
z_{4}=\frac1{24(2\pi)^{4}}\int\left\langle \varepsilon(x)\varepsilon
(0)\varepsilon(1)\varepsilon(\infty)\right\rangle
_{\text{ff}}\left\langle \left\langle
V_{a}(x)V_{a}(0)V_{a}(1)V_{a}(\infty)\right\rangle \right\rangle
_{\text{L}}^{(A)}d^{2}x\label{z4int}%
\end{equation}
where the normalized fixed area Liouville correlation functions
\begin{equation}
\left\langle \left\langle \ldots\right\rangle \right\rangle _{\text{L}}%
^{(A)}=\frac{\left\langle \ldots\right\rangle _{\text{L}}^{(A)}}{Z_{\text{L}%
}^{(A)}}\label{Cnorm}%
\end{equation}
are defined through the fixed area partition function
\begin{equation}
Z_{\text{L}}^{(A)}=\left(  \frac{\pi\gamma(b^{2})}A\right)
^{Q/b}\frac
{\Gamma(2-b^{2})}{\pi^{3}b^{3}\gamma(b^{2})\Gamma(b^{-2})}\label{ZLA}%
\end{equation}
and the fixed area unnormalized correlations
\begin{equation}
\left\langle V_{a_{1}}\ldots V_{a_{n}}\right\rangle _{\text{L}}^{(A)}%
=A\int_{\uparrow}\left\langle V_{a_{1}}\ldots V_{a_{n}}\right\rangle
_{\text{L}}e^{\mu A}\frac{d\mu}{2\pi i}=\frac{\left\langle
V_{a_{1}}\ldots V_{a_{n}}\right\rangle _{\text{L}}|_{\mu\rightarrow
A^{-1}}}{\Gamma
(-b^{-2}-1+b^{-1}\sum_{i=1}^{n}a_{i})}\label{CLA}%
\end{equation}

In the four-point case the Liouville correlation function can be
constructed almost explicitly as the holomorphic-antiholomorphic
decomposition \cite{AAl3}. We quote the relevant formulas explicitly
below in sect.10, where the problem of the fourth-order coefficient
is discussed in some more detail. Here we present only the matter
four-point function $\left\langle
\varepsilon(x)\varepsilon(0)\varepsilon(1)\varepsilon(\infty)\right\rangle
_{\text{ff}}$ . It is made of the free fermion propagators and has a
factorized form
\begin{equation}
\left\langle \varepsilon(x)\varepsilon(0)\varepsilon(1)\varepsilon
(\infty)\right\rangle
_{\text{ff}}=F_{\text{ff}}(x)F_{\text{ff}}(\bar
x)\label{e4}%
\end{equation}
where
\begin{equation}
F_{\text{ff}}(x)=\frac{1-x+x^{2}}{x(1-x)}\label{Fff}%
\end{equation}

We postpone the analysis of the more complicated four point term
and, as a preliminary step, restrict our perturbative information to
what we have explicitly in eq.(\ref{z0z2}). At the first sight this
pity information is by no means enough to make any conclusions about
the whole scaling function. We will see in a moment that when
supplemented with some analytic properties of the entire scaling
function (which, in order, follow from natural physical arguments)
and also with an additional hypothesis about the location of its
zeros, a surprisingly good description of $z(t)$ can be achieved
even with these restricted data.

\section{Criticality and $t\rightarrow\infty$ asymptotic}

If the area of the surface is very large as compared to the
characteristic scale $m^{-\rho}$ of the perturbed matter, we expect
the following asymptotic
\begin{equation}
Z_{A}(m)\sim A^{Q^{\prime}/b^{\prime}}\exp\left(
-\mathcal{E}_{0}A\right)
\label{Aass}%
\end{equation}
Here $\mathcal{E}_{0}$ is the specific (per unit area) free energy
of the perturbed matter interacting with the quantized gravity. We
introduced also a convenient parameter
\begin{equation}
\rho=(2s)^{-1}\label{rho}%
\end{equation}
For the dimensional reasons
\begin{equation}
\mathcal{E}_{0}=-m^{2\rho}f_{0}\label{f0}%
\end{equation}
where $f_{0}$ is some numerical constant, which depends in our model
on the Liouville parameter $b^{2}$. As for the power correction
$A^{Q^{\prime }/b^{\prime}}$, it can figured out through the
following speculation. It is almost obvious that in our model the
massive fermion develops a final correlation length $\sim m^{-\rho}$
and therefore, at the asymptotic scales we are interested in, it
doesn't influence the dynamics contributing only to the local
quantities like $\mathcal{E}_{0}$. Thus, the large scale
fluctuations of the surface are governed by the residual spectator
matter (and the ghosts) and therefore the effective infrared
Liouville theory is characterized by another parameters $b^{\prime}$
and $Q^{\prime}=1/b^{\prime}+b^{\prime}$. The latter can be found
from the IR central charge balance \cite{LYgrav}
\begin{equation}
1+6(Q^{\prime})^{2}+c_{\text{sp}}=26\label{IRc}%
\end{equation}
or
\begin{equation}
Q^{\prime}=\sqrt{Q^{2}+\frac1{12}}\label{Qp}%
\end{equation}

On all these physical accounts we assume the following
$t\rightarrow\infty$ asymptotic of the scaling function $z(t)$
\begin{equation}
z(t)\sim\exp\left(  \pi f_{0}t^{\rho}+\rho\left(
Q/b-Q^{\prime}/b^{\prime
}\right)  \log t+O(1)\right) \label{zass}%
\end{equation}
Note that as usual $\mu_{\text{c}}=m^{2\rho}f_{0}$ is interpreted as
the position of the critical singularity in the ``grand'' partition
function
(\ref{Fsphere}). In other words, the critical value of $m$ is $m_{\text{c}%
}=\left(  \mu/f_{0}\right)  ^{s}$.

\section{Basic conjecture}

Our basic conjecture states that in the region of interest $0\leq
b^{2}\leq1$ the scaling function $z(t)$ always enjoys the asymptotic
(\ref{zass}) in the whole complex plane except for the negative part
of the real axis, where \textit{all} its zeros are located. This is
very restrictive requirement for an entire function and it is this
property, together with some numerical fortuity, which makes even
very restricted perturbative input miraculously efficient.

At $t\rightarrow-\infty$ the asymptotic of $z(t)$ is a result of
competition of two terms
\begin{align}
z(t) &  =\exp\left(  \pi f_{0}t^{\rho}e^{i\pi\rho}-\left(
1/2+\delta\right)
\log e^{i\pi}t+\ldots\right)  +\text{c.c.}\label{zneg}\\
\  &  =2\exp\left(  \pi f_{0}(-t)^{\rho}\cos\pi\rho-\left(
1/2+\delta\right) \log(-t)\right)  \cos\left(  \pi
f_{0}(-t)^{\rho}\sin\pi\rho-\pi\left( 1/2+\delta\right)  \right)
\nonumber
\end{align}
where for convenience we have denoted
\begin{equation}
\rho\left(  Q/b-Q^{\prime}/b^{\prime}\right)  =-\frac12-\delta\label{del}%
\end{equation}
Asymptotic position of $n$-th zero $t_{n}$, $n=1,2,\ldots$ at
$n\rightarrow \infty$ follows
\begin{equation}
-t_{n}=\left(  \frac{n+\delta}{-f_{0}\sin\pi\rho}\right)  ^{2s}\label{tn}%
\end{equation}

If the zeros of an entire function are known we have (see below
about the convergence problem)
\begin{equation}
z(t)=\prod_{n=1}^{\infty}\left(  1-\frac t{t_{n}}\right) \label{zprod}%
\end{equation}
and therefore
\begin{equation}
\log z(t)=\sum_{n=1}^{\infty}\log\left(  1-\frac t{t_{n}}\right)
=t\sum
_{n=1}^{\infty}\frac1{-t_{n}}-\frac{t^{2}}2\sum_{n=1}^{\infty}\frac
1{(-t_{n})^{2}}-\ldots\label{t2t4}%
\end{equation}
From the first two terms we find
\begin{equation}
z_{2}=\sum_{n=1}^{\infty}\frac1{-t_{n}}\label{z2sum}%
\end{equation}
and
\begin{equation}
z_{4}-\frac{z_{2}^{2}}2=r_{2}=-\frac12\sum_{n=1}^{\infty}\frac1{(-t_{n})^{2}%
}\label{z4sum}%
\end{equation}

In our particular case the product (\ref{zprod}), as well as the sum
(\ref{z2sum}), are divergent since the order $\rho$ of our function
is always $1\leq\rho<(2-\sqrt{2})^{-1}=1.70711\ldots$. It can be
shown, however (see Appendix), that the sum rule (\ref{z2sum})
remains valid if the right hand side is understood in terms of the
zeta function of the zeros $t_{n}$. A mathematically correct
canonical product for the function of order $\rho<2$ with zeros at
$t_{n}$ reads
\begin{equation}
z(t)=\exp\left(  z_{2}t\right)  \prod_{n=1}^{\infty}\left(  1-\frac t{t_{n}%
}\right)  \exp\left(  \frac t{t_{n}}\right) \label{zweis}%
\end{equation}

\section{Numerics and approximations}%

\begin{table}[htb] \centering
\begin{tabular}
[c]{|llllllll|}\hline \multicolumn{1}{|l|}{$b^{2}$} &
\multicolumn{1}{l|}{$s$} & \multicolumn{1}{l|}{$z_{2}$} &
\multicolumn{1}{l|}{$\delta$} &
\multicolumn{1}{l|}{$-f_{0}$} & \multicolumn{1}{l|}{$-f_{0}^{\text{(exact)}}$}%
& \multicolumn{1}{l|}{$z_{4}^{\text{(est)}}$} & $z_{4}$\\\hline
\multicolumn{1}{|l|}{$0.05$} & \multicolumn{1}{l|}{$0.488$} &
\multicolumn{1}{l|}{$-5.26967$} & \multicolumn{1}{l|}{$-0.414$} &
\multicolumn{1}{l|}{$1.61954$} & \multicolumn{1}{l|}{} &
\multicolumn{1}{l|}{$13.8494$} & \\\hline
\multicolumn{1}{|l|}{$0.10$} & \multicolumn{1}{l|}{$0.475$} &
\multicolumn{1}{l|}{$-2.7913$} & \multicolumn{1}{l|}{$-0.411$} &
\multicolumn{1}{l|}{$0.831718$} & \multicolumn{1}{l|}{} &
\multicolumn{1}{l|}{$3.85341$} & \\\hline
\multicolumn{1}{|l|}{$0.15$} & \multicolumn{1}{l|}{$0.463$} &
\multicolumn{1}{l|}{$-1.9818$} & \multicolumn{1}{l|}{$-0.408$} &
\multicolumn{1}{l|}{$0.575331$} & \multicolumn{1}{l|}{} &
\multicolumn{1}{l|}{$1.91275$} & \\\hline
\multicolumn{1}{|l|}{$0.20$} & \multicolumn{1}{l|}{$0.450$} &
\multicolumn{1}{l|}{$-1.59146$} & \multicolumn{1}{l|}{$-0.404$} &
\multicolumn{1}{l|}{$0.452694$} & \multicolumn{1}{l|}{} &
\multicolumn{1}{l|}{$1.20427$} & \\\hline
\multicolumn{1}{|l|}{$0.25$} & \multicolumn{1}{l|}{$0.438$} &
\multicolumn{1}{l|}{$-1.37065$} & \multicolumn{1}{l|}{$-0.399$} &
\multicolumn{1}{l|}{$0.384486$} & \multicolumn{1}{l|}{} &
\multicolumn{1}{l|}{$0.863088$} & \\\hline
\multicolumn{1}{|l|}{$0.30$} & \multicolumn{1}{l|}{$0.427$} &
\multicolumn{1}{l|}{$-1.23648$} & \multicolumn{1}{l|}{$-0.393$} &
\multicolumn{1}{l|}{$0.344516$} & \multicolumn{1}{l|}{} &
\multicolumn{1}{l|}{$0.670052$} & \\\hline
\multicolumn{1}{|l|}{$0.35$} & \multicolumn{1}{l|}{$0.415$} &
\multicolumn{1}{l|}{$-1.15384$} & \multicolumn{1}{l|}{$-0.386$} &
\multicolumn{1}{l|}{$0.321859$} & \multicolumn{1}{l|}{} &
\multicolumn{1}{l|}{$0.547863$} & \\\hline
\multicolumn{1}{|l|}{$0.40$} & \multicolumn{1}{l|}{$0.404$} &
\multicolumn{1}{l|}{$-1.10562$} & \multicolumn{1}{l|}{$-0.378$} &
\multicolumn{1}{l|}{$0.311425$} & \multicolumn{1}{l|}{} &
\multicolumn{1}{l|}{$0.462962$} & \\\hline
\multicolumn{1}{|l|}{$0.45$} & \multicolumn{1}{l|}{$0.393$} &
\multicolumn{1}{l|}{$-1.08291$} & \multicolumn{1}{l|}{$-0.368$} &
\multicolumn{1}{l|}{$0.310866$} & \multicolumn{1}{l|}{} &
\multicolumn{1}{l|}{$0.398285$} & \\\hline
\multicolumn{1}{|l|}{$0.50$} & \multicolumn{1}{l|}{$0.382$} &
\multicolumn{1}{l|}{$-1.08110$} & \multicolumn{1}{l|}{$-0.357$} &
\multicolumn{1}{l|}{$0.319416$} & \multicolumn{1}{l|}{} &
\multicolumn{1}{l|}{$0.343762$} & \\\hline
\multicolumn{1}{|l|}{$0.55$} & \multicolumn{1}{l|}{$0.372$} &
\multicolumn{1}{l|}{$-1.09830$} & \multicolumn{1}{l|}{$-0.343$} &
\multicolumn{1}{l|}{$0.337485$} & \multicolumn{1}{l|}{} &
\multicolumn{1}{l|}{$0.292372$} & \\\hline
\multicolumn{1}{|l|}{$0.60$} & \multicolumn{1}{l|}{$0.362$} &
\multicolumn{1}{l|}{$-1.13467$} & \multicolumn{1}{l|}{$-0.327$} &
\multicolumn{1}{l|}{$0.366663$} & \multicolumn{1}{l|}{} &
\multicolumn{1}{l|}{$0.2381$} & \\\hline
\multicolumn{1}{|l|}{$0.65$} & \multicolumn{1}{l|}{$0.352$} &
\multicolumn{1}{l|}{$-1.19248$} & \multicolumn{1}{l|}{$-0.306$} &
\multicolumn{1}{l|}{$0.41011$} & \multicolumn{1}{l|}{} &
\multicolumn{1}{l|}{$0.174833$} & \\\hline
\multicolumn{1}{|l|}{$0.70$} & \multicolumn{1}{l|}{$0.342$} &
\multicolumn{1}{l|}{$-1.27680$} & \multicolumn{1}{l|}{$-0.281$} &
\multicolumn{1}{l|}{$0.473557$} & \multicolumn{1}{l|}{} &
\multicolumn{1}{l|}{$0.094298$} & \\\hline
\multicolumn{1}{|l|}{$0.75$} & \multicolumn{1}{l|}{$0.333$} &
\multicolumn{1}{l|}{$-1.39733$} & \multicolumn{1}{l|}{$-0.25$} &
\multicolumn{1}{l|}{$0.567684$} & \multicolumn{1}{l|}{$0.563124$} &
\multicolumn{1}{l|}{$-0.016269$} & $0$\\\hline
\multicolumn{1}{|l|}{$0.80$} & \multicolumn{1}{l|}{$0.325$} &
\multicolumn{1}{l|}{$-1.57298$} & \multicolumn{1}{l|}{$-0.21$} &
\multicolumn{1}{l|}{$0.714273$} & \multicolumn{1}{l|}{} &
\multicolumn{1}{l|}{$-0.18167$} & \\\hline
\multicolumn{1}{|l|}{$0.85$} & \multicolumn{1}{l|}{$0.316$} &
\multicolumn{1}{l|}{$-1.84474$} & \multicolumn{1}{l|}{$-0.16$} &
\multicolumn{1}{l|}{$0.964987$} & \multicolumn{1}{l|}{} &
\multicolumn{1}{l|}{$-0.462053$} & \\\hline
\multicolumn{1}{|l|}{$0.90$} & \multicolumn{1}{l|}{$0.308$} &
\multicolumn{1}{l|}{$-2.32185$} & \multicolumn{1}{l|}{$-0.10$} &
\multicolumn{1}{l|}{$1.47712$} & \multicolumn{1}{l|}{} &
\multicolumn{1}{l|}{$-1.05291$} & \\\hline
\multicolumn{1}{|l|}{$0.95$} & \multicolumn{1}{l|}{$0.300$} &
\multicolumn{1}{l|}{$-3.45523$} & \multicolumn{1}{l|}{$-0.023$} &
\multicolumn{1}{l|}{$3.04496$} & \multicolumn{1}{l|}{} &
\multicolumn{1}{l|}{$-3.04188$} & \\\hline
\multicolumn{1}{|l|}{$0.99$} & \multicolumn{1}{l|}{$0.294$} &
\multicolumn{1}{l|}{$-8.75236$} & \multicolumn{1}{l|}{$0.049$} &
\multicolumn{1}{l|}{$15.7725$} & \multicolumn{1}{l|}{} &
\multicolumn{1}{l|}{$-23.4948$} & \\\hline
\end{tabular}
\caption{First order approximation for $f_0$ at different values of
$b^2$. We present also our analytic estimate for the forth-order
perturbative coefficient $z_4$ through eq.(\ref{z4as})\label
{table1}}%
\end{table}

\begin{figure}
[tbh]
\begin{center}
\includegraphics[
height=3.8493in, width=4.8585in
]%
{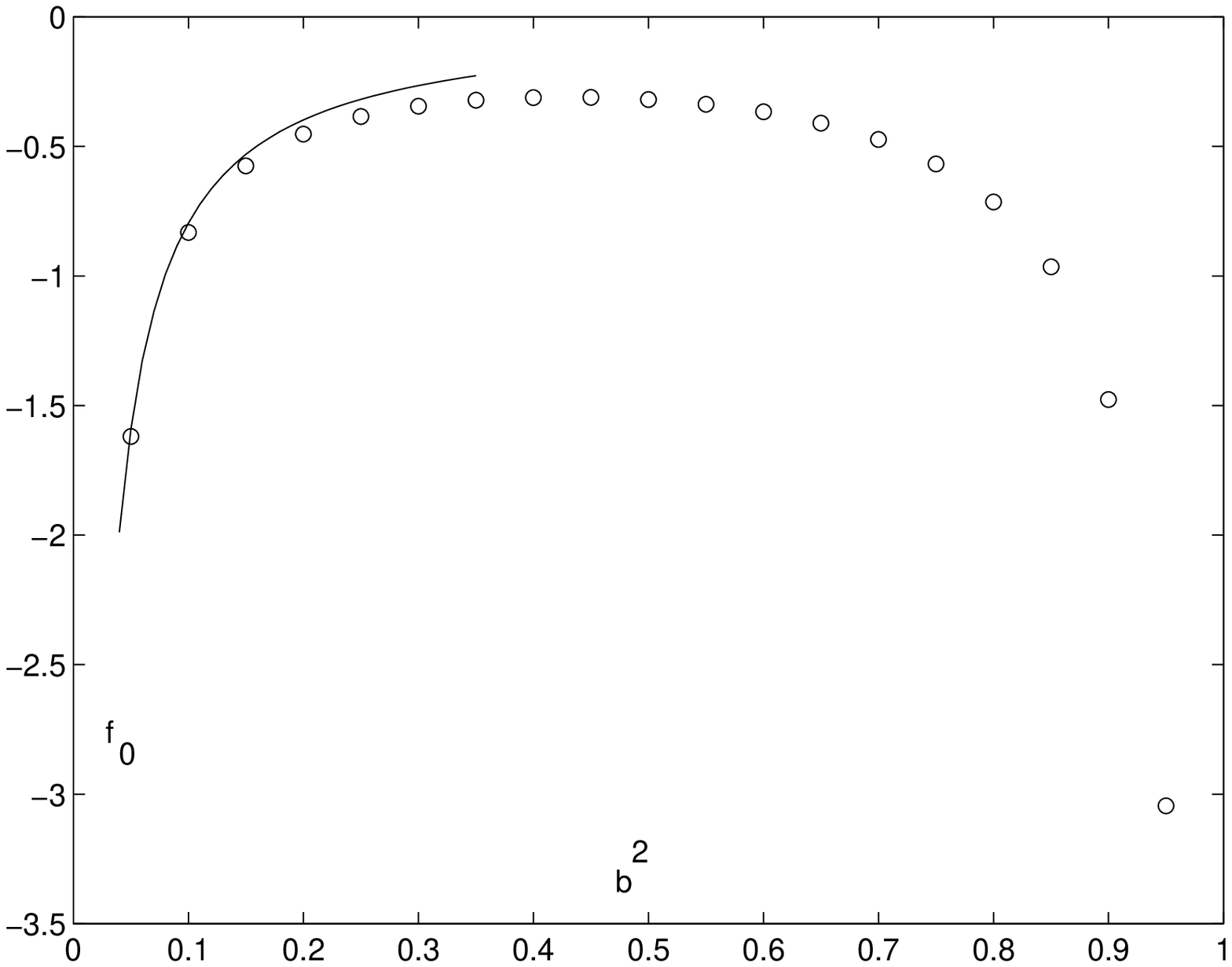}%
\caption{First order approximation for $f_{0}$ at different values
of $b^{2}$
(circles). Semiclassical limit is drawn as a continuous curve.}%
\label{f0plot}%
\end{center}
\end{figure}

Suppose that the asymptotic (\ref{tn}) gives a reasonably good
approximation for $t_{n}$ even at small $n$. Then the sum rule
(\ref{z2sum}) allows to relate approximately the parameter $f_{0}$
in the asymptotic (\ref{tn}) to the first perturbative coefficient
$z_{2}$
\begin{equation}
z_{2}=\left(  f_{0}\sin\pi\rho\right)
^{2s}\sum_{n=1}^{\infty}\frac1{\left( n+\delta\right)  ^{2s}}=\left(
f_{0}\sin\pi\rho\right)  ^{2s}\zeta
(2s,\delta+1)\label{z2zeta}%
\end{equation}
The divergent sum is evaluated as the analytic continuation through
the ordinary Riemann $\zeta$-function. The first estimate for
$f_{0}$ is then
\begin{equation}
f_{0}=-\frac1{\sin\pi\rho}\left(
\frac{z_{2}}{\zeta(2s,\delta+1)}\right)
^{\rho}\label{f0zeta}%
\end{equation}
For different values of the spectator parameter $b^{2}$ these
estimates are presented in Table \ref{table1} and plotted in
fig.\ref{f0plot}. Two examinations allow us to evaluate the accuracy
of the approximation. First is the exactly solvable point
$b^{2}=3/4$ which corresponds to pure Ising (no spectator matter)
and is believed to be related to a matrix model solution.
Accidentally or not, the comparison of the exact matrix model result
(see sect. 9 below) with our prediction for this point is impressive
(table \ref{table1}). Another one is the classical region
$b^{2}\rightarrow0$. Here we are dealing with the rigid sphere with
the quantum gravity effects suppressed. The specific vacuum energy
of the free massive fermion of mass $m $ in the flat classical
background is
\begin{equation}
\mathcal{E}_{0}=\frac{m^{2}}{4\pi}\log m\label{E0cl}%
\end{equation}
From the scaling (\ref{f0}) and (\ref{acl}) we interpret this as the
following
leading behavior of $f_{0}(b^{2})$ at $b^{2}\rightarrow0$%
\begin{equation}
f_{0}=-\frac1{4\pi b^{2}}+O(1)\label{f0scl}%
\end{equation}
This asymptotic is also plotted in fig.\ref{f0plot}. Further
semiclassical corrections to the vacuum energy also can be evaluated
systematically in the field theory framework. The authors plan to
address this interesting topic elsewhere.

The second sum rule (\ref{z4sum}) allows to obtain an estimate for
the next perturbative coefficient $z_{4}$
\begin{equation}
z_{4}\approx z_{4}^{\text{(est)}}=\frac{z_{2}^{2}}2-\frac12\left(  -f_{0}%
\sin\pi\rho\right)  ^{4s}\zeta(4s,\delta+1)\label{z4as}%
\end{equation}
These numbers for $z_{4}^{\text{(est)}}$ are also presented in the
Table. In principle they can be compared with the direct evaluation
of the integral in (\ref{z4int}). This would be a crucial
examination for our basic conjecture and also an estimate of the
convergence of the analytic-numeric procedure. Unfortunately this
direct calculation is more difficult technically. A
preliminary discussion will be given in sect.10.%

\begin{figure}
[tbh]
\begin{center}
\includegraphics[
height=3.7092in, width=4.7495in
]%
{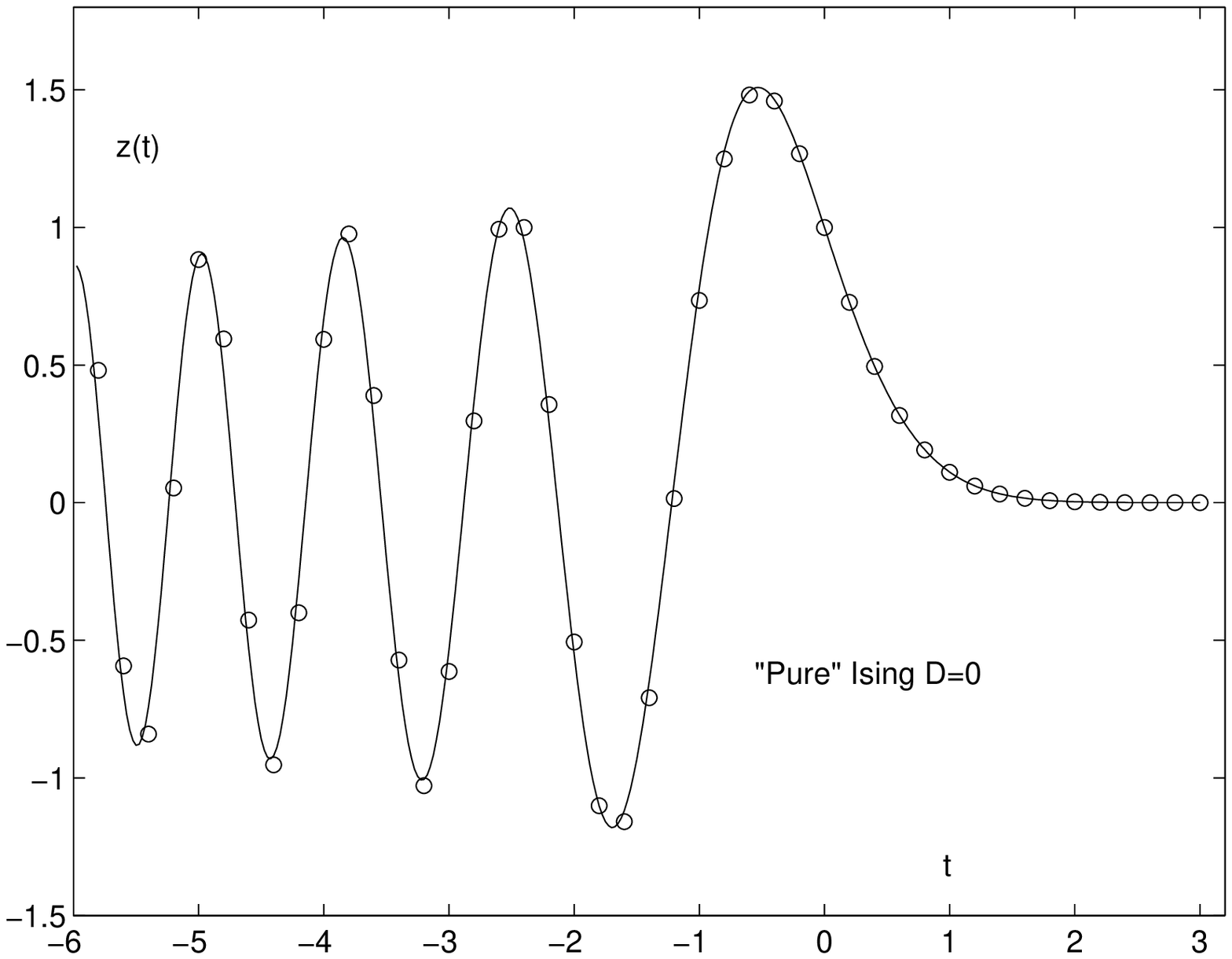}%
\caption{Scaling function $z(t)$ at the solvable point $b^{2}=0.75$.
Our approximation (\ref{fast}) (circles) is compared with the exact
matrix model
function (\ref{fsphAiry}) (continuous curve).}%
\label{plot34}%
\end{center}
\end{figure}

Once the parameter $f_{0}$ is estimated the zeros $t_{n}$ are
evaluated through the asymptotic formula (\ref{tn}). The approximate
scaling function $z(t)$ is then constructed as the product
(\ref{zweis}). Even better convergent expression, more suitable for
the numerics, is
\begin{equation}
z_{\text{est}}(t)=\exp\left(  z_{2}t+\left(  z_{4}^{\text{(est)}}-z_{2}%
^{2}/2\right)  t^{2}\right)  \prod_{n=1}^{\infty}\left(  1-\frac t{t_{n}%
}\right)  \exp\left(  \frac t{t_{n}}+\frac{t^{2}}{2t_{n}^{2}}\right)
\label{fast}%
\end{equation}
For the ``pure Ising'' case $b^{2}=0.75$ this scaling function is
plotted in fig.\ref{plot34} and compared with the exact scaling
function available from the matrix model solution (see below).

\section{Random lattice Ising model}

Now we consider a model of discrete (random lattice) gravity which
seems to be directly related (in its continuous limit) to the field
theory of the previous sections. Although the continuous limit is
not supposed to depend essentially on the details of the microscopic
lattice description, here we'll formulate a very particular system,
the Ising model on the triangular random lattice. This is basically
the standard random lattice Ising system as e.g., considered in
refs.\cite{KIsing, Boulatov}. In this standard formulation the model
is exactly solvable by the matrix model technique, and we're going
to use the essence of this solution in what follows. However, in
order to match the more general pattern introduced in sect.2 with
the spectator conformal matter (and in fact in order to make the
situation less vulgar) we decorate this model by a $D$-component
massless boson, or, in the string theory language, by immersing the
system to $D$-dimensional target space. We expect that in the
continuous limit this additional lattice boson simulates our
``spectator matter'', at least for the spherical topology, with
$c_{\text{sp}}=D$.

\begin{figure}
[tbh]
\begin{center}
\includegraphics[
natheight=3.358900in, natwidth=4.170100in, height=3.3589in,
width=4.1701in
]%
{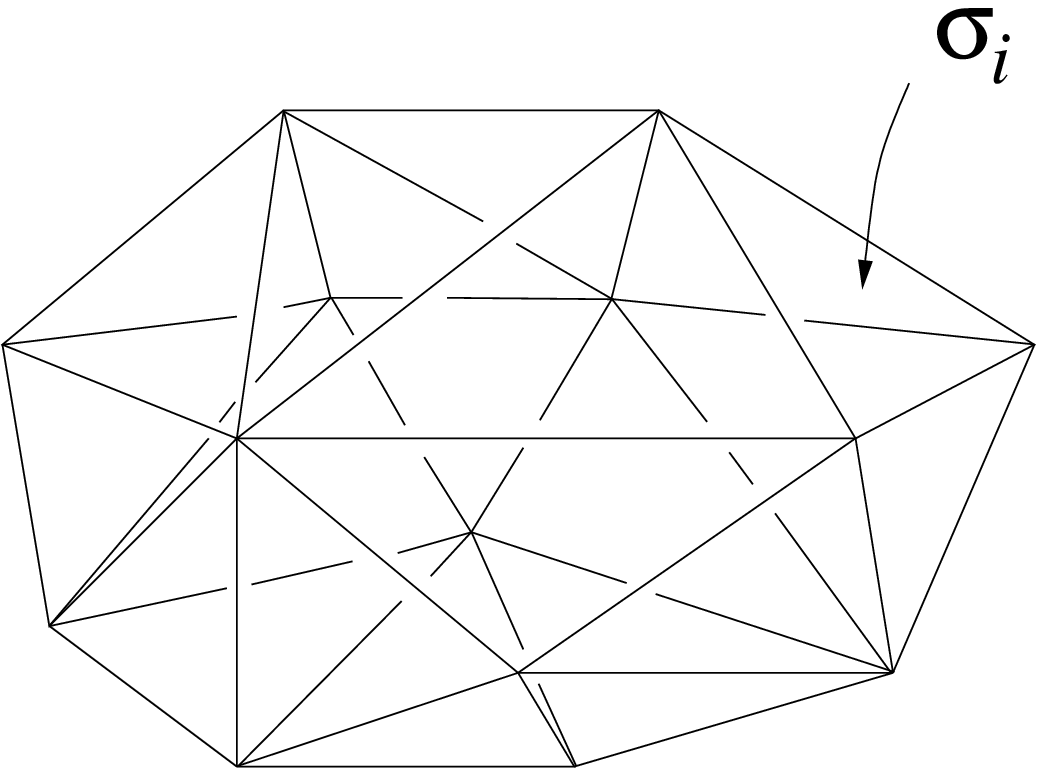}%
\caption{A sample element of $\left\{  G_{N}\right\}  $ for $N=24$.
Ising
spins $\sigma_{i}$ are attached to the faces of the triangulation.}%
\label{spinsw}%
\end{center}
\end{figure}

Take an irregular planar lattice (called also a graph or a
triangulation) constructed from $N$ triangles. Planar means here
that the graph can be drawn on a sphere without intersections. An
example is given in fig.\ref{spinsw}. Let $\{G_{N}\}$ be the
ensemble of such topologically different graphs. Let's also
enumerate the triangles of a particular graph by index
$i=1,\ldots,N$ and attach to each one a spin variable
$\sigma_{i}=\pm1$. Finally, let $I_{ij}$ be
the incidence matrix of the corresponding triangulation%
\begin{equation}
I_{ij}=%
\genfrac{\{}{.}{0pt}{}{1\;\;\text{ if triangles }i\;\text{and
}j\text{ have
common edge}}{0\;\;\text{otherwise\ }}%
\label{Iij}%
\end{equation}
The familiar Ising model corresponds to the spin configurations
distributed with the Gibbs weights $W[\left\{  \sigma_{i}\right\}
]=\exp\left( -\mathcal{H}[\left\{  \sigma_{i}\right\}  ]\right)  $
where
\begin{equation}
\mathcal{H}[\left\{  \sigma_{i}\right\}  ]=K\sum_{i,j}I_{ij}\sigma_{i}%
\sigma_{j}+H\sum_{i}\sigma_{i}\label{Hising}%
\end{equation}
and $K$ and $H$ are standard ``temperature'' and ``magnetic field''
parameters. Although a non-vanishing magnetic field makes the
physics of the model much more rich and interesting, to avoid
additional complications presently we restrict ourselves by the zero
magnetic field $H=0$. Thermodynamic information is encoded in the
microcanonic partition sum
\begin{equation}
Z_{N}(K)=N\sum_{^{\left\{  G_{N}\right\}
}}\det\nolimits^{-D/2}\left(
\Delta_{ij}^{\text{lat}}\right)  \sum_{\left\{  \sigma_{i}\right\}  }%
\exp\left(  -\mathcal{H}[\left\{  \sigma_{i}\right\}  ]\right) \label{ZN}%
\end{equation}
Notice unusual factor of $N$ in the definition of the partition sum,
which is introduced to facilitate the subsequent comparisons with
the continuous theory. The determinant factor represents the
discrete version of the Laplace operator
\begin{equation}
\Delta_{ij}^{\text{lat}}=\left(  I_{ij}-3\delta_{ij}\right)
^{\prime
}\label{Dij}%
\end{equation}
and introduces new ``spectator central charge'' parameter $D$. This
discrete operator always have zero eigenvalue, which should be
projected out in the definition (\ref{ZN}), the prime in the last
formula indicating this zero mode prescription.

As it is, the problem (\ref{ZN}) does not seem easier then the
continuous path integral over $g_{ab}$. To simulate a continuous
surface one should take a kind of thermodynamic limit, where the
size of the graph $N$ goes to infinity. In this limit direct
calculation of the partition sum seems a complicated problem.
Fortunately, for certain choices of statistical weights, there is a
powerful machinery, which permits to calculate effectively the
thermodynamic limit of the sums like (\ref{ZN}). This is famous
matrix model technique. Here we do not go into any details of this
interesting theory, referring for example to the review
\cite{matrix}. We'd only like to mention that a particular case of
(\ref{ZN}) with $D=0$ (pure RLIM) is solved exactly in \cite{KIsing}
and \cite{Boulatov} (see also \cite{Burda} where precisely our model
of random triangulations with $D=0$ is treated). Some important
things can be learned from this exact solution\footnote{Here we
discuss only the case $H=0$, although the model is also solvable in
a non-zero magnetic field.}. We consider the asymptotic of large $N$
which only can be related to a continuous theory.

\textbf{1. At generic values} of the temperature parameter $K$ the
$N\rightarrow\infty$ asymptotic has the form
\begin{equation}
Z_{N}(K)\sim\mathcal{Z}(K)N^{-5/2}e^{-E(K)N}\label{Pass}%
\end{equation}
Here $E(K)$ is the specific (per a triangle) free energy. This
function is non-universal and depends on the details of the random
lattice model. For our very particular realization it can be found
explicitly in \cite{Burda}. On the contrary, the power-like
pre-exponential dependence $N^{-5/2}$ is universal, the index $-5/2$
is known as the ``pure gravity'' critical exponent\footnote{It
differs from the famous $-7/2$ of ref.\cite{Brezinetal} just because
of extra factor of $N$ in the definition (\ref{ZN}).}.

\textbf{2. A criticality,} like in the standard Ising model, occurs
at certain value $K=K_{\text{c}}$ (the critical temperature). This
value again depends on the microscopic details of the model. Our
particular triangular system has \cite{Burda}
\begin{equation}
K_{\text{c}}=\frac12\log\frac{23}{108}\label{Kc}%
\end{equation}
The asymptotic (\ref{Pass}) changes here to
\begin{equation}
\,Z_{N}(K_{\text{c}})\sim\mathcal{Z}_{\text{c}}N^{-7/3}e^{-E(K_{\text{c}}%
)N}\label{Iass}%
\end{equation}
The power pre-exponential behavior is changed, the exponent $-7/3$
being familiar as the ``gravitational Ising'' critical index. In
particular this change of the asymptotic means that $\mathcal{Z}(K)$
is singular at
$K=K_{\text{c}}$ and in any case $\mathcal{Z}_{\text{c}}\neq\mathcal{Z}%
(K_{\text{c}})$.

\textbf{3. The crossover scaling} behavior at $K$ near the critical
point is expected to be universal and a subject of the field theory
application. Let
\begin{equation}
\tau=\frac{K-K_{\text{c}}}{K_{\text{c}}}\label{tau}%
\end{equation}
and $\left|  \tau\right|  \ll1$. In this region it turns out
relevant to introduce the ``correlation size'' $N_{\text{c}}$ of the
lattice. While generically $N_{\text{c}}\sim1$, near the critical
point $N_{\text{c}}$ diverges. Exact solution of the RLIM shows that
\begin{equation}
N_{\text{c}}\sim\frac{L_{0}}{\left|  \tau\right|  ^{3}}\gg1\label{Nc}%
\end{equation}
where $L_{0}$ is some (non-universal) constant dependent on the
choice of the scale. In the large $N$ asymptotic of the partition
function $Z_{N}(K)$ we have to distinguish the following regions:

i) At $1\ll N\ll N_{\text{c}}$ the ``Ising'' behavior (\ref{Iass})
is observed.

ii) At $N\gg N_{\text{c}}$ the Ising spins correlate locally,
contributing only to $E(K)$. The ``pure gravity'' asymptotic
(\ref{Pass}) appears with
certain singular in $\tau$ contribution to $E(K)=E_{\text{reg}}%
(K)+E_{\text{sing}}(\tau)$, the singularity being caused by the long
range (as compared to the lattice scale) correlation of the Ising
spins
\begin{equation}
\,Z_{N}(K)\sim\mathcal{Z}(K_{\text{c}})N^{-5/2}e^{-E_{\text{reg}%
}(K)N-E_{\text{sing}}(\tau)N}\label{Psing}%
\end{equation}
The singular contribution $E_{\text{sing}}(\tau)$ in the exactly
solvable RLIM reads
\begin{equation}
E_{\text{sing}}(\tau)=e_{0}\left|  \tau\right|  ^{3}\label{e0}%
\end{equation}
where the amplitude of critical singularity $e_{0}$ again depends on
the choice of the scale.

iii) Crossover scaling function $F_{\text{sphere}}(y)$ appears at
$N\sim
N_{\text{c}}$%
\begin{equation}
Z_{N}(K)\sim F_{\text{sphere}}\left(  \frac N{N_{\text{c}}}\right)
N^{-7/3}e^{-E_{\text{reg}}(K)N}\label{Fscaling}%
\end{equation}
This function has the following asymptotic properties at small and
large values of its argument
\begin{align}
F_{\text{sphere}}(y) &  \sim\mathcal{Z}_{\text{c}}%
\;\;\;\;\;\;\;\;\;\;\;\;\;\;\;\;\;\;\;\;\;\;\;\;\;
\;\;\;\;\;\;\;\;\;\text{at}\;\;\;y\ll1\label{Fscass}\\
F_{\text{sphere}}(y) &  \sim F_{\infty}y^{-1/6}\exp\left(  -L_{0}%
e_{0}y\right)  \;\;\;\text{at\ \ \ \ }y\gg1\nonumber
\end{align}
where apparently
\begin{equation}
F_{\infty}=\mathcal{Z}(K_{\text{c}})N_{\text{c}}^{-1/6}\label{Finf}%
\end{equation}

Moreover, the exact solution of the matrix model gives the scaling
function $F_{\text{scaling}}(y)$ of RLIM in explicit form. Before
discussing it, let us conjecture natural generalizations of the
above scaling relations for model (\ref{ZN}) with
$-\infty<D\leq1/2$, which we find natural to name the generalized
random lattice Ising model (GRLIM). First, the assumed relation to
the continuous quantum gravity described in sect.2 suggests to
replace the generic asymptotic (\ref{Pass}) by
\begin{equation}
Z_{N}(K,D)\sim\mathcal{Z}(K,D)N^{-d_{\text{g}}(D)}e^{-E(K,D)N}\label{Dass}%
\end{equation}
where the familiar $c_{\text{sp}}=D$ KPZ \cite{KPZ} scaling is
expected\footnote{This exponent is related to the famous index
$\gamma _{\text{string}}$ (which string?) \cite{KMK} as
$d_{\text{g}}(D)=2-\gamma _{\text{string}}$.}
\begin{equation}
d_{\text{g}}(D)=\frac1{12}\left(  25-D+\sqrt{\left(  25-D\right)
\left(
1-D\right)  }\right) \label{dIR}%
\end{equation}
A critical point $K=K_{\text{c}}(D)$ is expected to exist in
general, where the asymptotic is different
\begin{equation}
Z_{N}(K_{\text{c}},D)\sim\mathcal{Z}_{\text{c}}(D)N^{-d_{\text{c}}%
(D)}e^{-E(K_{\text{c}},D)N}\label{IDass}%
\end{equation}
Now
\begin{equation}
d_{\text{c}}(D)=\frac1{24}\left(  49-2D+\sqrt{\left(  49-2D\right)
\left(
1-2D\right)  }\right) \label{dc}%
\end{equation}
Again $E(K,D)$ is interpreted as the specific (per triangle) free
energy of
the infinite system, which develops a singularity at $K=K_{\text{c}}$%
\begin{equation}
E(K,D)=E_{\text{reg}}(K,D)+e_{0}(D)\left|  \tau\right|  ^{2\rho(D)}%
\label{EDsing}%
\end{equation}
where
\begin{equation}
\rho(D)=\frac{\left(  \sqrt{49-2D}-\sqrt{1-2D}\right)  \left(  \sqrt
{49-2D}+\sqrt{25-2D}\right)  }{48}\label{rD}%
\end{equation}
$\tau$ is defined in eq.(\ref{tau}) and $e_{0}(D)$ is the amplitude
of critical singularity. Finally, in the scaling region $\left|
\tau\right| \ll1$ the correlation size scales as
\begin{equation}
N_{\text{c}}(D)\sim L_{0}(D)\left|  \tau\right|  ^{-2\rho(D)}\label{NcD}%
\end{equation}
and becomes large as compared to the lattice scales. In this region
and at $N\sim N_{\text{c}}$ equation (\ref{Fscaling}) is generalized
to
\begin{equation}
Z_{N}(K,D)\sim F_{\text{sphere}}\left(  \frac
N{N_{\text{c}}},D\right)
N^{-d_{\text{c}}(D)}e^{-E_{\text{reg}}(K,D)N}\label{FD}%
\end{equation}
where the scaling function now depends on $D$ and
\begin{align}
F_{\text{sphere}}(y) &  \sim\mathcal{Z}_{\text{c}}(D)%
\;\;\;\;\;\;\;\;\;\;\;\;\;\;\;\;\;\;\;\;\;\;\;\;\;\;\;\;\;\;\;\;\;
\;\;\;\;\;\;\;\;\;\;\;\;\;\;\;\text{at}%
\;\;\;y\ll1\label{FDass}\\
F_{\text{sphere}}(y) &  \sim F_{\infty}y^{d_{\text{c}}(D)-d_{\text{g}}(D)}%
\exp\left(  -e_{0}(D)L_{0}(D)y\right)  \;\;\;\text{at\ \ \ \
}y\gg1\nonumber
\end{align}

Notice, that we expect certain symmetry $\tau\rightarrow-\tau$ of
the singular part. This can be argued form the Ising model duality.
This symmetry holds also over a random lattice, mapping the model on
a triangulation on a similar model on a $\phi^{3}$ graph. Of course
these arguments rely strongly on the universality of the critical
behavior and support only the symmetry of the scale independent
characteristics. For example, the scale-dependent factors like
$L_{0}(D)$, $e_{0}(D)$ as well as the overall normalization of the
scaling function, can well be different in the low and high
temperature phases $\tau>0$ and $\tau<0$ respectively. We didn't
make this fact explicit above to render the equations more
transparent.

It is easy to relate all these scaling characteristics to the
observables of the field theory problem discussed in sect.2. Of
course the Liouville parameters $b$ and $a$ in (\ref{Lm}) are
\begin{align}
b &  =\sqrt{\frac{49-2D}{48}}-\sqrt{\frac{1-2D}{48}}\label{baD}\\
a &  =\sqrt{\frac{49-2D}{48}}-\sqrt{\frac{25-2D}{48}}\nonumber
\end{align}
The scaling function $z(t)$ defined in eq.(\ref{zt}) is, up to the
overall normalization and the normalization of the argument,
nothing but $F_{\text{sphere}}(y,D)$ in GRLIM. More precisely
\begin{equation}
f_{\text{sphere}}(y,D)=\frac{F_{\text{sphere}}(y,D)}{F_{\text{sphere}}%
(0,D)}=z(t)\label{fnorm}%
\end{equation}
where $y\sim t^{\rho}$ while the coefficient depends on the choice
of the scale (or precise definition of $N_{\text{c}}$). To be
slightly more explicit, let us introduce dimensional parameters
$a_{0}$ and $m_{0}$ which relate the area $A$ and the coupling
constant $m$ as defined in sect.2 to $N$ and $\tau$ of the
microscopic model
\begin{align}
A &  =a_{0}N\label{AN}\\
m &  =m_{0}\tau\nonumber
\end{align}
Let us also fix unambiguously the correlation size of the surface by
the relation
\begin{equation}
A_{\text{c}}=\pi m^{-2\rho}\label{Ac}%
\end{equation}
so that
\begin{equation}
y=\frac A{A_{\text{c}}}=t^{\rho}\label{yt}%
\end{equation}
and
\begin{equation}
L_{0}=\frac\pi{a_{0}m_{0}^{2\rho}}\label{L0}%
\end{equation}
In particular, the amplitude $e_{0}(D)$ in eq.(\ref{EDsing}) is
related to the universal parameter $f_{0}$ of eq.(\ref{f0})
\begin{equation}
L_{0}(D)e_{0}(D)=\pi f_{0}\label{e0f0}%
\end{equation}

Once the normalization is fixed we can compare the normalized GRLIM
scaling function $f_{\text{sphere}}(y,D)$ with our perturbative
prediction (\ref{zt}), (\ref{z0z2})
\begin{equation}
f_{\text{sphere}}(y,D)=1+z_{2}y^{2s}+z_{4}y^{4s}+\ldots\label{fexpy}%
\end{equation}
as well as with the asymptotic, controlled mainly by the parameter
$f_{0}$ from table \ref{table1}. In particular, in the exactly
solvable case $D=0$ this scaling function is extracted explicitly
from the matrix model treatment (see next section for more details)
\begin{equation}
f_{\text{sphere}}(y,0)=3^{2/3}\Gamma\left(  \frac23\right)  \operatorname*{Ai}%
\left(  \frac{3^{2/3}l_{\text{eg}}^{2}}4y^{2/3}\right) \label{fsphAiry}%
\end{equation}
This is the function plotted in the continuous line in
fig.\ref{plot34}. In (\ref{fsphAiry}) $l_{\text{eg}}$ is a numeric
constant
\begin{equation}
l_{\text{eg}}=2\gamma(1/3)\gamma^{2/3}(3/4)=1.919868134043972\ldots\label{leg}%
\end{equation}
caused by our particular definition of $N_{\text{c}}$ and $\operatorname*{Ai}%
(x)$ is the Airy function
\begin{equation}
\text{
}\operatorname*{Ai}(x)=\frac1{3^{2/3}}\sum_{n=0}^{\infty}\frac{\left(
-3^{1/3}x\right)  ^{n}}{n!\Gamma(2/3-n/3)}\label{Aiseries}%
\end{equation}
Notice that according to this power expansion of the Airy function,
$z_{4}=0$ at the pure Ising point $b^{2}=0.75$. In the last section of
this paper we are going to support this conclusion from the point of
view of the Liouville
gravity integral (\ref{z4int}).%

\begin{figure}
[tbh]
\begin{center}
\includegraphics[
height=3.7697in, width=4.6596in
]%
{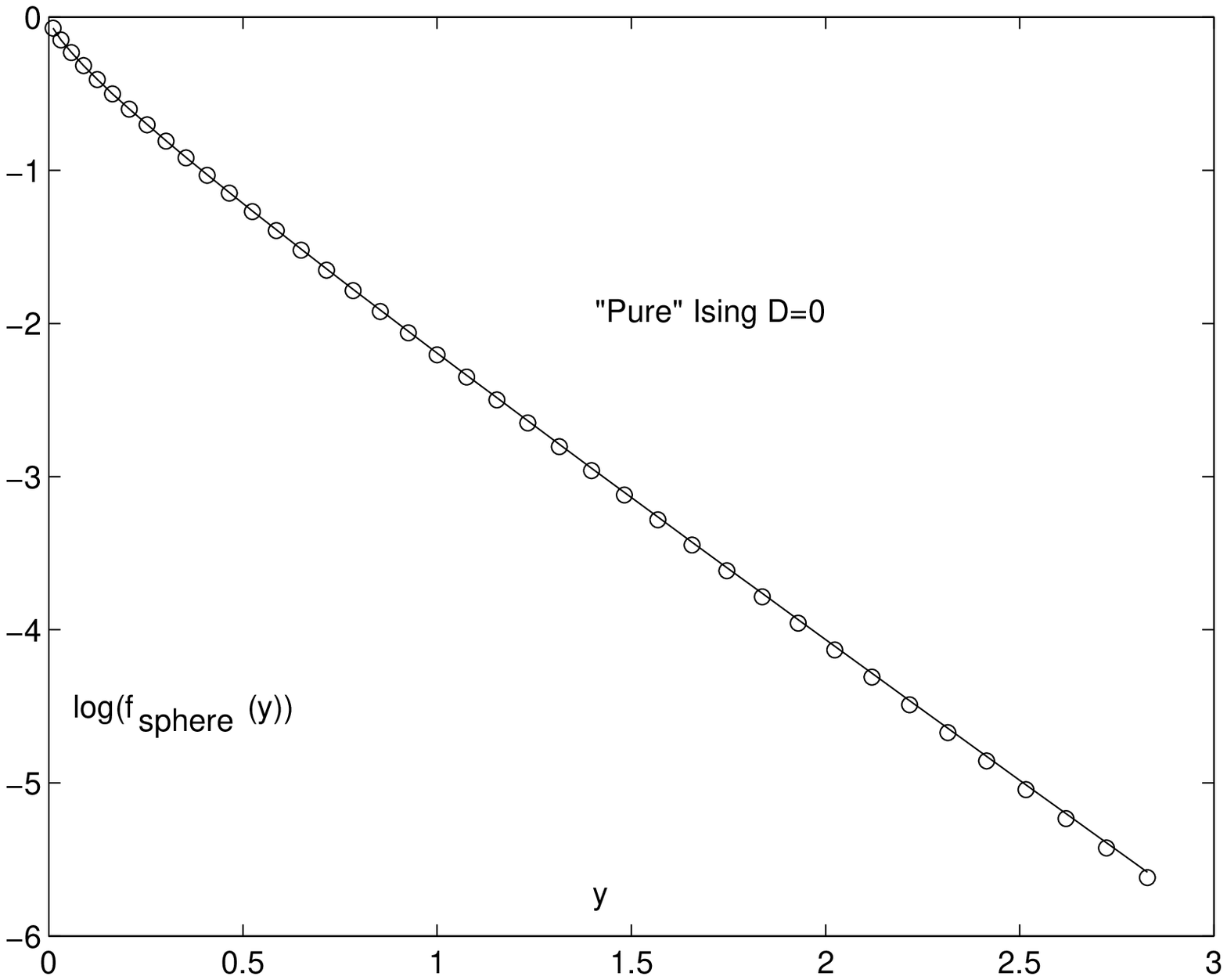}%
\caption{Logarithmic plot of $f_{\text{sphere}}(y)$ for the ``pure
Ising'' model $D=0$. Circles -- our approximation, continuous --
exact matrix model
function. }%
\label{plot34log}%
\end{center}
\end{figure}

\begin{figure}
[tbh]
\begin{center}
\includegraphics[
height=3.819in, width=4.8187in
]%
{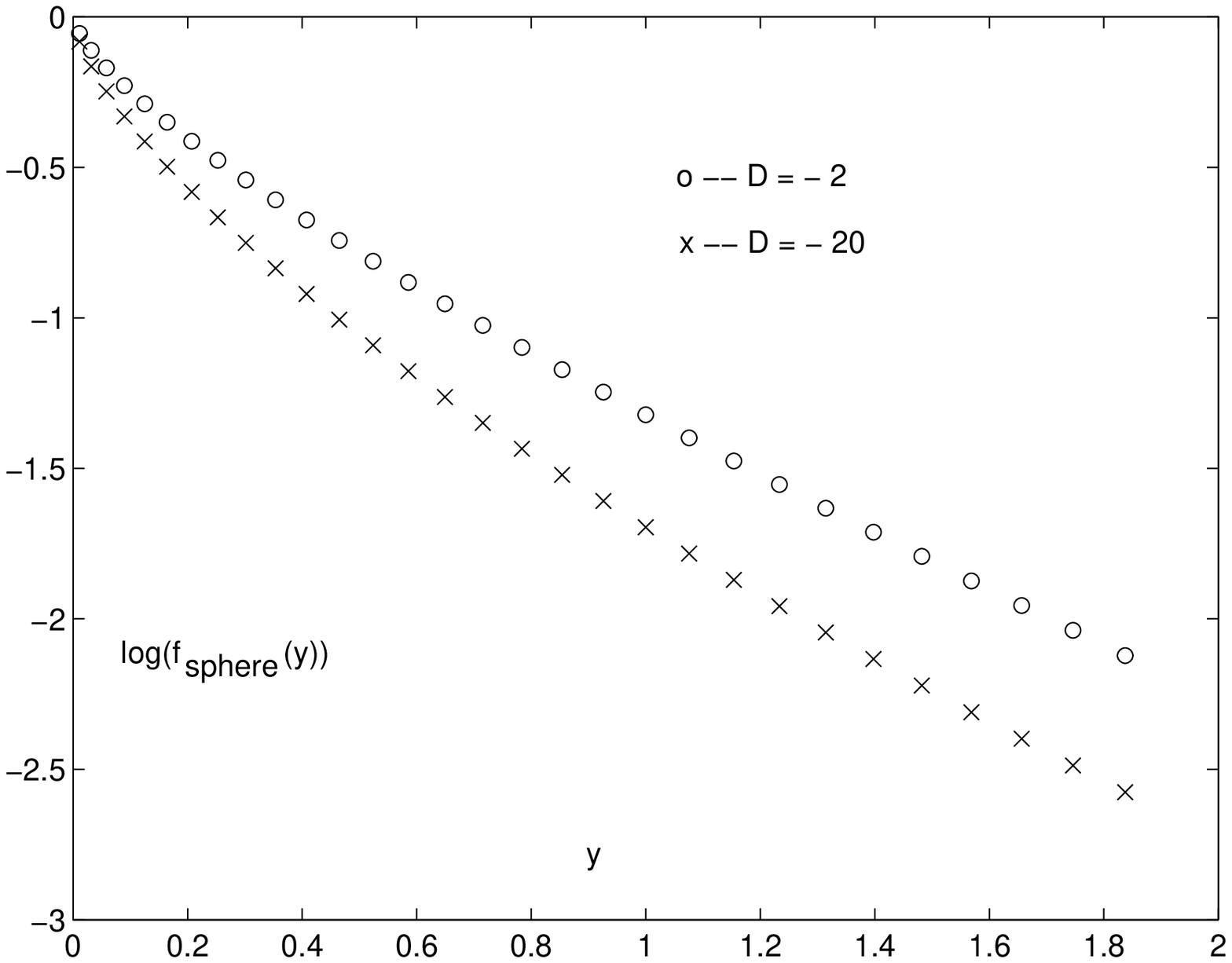}%
\caption{Our approximation for the scaling functions
$f_{\text{sphere}}(y,D)$
at $D=-2$ and $D=-20$.}%
\label{plot24}%
\end{center}
\end{figure}

The essence of the exact solution to the matrix model in the
continuous limit will be discussed in the next section. What we
would like to stress here, is that the generalized random lattice
model at generic $D\leq1/2$ is still an interesting model of the
dynamical lattice statistics. Although our field theoretic approach
involves many poorly justified conjectures and assumptions, it
provides very definite predictions for measurable observables in
GRLIM, equally well at $D=0$ and $D\neq0$, where the famous matrix
models (sometimes traded as a substitute for the field theory) fail
to answer any questions. Numerical coincidence or not, our data
presented in fig.\ref{plot34} give astonishingly good and detailed
description of the exact matrix model scaling function.
Fig.\ref{plot34log} presents similar comparison for
$f_{\text{sphere}}(y,0)$ for physically relevant positive $t$ and
for much larger values of $y$ where the exponential asymptotic
(\ref{Aass}) is clearly seen. Notice that this plot (in the
logarithmic scale) follows the falloff of $f_{\text{sphere}}(y,0)$
in at least $3$ orders of magnitude. In fig.\ref{plot24} we also
present our prediction for the scaling function
$f_{\text{sphere}}(y,D)$ for $D=-2$ ($b^{2}=0.530049\ldots$) and
$D=-20$
($b^{2}=0.191377\ldots$).%

\section{Minimal gravity point (pure Ising)}

This point corresponds to $b^{2}=3/4$ (or, in the spectator matter
language, to $c_{\text{sp}}=D=0$). The situation is supposed to be
described by the two matrix model by Kazakov \cite{KIsing,
Boulatov}. In the double scaling limit of this matrix model the
genus $0$ partition function $Z(T,x)$ is determined
through\footnote{At this point we, as well as other authors working
on the matrix models, are rather arbitrary about the precise
normalization of the partition function and the coupling constants.
Only the scale invariant combinations will matter.}
\begin{equation}
u(T,x)=Z_{xx}(T,x)\label{ZT}%
\end{equation}
where $u(T,x)$ is the solution to the following (renormalized)
algebraic equation \cite{KIsing}
\begin{equation}
x=u^{3}-\frac34T^{2}u\label{uxT}%
\end{equation}
Here $x$ is interpreted as the cosmological constant while $T$
corresponds to the off-critical temperature of the Ising system.
Precise relation between the parameters $T$ and $x$ in this section
and, respectively, $\tau$ and the activity conjugated to $N$ in the
previous one, depends on the microscopic realization of the RLIM. In
the particular formulation given in sect.8 these parameters can be
related using the explicit expressions of \cite{Burda}.

From (\ref{ZT}) and (\ref{uxT}) the expansion in $T^{2}$ follows
\begin{align}
Z(T,x) &  =-\frac{x^{7/3}}3\sum\frac{\Gamma(2n/3-7/3)}{n!\Gamma(2/3-n/3)}%
\left(  -\frac{3T^{2}}{4x^{2/3}}\right)  ^{n}\label{ZTexp}\\
&  =\frac{9x^{7/3}}{28}+\frac{9T^{2}x^{5/3}}{40}+\frac{3T^{6}x^{1/3}}%
{128}+\frac{3T^{8}}{1024x^{1/3}}+\ldots\nonumber
\end{align}
The fixed area $a$ partition function (we introduce new letter $a$
for this area instead of $A$, because the normalization of the
conjugated cosmological constant $x$ differs from that of $\mu$)
\begin{equation}
Z_{a}(T)=a\int_{\uparrow}Z(T,x)e^{ax}\frac{dx}{2\pi i}\label{Za}%
\end{equation}
where the contour $\uparrow$ is as in (\ref{ZA}). This is
transformed to
\begin{equation}
Z_{a}(T)=-\frac13a^{-7/3}\int_{\uparrow}\exp\left(  v^{2}-\frac34a^{2/3}%
T^{2}v^{2/3}\right)  \frac{dv}{\pi iv^{1/3}}\label{ZaT}%
\end{equation}
and then expressed in terms of the Airy function
\begin{equation}
Z_{a}(T)=-\frac{a^{-7/3}}{3^{1/3}}\operatorname*{Ai}\left(  \frac
{(3a)^{2/3}T^{2}}4\right) \label{Zairy}%
\end{equation}
The relation between the scales
\begin{equation}
Ta^{1/3}=l_{\text{eg}}m\left(  \frac A\pi\right)  ^{1/3}\label{Tm}%
\end{equation}
($l_{\text{eg}}$ is from eq.(\ref{leg})) is easily recovered
comparing the $T^{2}$ order of (\ref{ZTexp}) with $z_{2}$ of
(\ref{z0z2}). Finally (\ref{Zairy}) is reduced to
\begin{equation}
Z_{a}(T)=-3^{-1/3}A^{-7/3}\operatorname*{Ai}\left(  \left(  m\left(
3A/\pi\right)  ^{1/3}l_{\text{eg}}/2\right)  ^{2}\right) \label{Zleg}%
\end{equation}
Up to normalization this is equivalent to (\ref{fsphAiry}). In
particular
\begin{equation}
Z_{a}\sim\exp\left(  -l_{\text{eg}}^{3}m^{3}\frac A{4\pi}\right) \label{Zasim}%
\end{equation}
From this asymptotic we get
\begin{equation}
-f_{0}^{\text{(exact)}}=\frac{l_{\text{eg}}^{3}}{4\pi}=\frac{2\gamma^{3}%
(1/3)}{\pi\gamma^{2}(1/4)}=0.563124\ldots\label{f0exact}%
\end{equation}
the number quoted in table \ref{table1}.

\section{Four-point integral}

Here we consider on a preliminary footing the problem of the forth
order correction (\ref{z4int}). The matter ingredient of the
integrand is quite simple, see eq.(\ref{e4}). An explicit
representation of the Liouville four-point function can be taken
from \cite{AAl3}. Here we need only the ``symmetric'' case of this
function with all four external dimensions equal (and in fact equal
to $1/2$). For numerical purposes it is convenient to reduce the
holomorphic-antiholomorphic (spectral) integral to the form
\cite{LYgrav}
\begin{align}
\left\langle V_{a}(x)V_{a}(0)V_{a}(1)V_{a}(\infty)\right\rangle _{\text{L}%
}^{(A)}  & =\mathcal{R}_{a}^{(A)}\times\label{Va4}\\
& \int^{\prime}\frac{dP}{4\pi}r_{a}(P)\mathcal{F}_{P}\left(  \left.
\begin{array}
[c]{cc}%
1/2 & 1/2\\
1/2 & 1/2
\end{array}
\right|  x\right)  \mathcal{F}_{P}\left(  \left.
\begin{array}
[c]{cc}%
1/2 & 1/2\\
1/2 & 1/2
\end{array}
\right|  \bar x\right) \nonumber
\end{align}
where the overall factor is
\begin{equation}
\mathcal{R}_{a}^{(A)}=\frac{\left(
\pi\gamma(b^{2})b^{2-2b^{2}}A^{-1}\right)
^{(Q-4a)/b}}{\pi^{2}\Gamma(-b^{-2}-1+4b^{-1}a)}\frac{\Upsilon_{b}%
^{4}(b)\Upsilon_{b}^{4}(2a)}{\Upsilon_{b}^{4}(2a-Q/2)}\label{Ra3}%
\end{equation}
The weight function
\begin{equation}
r_{a}(P)=\frac{\pi^{2}\Upsilon_{b}(2iP)\Upsilon_{b}(-2iP)\Upsilon_{b}%
^{4}(2a-Q/2)}{\Upsilon_{b}^{2}(b)\Upsilon_{b}^{2}(2a-Q/2-iP)\Upsilon_{b}%
^{2}(2a-Q/2+iP)\Upsilon_{b}^{2}(Q/2-iP)\Upsilon_{b}^{2}(Q/2+iP)}\label{ra}%
\end{equation}
admits the following convenient integral representation
\begin{align}
r_{a}(P)  & =\sinh2\pi b^{-1}P\sinh2\pi bP\times\label{rint}\\
& \exp\left(  -8\int_{0}^{\infty}\frac{dt}t\frac{\sin^{2}Pt\left(
\cosh ^{2}(Q-2a)t-e^{-Qt}\cos^{2}Pt\right)  }{\sinh bt\sinh
b^{-1}t}\right) \nonumber
\end{align}
The prime near the integral sign in (\ref{Va4}) indicates possible
discrete terms. The general block function
\begin{equation}
\mathcal{F}_{P}\left(  \left.
\begin{array}
[c]{cc}%
\Delta_{1} & \Delta_{3}\\
\Delta_{2} & \Delta_{4}%
\end{array}
\right|  x\right)  =%
\raisebox{-0.4696in}{\includegraphics[ height=1.0395in,
width=2.2693in
]%
{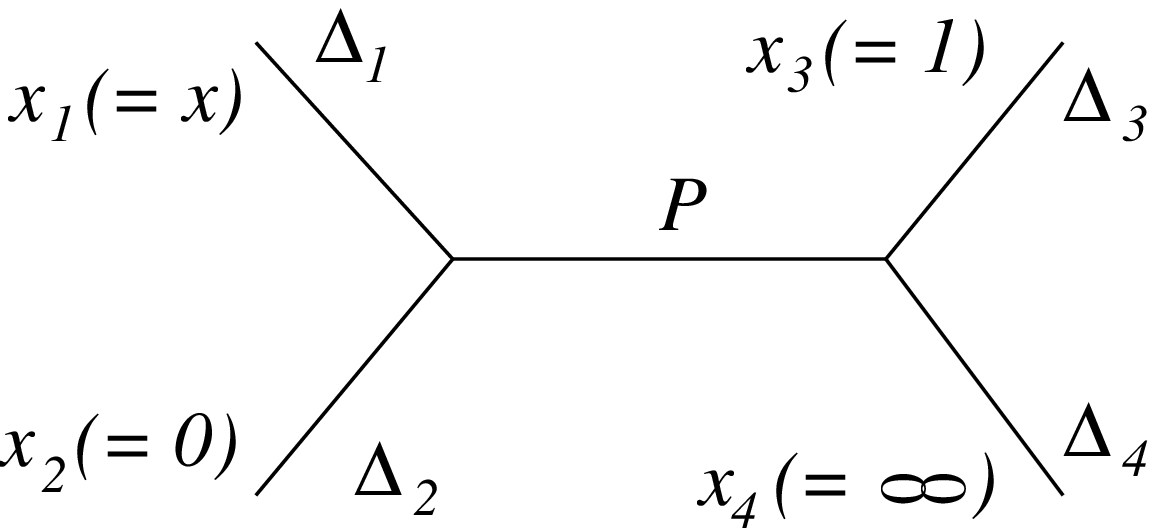}%
}%
\label{block3}%
\end{equation}
is effectively computable through the recursive relation of
\cite{Block}. In our problem we are interested in the particular
case $\Delta_{1}=\Delta _{2}=\Delta_{3}=\Delta_{4}=1/2$ but general
$P$.

It is easy to see that in our example of free fermion at least one
discrete term is always present in the r.h.s of (\ref{Va4}). It is
located at $iP=\pm(Q/2-2a)$. Moreover, if $b^{2}<b_{1}^{2}$, where
$b_{1}^{2}=\left( \sqrt{8/3}+1\right)  ^{-1}\approx0.38$, a second
discrete term appears at $iP=\pm(Q/2-2a-b)$, and so on. For the
values of $b^{2}$ below $b_{1}^{2}$ there are more discrete terms
and in the classical limit only these terms count. All these
complications temper the straightforward numerical analysis of the
integral (\ref{z4int}). The problem will be studied in more details
in a separate publication.

In the ``matrix model'' case $b^{2}=0.75$ the situation is
essentially simplified. First, it is easy to see that the discrete
term is singular near this point. Here $4a=Q-b$ and we are at the
first order resonance. However, in the fixed area correlation
function (\ref{CLA}) this pole is canceled out by the
$\Gamma$-function in the denominator and we are left with a finite
expression. Moreover, this denominator kills completely the residual
integral part and the only remaining contribution is the discrete
term. In this case it is not difficult to evaluate explicitly the
fixed area Liouville four-point function (see e.g., \cite{LYgrav})
\begin{equation}
\left\langle V_{a}(x)V_{a}(0)V_{a}(1)V_{a}(\infty)\right\rangle _{\text{L}%
}^{(A)}=\frac{4\left[  K(x)K(1-\bar x)+K(\bar x)K(1-x)\right]
}{b\left[
x\bar x(1-x)(1-\bar x)\right]  ^{1/6}}\label{Lres}%
\end{equation}
where
\begin{equation}
K(x)=\frac12\int\left[  t(1-t)(1-xt)\right]  ^{-1/2}dt\label{K3}%
\end{equation}
is the complete elliptic integral of the first kind. Thus, with the
matter four point function (\ref{e4}) our problem is reduced to the
integral
\begin{equation}
I(1/6)=\int\frac{\left(  1-x+x^{2}\right)  \left(  1-\bar x+\bar
x^{2}\right) }{x\bar x(1-x)(1-\bar x)}\frac{K(x)K(1-\bar x)+K(\bar
x)K(1-x)}{\left[  x\bar
x(1-x)(1-\bar x)\right]  ^{1/6}}d^{2}x\label{I16}%
\end{equation}
This integral is divergent and we want to evaluate its finite part.
In fact, our intention is to demonstrate that this finite part
vanishes. If it doesn't, this resonant contribution will show up as
the forth order term in the grand partition function (\ref{Zexp}) of
the form $m^{4}\mu\log\mu$. Such logarithmic contributions never
appear in the (mean-field-like) machinery of the matrix models. If
we believe that this $b^{2}=3/4$ point is indeed related to the
solvable matrix model, it is natural to expect a cancelation of this
logarithm as the result of the vanishing of the integral
(\ref{I16}).

For not to mess with the divergent contributions it is convenient to
consider
a more general parametric family of integrals $I(\nu)$%
\begin{equation}
I(\nu)=\int\frac{\left(  1-x+x^{2}\right)  \left(  1-\bar x+\bar
x^{2}\right) }{x\bar x(1-x)(1-\bar x)}\frac{K(x)K(1-\bar x)+K(\bar
x)K(1-x)}{\left[  x\bar
x(1-x)(1-\bar x)\right]  ^{\nu}}d^{2}x\label{Inu}%
\end{equation}
Although this integral is always divergent, at non-integer $\nu$ it
is safe to handle it formally. In the standard way we reduce it to
the contour integrals
\begin{equation}
I(\nu)=\frac1{2i}\int_{0}^{1}\frac{\left(  1-x+x^{2}\right)
K(x)}{x^{1+\nu }(1-x)^{1+\nu}}dx\int_{C}\frac{\left(
1-y+y^{2}\right)  K(y)}{y^{1+\nu
}(1-y)^{1+\nu}}dy\label{Icont}%
\end{equation}
where $C$ goes from $-\infty$ to $-\infty$ around $0$
counterclockwise. Using the integration formulas
\begin{align}
\int_{0}^{1}\frac{\left(  1-x+x^{2}\right)
K(x)}{x^{1+\nu}(1-x)^{1+\nu}}dx  &
=\frac{\left(  1-2\nu\right)  \left(  1-6\nu\right)  }{(1-4\nu)^{2}}%
\frac{2^{2\,\nu}\pi^{2}\,\Gamma^{2}(-\nu)}{\,\Gamma^{2}(3/4)\,\Gamma
^{2}(1/4-\nu)}\label{Kint}\\
\int_{0}^{1}\frac{(1-t+t^{2})K(t)}{t^{1+\nu}(1-t)^{3/2-2\nu}}dt  &
=\frac{\left(  1-2\nu\right)  \left(  1-6\nu\right)  }{(1-4\nu)^{2}}%
\frac{2^{2\nu-1}\pi\gamma(-\nu)\Gamma^{2}(3/4+\nu)}{\Gamma^{2}(3/4)}%
\;\nonumber
\end{align}
we arrive at the following compact expression
\begin{equation}
I(\nu)=\frac{2^{4\nu-9}\pi^{2}\,\left(  1-2\nu\right)  ^{2}\left(
1-6\nu\right)  ^{2}\gamma^{2}(\nu-1/4)}{\gamma^{2}(1+\nu)\gamma^{2}%
(3/4)}\label{Icompact}%
\end{equation}
Among many other interesting features, this integral shows a double
zero at $\nu=1/6$, which justifies the expected vanishing of
(\ref{I16}). We'd like to remind in this relation that our estimate
of the forth order coefficient $z_{4}$ through the sum rules
(\ref{z4as}) is also numerically quite close to zero, as it can be
seen in table \ref{table1}.

\textbf{Acknowledgments}

The authors thank H.Kawai for discussions and interest to their
work. Y.I. is grateful to I.Kostov and SPTh/CEA Saclay for their
help and hospitality. The same to M.Staudacher and the members of
Albert-Einstein Institute for their discussions and hospitality
during his stay in Potsdam. As usual, Al.Z was guided by the
Galchenok Lodestar. The work is partially a result of discussions
held between the authors during the visit of one of them at the
Theoretical Physics Laboratory of RIKEN. Thus Al.Z acknowledges the
hospitality and stimulating scientific atmosphere this theory group.
He was also supported by the European Committee under contract
EUCLID HRPN-CT-2002-00325.

\appendix

\section{Zeta regularization}

Let
\begin{equation}
z(t)=\sum_{n=0}^{\infty}z_{2n}t^{n}\label{zap}%
\end{equation}
be an entire function of order $\rho$ and $t_{n}$, $n=1,2,\ldots$
its zeros, which are supposed to be all real and negative. We
normalize it in the way that $z(0)=1$. To simplify the
considerations we assume that $1<\rho<2$, although the arguments are
straightforwardly extended to any finite order. The convergent
canonical product reads
\begin{equation}
z_{\text{can}}(t)=\prod_{n=1}^{\infty}\left(  1-\frac
t{t_{n}}\right)
\exp\left(  \frac t{t_{n}}\right) \label{zcan}%
\end{equation}
while apparently
\begin{equation}
z(t)=\exp(z_{2}t)z_{\text{can}}(t)\label{zzcan}%
\end{equation}
The zeta function of the zeros is defined as the
series\footnote{Please do not to mix $s$ here with $ab^{-1}$ in the
body of the paper.}
\begin{equation}
\zeta_{z}(s)=\sum_{n=1}^{\infty}\frac1{(-t_{n})^{s}}\label{zetaz}%
\end{equation}
convergent at $\operatorname*{Re}s>\rho$, otherwise being understood
as the analytic continuation. At $2>\operatorname*{Re}s>\rho$ it can
also be defined through the canonical product as
\begin{equation}
\zeta_{z}(s)=\frac{\sin\pi s}\pi\int_{0}^{\infty}t^{-s}\frac{d\log
z_{\text{can}}(t)}{dt}dt\label{zint}%
\end{equation}
(we need to take here $z_{\text{can}}(t)$ instead of $z(t)$ in order
to avoid a divergency at $t=0$). Inversely
\begin{equation}
\log z_{\text{can}}(t)=\int_{\uparrow}\frac{\pi\zeta_{z}(s)}{\sin\pi s}%
t^{s}\frac{ds}{2i\pi s}\label{zcint}%
\end{equation}
where the contour of integration $\uparrow$ goes along the imaginary
axis to the right from the pole at $s=\rho$ and to the left from the
pole of the sine function at $s=2$. At the same time, similar pole
at $s=1$ contributes to the asymptotic of $\log z_{\text{can}}(t)$
at $t\rightarrow\infty$. Comparing with (\ref{zzcan}) we find that
\begin{equation}
z_{2}=\zeta_{z}(1)\label{z2z}%
\end{equation}
The divergent sum in eq.(\ref{z2sum}) should be understood as the
analytic continuation of (\ref{zetaz}).

\end{document}